%% file: astroph.tex
\documentclass[twocolumn]{aastex631}

\shortauthors{M. G. Walker et al.}
\usepackage{amsmath}
\usepackage{natbib}
\usepackage{graphicx}
\usepackage{longtable}
\usepackage[caption=false]{subfig}
\newcommand{\kms}{km s$^{-1}$}
\newcommand{\vlos}{$V_{\rm LOS}$}
\newcommand{\teff}{$T_{\rm eff}$}
\newcommand{\logg}{$\log g$}
\newcommand{\feh}{[Fe/H]}
\newcommand{\mgfe}{[Mg/Fe]}
\newcommand{\lsf}{$\sigma_{\rm LSF}$}
\newcommand{\gaia}{\textit{Gaia}$\,$}
\newcommand{\vt}{\mbox{$v_{\rm t}$}}
\newcommand{\loggf}{\mbox{$\log(gf)$}}

\input{m2fs_error_adjust}%updated for medres
\input{hecto_error_adjust}%updated for medres
\input{offsets_newcommands}%updated for medres
\input{summary_stats}%updated for medres
\input{sspp_offsets}%updated for medres
\input{m2fs_obs_newcommands}%updated for medres

\input{hecto_obs_newcommands}% updated for medres}
\input{flags_newcommands}%updated for medres
\input{nmember}

\DeclareCaptionFormat{cont}{#1 #2#3\par}

\begin{document}

\title{Magellan/M2FS and MMT/Hectochelle  Spectroscopy of Dwarf Galaxies and Faint Star Clusters within the Galactic Halo\footnote{This paper presents data gathered with the Magellan Telescopes at Las Campanas Observatory, Chile, and the MMT Observatory, a joint
facility of the Smithsonian Institution, and the University of Arizona.}}

\correspondingauthor{Matthew G. Walker}
\email{mgwalker@cmu.edu}

\author[0000-0003-2496-1925]{Matthew G. Walker}
\affiliation{McWilliams Center for Cosmology, Carnegie Mellon University, 5000 Forbes Ave, Pittsburgh, PA 15213, USA }

\author[0000-0003-2352-3202]{Nelson Caldwell}
\affiliation{Harvard-Smithsonian Center for Astrophysics, 60 Garden Street, MS-15, Cambridge, MA 02138, USA}

\author[0000-0002-3856-232X]{Mario Mateo}
\affiliation{Department of Astronomy, University of Michigan, Ann Arbor, MI 48109, USA}

\author[0000-0002-7157-500X]{Edward W. Olszewski}
\affiliation{Steward Observatory, The University of Arizona, 933 N. Cherry Avenue, Tucson, AZ 85721, USA}

\author[0000-0002-6021-8760]{Andrew B. Pace}
\affiliation{McWilliams Center for Cosmology, Carnegie Mellon University, 5000 Forbes Ave, Pittsburgh, PA 15213, USA }

\author[0000-0002-4272-263X]{John I. Bailey, III}
\affiliation{Department of Physics, University of California Santa Barbara, Santa Barbara, CA 93016, USA}

\author[0000-0003-2644-135X]{Sergey E. Koposov}
\affiliation{Institute for Astronomy, University of Edinburgh, Royal Observatory, Blackford Hill, Edinburgh EH9 3HJ, UK}
\affiliation{Institute of Astronomy, University of Cambridge, Madingley Road, Cambridge CB3 0HA, UK}
\affiliation{Kavli Institute for Cosmology, University of Cambridge, Madingley Road, Cambridge CB3 0HA, UK}
\affiliation{McWilliams Center for Cosmology, Carnegie Mellon University, 5000 Forbes Ave, Pittsburgh, PA 15213, USA }

\author[0000-0001-5107-8930]{Ian U. Roederer}
\affiliation{Department of Astronomy, University of Michigan, Ann Arbor, MI 48109, USA}
\affiliation{Joint Institute for Nuclear Astrophysics---Center for the Evolution of the Elements (JINA-CEE), USA }

\begin{abstract}
We present spectroscopic data for \allgoodstar\ stellar targets within and/or toward \allsystems\ dwarf spheroidal galaxies and faint star clusters within the Milky Way halo environment.  All spectra come from observations with the multi-object, fiber-fed echelle spectrographs M2FS at the Magellan/Clay telescope or Hectochelle at the MMT, reaching a typical limiting magnitude $G\la 21$.  Data products include processed spectra from all observations and catalogs listing estimates---derived from template model fitting---of line-of-sight velocity (median uncertainty \vlosmedianerror\ \kms) effective temperature (\teffmedianerror\ K), (base-10 logarithm of)  surface gravity (\loggmedianerror\ dex in cgs units), \feh\ (\fehmedianerror\ dex) and \mgfe\ (\mgfemedianerror\ dex) abundance ratios.  The sample contains multi-epoch measurements for \allrepeated\ sources, with up to \allmaxgoodobs\ epochs per source, enabling studies of intrinsic spectroscopic variability.  The sample contains \ngianttotal\ likely red giant stars (based on surface gravity), and \nmembertotal\ likely members (based on line-of-sight velocity and \textit{Gaia}-measured proper motion) of the target systems.  The number of member stars per individual target system ranges from a few, for the faintest systems, to $\sim 850$ for the most luminous.  For most systems, our new samples extend over wider fields than have previously been observed; of the likely members in our samples, \nmemberfartotal\ lie beyond $2\times$ the projected halflight radius of their host system, and \nmemberfarfivetotal\ lie beyond $5R_{\rm half}$.
\end{abstract}

\section{Introduction}

The Galactic halo teems with stellar substructure.  This local environment provides our clearest window onto the processes of galaxy formation and the nature of dark matter.  The hierarchy of surviving Halo substructures stretches from the smallest scales, where diffuse star clusters overlap in luminosity with the faintest, most primitive dwarf galaxies \citep{gilmore07,martin08}, to the readily-visible and star-forming Magellanic Clouds.  All of these objects are in various stages of dissolution within the Galactic Halo, where ghosts of their earlier-infalling cousins remain detectable by their stellar-orbital configurations and chemical composition \citep[e.g.,][]{belokurov18,helmi18b,naidu20}.    

\smallskip
Known Halo substructures exhibit a wide range of properties that reveal details of their own formation, internal structure, and chemical evolution \citep{helmi20}.  The abundance and systemic motions of Halo substructures can be used to trace and characterize the Galaxy's extended dark matter halo.  The internal kinematics of individual substructures---dwarf galaxies, stellar streams and stellar overdensities---trace dark matter on the smallest scales where it is known to exist \citep{aaronson83,mateo93,willman11}.  The chemical abundance patterns of constituent stars reflect the processes at work in the earliest stages of cosmic star formation  \citep{tolstoy09,weisz17}.

\smallskip
Over the past several decades, spectroscopic studies of individual stars within the Milky Way's surviving satellites have developed in fits and starts.  Early campaigns used 2-4m class telescopes to target red giant candidates in the Milky Way's $\sim 10$ `classical' dwarf spheroidal companions, building line-of-sight velocity samples for a few to several tens of member stars per system \citep[e.g.,][]{aaronson83,olszewski85,mateo91,mateo93,hargreaves94a,hargreaves94b,hargreaves96a,olszewski95}.  With the advent of multi-object fiber spectrographs, samples grew to $\sim 100$ members per system \citep[e.g.,][]{kleyna02,wilkinson04}.  Ultimately, multi-object spectrographs at 6-10m class telescopes enabled samples not only of line-of-sight velocity, but also chemical composition for several hundreds to a few thousand members per system \citep[e.g.,][]{tolstoy04,koch06,battaglia06,koch07,koch07b,walker09a,kirby10}.; for a few bright confirmed member stars, higher-resolution followup could then measure detailed abundance patterns \citep[e.g.,][]{shetrone01,letarte07,aoki09,cohen09,lucchesi20}  Meanwhile, the same instrumentation provided samples reaching a few to tens of members per each of the low-luminosity ($M_V\ga -6$), `ultra-faint' Milky Way satellites that were revealed by, e.g., the Sloan Digital Sky Survey, Pan-STARRs and the Dark Energy Survey \citep[e.g.,][]{kleyna05,munoz06b,martin07,simon07,koposov11}.  

\smallskip
These observational datasets have delivered a wealth of information about the systemic motions and internal chemo-dynamical properties of the Milky Way satellite population; for review articles, see \citet{mateo98,tolstoy09,mcconnachie12,simon19,battaglia22,belokurov22}.  However, the available datasets leave room for substantial improvement.  First there is the obvious statistical improvement that would come with even larger samples and higher (spectroscopic) resolution.  There is also the systematic improvement that would come with expanded spatial and temporal sampling.  Due to finite field sizes, the oldest, lowest-metallicity, most weakly bound outermost member stars are under-represented in nearly all existing spectroscopic samples of dwarf galaxies.  While recent observational campaigns are beginning to focus on outer regions \citep[e.g.][]{waller23,sestito23,tolstoy23}, most measurements of stellar velocity and metallicity distributions, as well as formation histories, remain biased toward central values where stellar populations skew younger, kinematically colder and more chemically evolved \citep{tolstoy09}.  Furthermore, the lack of multi-epoch observations for most stars precludes knowledge of intrinsic variability, limiting the accuracy with which, e.g., intrinsic velocity distributions (and hence dynamical masses) can be inferred \citep[e.g.,][]{mcconnachie09}.  

\smallskip
With the goal of overcoming these and other limitations, we are using wide-field, high-resolution, multi-object spectrographs at the 6.5m MMT and Magellan telescopes in the northern and southern hemispheres, respectively, to conduct a spectroscopic campaign that targets the known dwarf galaxies and faint star clusters within the Galactic halo.  Compared to previous efforts, our current observations provide higher spectroscopic resolution, wider spatial coverage and/or multi-epoch temporal coverage.  Here we describe the observations, data processing and quality, and release processed spectra and data catalogs from our ongoing programs.    

\section{Observations}
\label{sec:observations}
We present results from spectroscopic observations of \allsystems\ dwarf galaxies and star clusters within the Galactic Halo, conducted over portions of more than 200 clear nights during the years 2005 -- 2022.  All observations use multi-object, fiber-fed echelle spectrographs at one of two telescopes.  We observe northern targets using the Hectochelle spectrograph \citep{szentgyorgyi06} at the 6.5m MMT Observatory in Arizona, and southern targets using the M2FS spectrograph \citep{mateo12} at the 6.5m Magellan/Clay telescope at Las Campanas Observatory in Chile. 

\subsection{Target Selection}
The quantity and quality of imaging data available for selecting spectroscopic targets have evolved dramatically over the course of our observations.  Our earliest spectroscopic targets were chosen based on our own two-filter photometry, which was limited to relatively central regions of the most luminous dwarf galaxies \citep[e.g.,][]{mateo08}.  Others use more recent data sets from observational campaigns---e.g., the PRISTINE survey \citep{starkenburg17}---that target individual systems.   In one case---M2FS observations of the Reticulum II dwarf galaxy---we received targeting coordinates directly from the Dark Energy Survey's Milky Way working group, which had just discovered Reticulum~II based on  their then-proprietary photometric catalogs \citep{des15}. 
 Most recently, we select targets based on public data from large sky surveys---e.g., SDSS \citep{sdssdr9}, PanSTARRs \citep{ps1}, DES \citep{desdr2}---that provide multi-color photometry and, with the \gaia\ mission, precise and time-dependent astrometry over wide fields \citep{gaia16,gaiadr3}.  In the special case of recent observations of star clusters at low Galactic latitude, we select targets based entirely on photometry and astrometry from \textit{Gaia} \citep{pace23}.  

\smallskip
One consequence of this progress is that our spectroscopic targeting criteria are heterogeneous, varying not only from system to system, but also across different fields and/or different epochs within a given system.  Thus we cannot provide a rigorous and consistent selection function that accounts for the sampling that produced the spectroscopic data sets presented herein.  Instead, here we describe our general approach to selecting spectroscopic targets, and how that approach has evolved in response to advances in imaging surveys.  In any case, our data products include coordinates of all observed spectroscopic targets regardless of data quality, allowing users to infer effective selection functions where necessary.  

\smallskip
For nearly all of the stellar systems studied here, the member stars that are sufficiently bright for spectroscopy (magnitude $G\lesssim 21$) are post-main-sequence stars on the red giant, subgiant and horizontal branches.  At distances ranging from tens to hundreds of kpc, stars at these evolutionary stages have broad-band colors and magnitudes that are similar to those of late-type dwarf stars in the Galactic foreground.  Our general strategy for target selection is first to use available photometry to identify these sequences of evolved stars along the line of sight to the system of interest, then to use additional information (e.g., parallax and proper motion), where available, to filter out likely foreground contaminants.  

\smallskip
More specifically, since proper motion data became available with \textit{Gaia's} second data release \citep{gaiadr2}, we select spectroscopic targets according to the following procedure.  First, we use wide-field survey photometry (e.g., SDSS, DES, PanSTARRS, etc.) to identify red giant, horizontal and subgiant branch candidates as likely point sources (based on survey-specific criteria, e.g., requiring  TYPE=6 for SDSS photometry, $\mathrm {|wavg\_spread\_model\_r|}<0.003$ for DES data) having $g$-band magnitudes and $g-r$ colors within $\delta$ magnitudes of a best-fitting (by eye) theoretical isochrone \citep{dotter16}.  The tolerance $\delta=\sqrt{\delta^2_{\rm err}+\delta^2_{\rm min}}$ is set by the observational error, $\delta_{\rm err}$, associated with the photometric color, and a  minimum tolerance that takes a typical value of $\delta_{\rm min}=0.2$ mag.  Next we identify the photometrically-selected stars for which \gaia\ measures a parallax that is unresolved (parallax angle is smaller than 3 times its observational error), and a proper motion that is consistent, given observational errors, with the systemic mean \citep[e.g.,][]{helmi18,pace19}.  Given the list of prospective spectroscopic targets that pass these photometric and astrometric filters, we select randomly from those that lie within the available field of view of a given telescope pointing.  For systems that extend beyond a single telescope pointing, our choice of pointing is based on competing interests in 1) observing large numbers of high-probability member stars, which favors central fields, 2) fairly sampling across the target system, which requires outer fields where member stars can be scarce, and 3) obtaining sufficient repeat measurements to gauge observational errors and intrinsic  variability.  Finally, we note that photometric and/or astrometric filter tolerances can be adjusted based on target density in order to make use of available fibers.  

\smallskip
Figures \ref{fig:m2fs_cmdmap} and \ref{fig:hecto_cmdmap} of the Appendix display sky positions, color-magnitude diagrams (CMDs),  proper motion coordinates and our own measurements of metallicity, \feh,  vs. (heliocentric) line-of-sight velocity, \vlos, for spectroscopic targets toward each Galactic satellite that we observe.  As discussed above, our actual target selection used a variety of different photometric data sets; however, for uniformity of presentation the plotted CMDs all use \textit{Gaia's} $G$-band photometry and integrated BP-RP spectra, with extinction corrections applied according to the procedure described by \citet{babusiaux18}.  Overplotted in the CMDs are theoretical isochrones \citep{dotter16,isochrones} computed for old (age=10  Gyr) stellar populations and published values of metallicity for each object \citep[e.g.,][]{mcconnachie12}.  
Ellipses in the sky maps have semi-major axes $a=2R_{\rm half}/\sqrt{1-\epsilon}$, where $R_{\rm half}$ is the projected halflight radius and $\epsilon\equiv 1-b/a$ is the measured ellipticity.  In the proper motion and \feh\ vs. \vlos\ panels, dashed lines indicate previously-published values for systemic mean proper motions and velocities \citep{pace22}, where available.  

\subsection{Magellan/M2FS}
The Michigan/Magellan Fiber System (M2FS; \citealt{mateo12}) is a fiber-fed, double spectrograph operating at the f/11 Nasmyth port of the Magellan/Clay 6.5 m telescope at Las Campanas Observatory, Chile.  Each of the two M2FS spectrographs receives light from up to 128 fibers.  A wide field corrector provides good image quality over a field of diameter 30 arcmin.  Fibers can operate at wavelengths between 3700 - 9500 \AA, have entrance apertures of diameter 1.2 arcsec, and tolerate center-to-center target separations as small as 12 arcsec.  M2FS observers plug fibers by hand into masks that are machined at the Carnegie Observatories machine shop.  Depending on choice of diffraction grating and order-blocking filters, M2FS offers a wide range of observing configurations, with spectral resolution ranging from $\mathcal{R}\sim [0.2 - 34]\times 10^3$ and wavelength coverage ranging from tens to thousands of \AA. 

\smallskip
For the vast majority of M2FS observations reported here, we use the high-resolution (`HiRes' hereafter) grating with both spectrographs, and with filters selected to pass light over a single order at $5130 \lesssim \lambda\lesssim 5190$~\AA.  The most prominent feature in this region is the Mg~\textsc{i} `b'  triplet, with rest wavelengths of 5167.32~\AA, 5172.68~\AA, and 5183.60~\AA.  This region also contains many iron lines that enable a direct measurement of iron abundance.  With these choices, we acquire single-order spectra for up to 256 sources per pointing, with resolving power $\mathcal{R} \sim 24,000$.  We bin the detector at $2\times 2$ pixels$^2$, giving plate scale $\sim 0.065$ \AA/pixel over the useful wavelength range.  

\smallskip
For a small fraction of M2FS observations reported here, we use an alternative configuration that has at least one of the two spectrographs using a medium-resolution (henceforth `MedRes') grating that gives resolving power $\mathcal{R}\sim 7000$.  In order to cover the Mg triplet region, we use an order-blocking filter that passes light over the range 5115--5300~\AA.  Using the same $2\times 2$ binning that we use with the HiRes grating, the MedRes observations have plate scale $\sim 0.2$ \AA/pixel over the useful wavelength range.  

\smallskip
During a typical observing night with M2FS, we take 100-200 zero-second `exposures' in order to measure the bias levels of the detectors in both spectrographs.  We take between 3-10 exposures of the (scattered) solar spectrum during evening and/or morning twilight.  For a typical science field, we expose for 1-3 hours, broken into 2-5 sub-exposures.  Of the 256 available fibers, we assign $\sim 30$ to regions of blank sky.  
Immediately before and after science exposures, and often between sub-exposures, we acquire calibration spectra of an LED source and then a ThArNe arc lamp, both of which are located at the secondary cage and illuminate the fibers at the focal surface.  During daylight hours, we acquire sequences of hour-long `dark' exposures with both spectrographs' shutters closed.  

\smallskip
Table \ref{tab:m2fs_obs_table} lists the instrument configuration, central field coordinates, date, total exposure time and number of targets for all M2FS science fields observed for our program thus far.  Including repeat observations, we have observed a total of \mtwofsnfieldstot\ science fields with M2FS for this program---\mtwofsnfieldshires\ with both spectrographs using the HiRes grating, \mtwofsnfieldsmedres\ with both using the MedRes grating, and \mtwofsnfieldsmix\ with one spectrograph in HiRes mode and the other in MedRes mode---for a total science exposure time of \mtwofsexptimetot\ megaseconds (Ms).  We obtain acceptable M2FS HiRes spectroscopic measurements for $\sim$\mtwofshiresgoodstark k unique sources within \mtwofshiressystems\ different target systems, and we obtain acceptable M2FS MedRes measurements for $\sim$\mtwofsmedresgoodstar\ unique sources within \mtwofsmedressystems\ different systems.  For $\sim$\mtwofsrepeatedk k M2FS sources we have (up to \mtwofsmaxgoodobs\ per source) multiple independent measurements.  

\input{m2fs_obs_table_short}

\subsection{MMT/Hectochelle}
Hectochelle is a fiber-fed echelle spectrograph at the f/5 focal surface of the MMT Observatory on Mt. Hopkins, Arizona, United States \citep{szentgyorgyi06}.  Hectochelle's optical fibers have entrance apertures of diameter 1.5 arcsec, and are positioned robotically, allowing simultaneous observation of up to 240 distinct sources.  A wide field corrector, coupled with an atmospheric dispersion compensator, gives a field of view of diameter 1 degree.  Hectochelle spectra consist of a single diffraction order spanning $\sim 150$~\AA, with resolving power $\mathcal{R}\sim 32$,000 at wavelength $\lambda\sim 5200$~\AA.  We use Hectochelle's `RV31' order-blocking filter, which isolates the wavelength range 5150--5300~\AA.  We bin the detector by factors of 2 and 3 in the spectral and spatial dimensions, respectively, giving plate scale $\sim 0.10$ \AA/pixel.   

\smallskip
Our observing strategy with Hectochelle is similar to the one described above for M2FS.  On a typical night we acquire $\sim 100$ zero-second bias `exposures', plus exposures of the scattered solar spectrum during evening and/or morning twilight.  As with M2FS, for a given science field we acquire between 2-5 sub-exposures totalling 1-3 hours of integration time.  Before and after science exposures we acquire spectra of a ThAr arc lamp.  Either before or after science exposures, we acquire the spectrum of a quartz lamp.  The observatory staff acquires dark exposures regularly during daylight hours.  

\smallskip
Table \ref{tab:hecto_obs_table} lists the same information as Table \ref{tab:m2fs_obs_table}, but for Hectochelle observations.  With Hectochelle we have observed a total of \mtwofsnfieldstot\ (including repeat observations) science fields for a total science exposure time of \hectoexptimetot\ Ms.  We obtain acceptable measurements for $\sim$\hectogoodstark k unique sources within within \hectosystems\ target systems.  For $\sim$\hectorepeatedk k sources we have (up to \hectomaxgoodobs\ per source) multiple independent measurements.  

\input{hecto_obs_table_short}

\section{Processing of raw spectra}
\label{sec:processing}

All MMT/Hectochelle spectra are processed using the standard TDC pipeline\footnote{\url{https://lweb.cfa.harvard.edu/mmti/hectospec/hecto_pipe_report.pdf}} which is written in IDL.  
Briefly, the four channels from two CCDs are corrected for bias and merged. Cosmic rays are then detected and interpolated over, and individual exposures are coadded. Dark structure is subtracted depending on the exposure time, and spectra are extracted in the manner described in the next section.

\smallskip
The remainder of this section describes the set of Python-based modules that we have written for end-to-end processing of Magellan/M2FS spectra.  Where applicable and convenient, we incorporate modules that are publicly available as part of the  Astropy software package \citep{astropy1,astropy2,astropy3}. 

\subsection{Overscan/bias/dark/gain corrections and  uncertainties}
\label{subsec:initial}

\smallskip
We begin by using the Astropy-affiliated package `ccdproc' \citep{ccdproc} to perform standard corrections for overscan, bias, dark current and gain.  We apply all of these corrections independently to images from each of the two M2FS channels and, for a given channel, to each of the $1024\times1028$ (plus $128\times128$ overscan) image sections read out via each detector's four independent amplifiers.  `ccdproc' replicates the tasks performed by the original IRAF \citep{iraf} package of the same name, but also calculates and stores a 2D array containing an estimate of the variance at each pixel.  

\smallskip
For each amplifier on each detector and for each M2FS run individually, we generate an image of the master bias level, denoted $B$, by averaging (after iteratively discarding $3\sigma$ outliers at the pixel level) $\ga 100$ zero-second (overscan-corrected) exposures.  We generate an image of the master dark current rate, denoted $D$, by averaging  (again with iterative $3\sigma$ outlier pixel rejection) the $\approx 250$ 3600-second dark exposures (after performing overscan correction and subtracting the run-dependent master bias image) taken over all M2FS runs\footnote{We do not generate a new master dark frame for each run because a given run permits the acquisition of only a few long dark exposures.} involving observations presented here.  For all individual exposures of interest, we then use `ccdproc' to perform overscan correction to obtain an image of raw counts, denoted $C$, and then to subtract estimates of the master bias and dark counts.  Finally, `ccdproc' applies the appropriate gain correction (typically $g\approx 0.68$ e$^-$/ADU) to obtain an estimate of counts in units of electrons:
\begin{equation}
    \hat{N}=g(\hat{C}-\hat{B}-t_{\rm exp}\hat{D}),
    \label{eq:count_obs}
\end{equation}
where $t_{\rm exp}$ is exposure time and we adopt the convention under which $\hat{X}$ denotes the estimate of $X$.

\smallskip
The variance estimated by `ccdproc' is by default the sum of the estimated gain-corrected count, $\hat{N}$, and the square of the read noise.  One problem with this estimate is that for weak signals, read noise can  dominate such that $\hat{N}$ and hence the estimated variance can be  negative.  Another problem with weak signals is that---even in the absence of read noise---the observed count skews toward values smaller than the expected count, and hence the variance, of a Poisson distribution.  For example, for expected counts of $1,10,100$, random draws from Poisson distributions will be smaller than the expectation value with probability 0.37, 0.46, 0.49, respectively, and larger than the expectation value with probability 0.26, 0.42, 0.47.  

\begin{figure}
    \includegraphics[width=3.5in,trim=0.15in 2.5in 2in 0.6in, clip]{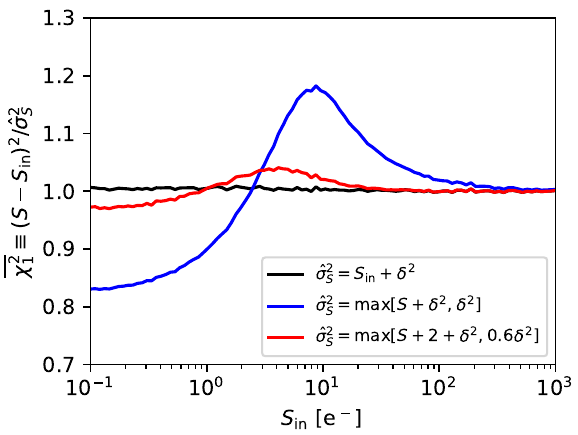}
    \caption{Mean value of $\chi^2_1\equiv (S-S_{\rm in})^2/\hat{\sigma}^2_N$ as a function of expected signal $S_{\rm in}$, from $10^6$ realizations at each input signal, assuming read noise $\sigma_r=2.6 e^-$; see Section \ref{subsec:initial} for details.  Curves show results for different estimators of the variance, $\hat{\sigma}^2$; black uses the true variance, blue uses a commonly-used estimator, red uses the estimator we use for real M2FS data (Equation \ref{eq:fudge}).
    \label{fig:variance_test}}
\end{figure}

\smallskip
Illustrating these problems and our ad hoc solution, Figure \ref{fig:variance_test} depicts the mean value, from $10^6$ trials over a range of input signals, of $\chi_1^2\equiv (S-S_{\rm in})^2/\hat{\sigma}^2_S$, where $S$ is a simulated observation, $\hat{\sigma}^2_S$ is an estimate of its variance, and $S_{\rm in}$ is the known input signal.  In each trial, the simulated observation is $S=S_0+\epsilon$, where\footnote{We use $\mathcal{P}_k(\lambda)$ to denote the Poisson distribution of number of occurrences, $k$, with expected value $\lambda$, and we use $\mathcal{N}_x(\mu,\sigma^2)$ to denote the normal distribution of random variable $x$, with expected value $\mu$ and variance $\sigma^2$.} $S_0\sim \mathcal{P}_k(S_{\rm in})$ is drawn from the Poisson distribution with expected value equal to the input signal, and $\epsilon\sim \mathcal{N}_x(0,\delta^2)$ is drawn from the normal distribution with mean 0 and variance $\delta^2$.  In our simulation, we set $\delta$ equal to the typical M2FS read noise of $\sigma_r=2.6\, e^{-}$, and we assume that any additional noise associated with, e.g., empirical estimation of bias and dark levels is negligible.  

\smallskip
The black curve in Figure \ref{fig:variance_test} indicates the mean values of $\chi_1^2$ that are calculated using the `true' variance, $\hat{\sigma}^2_S=\mathrm{Var}(S)=S_{\rm in}+\delta^2$.  As expected, use of the true variance gives mean $\chi^2_1$ values of unity; unfortunately, the true variance is inaccessible to the observer who does not know the input signal.  

\smallskip
The blue curve in Figure \ref{fig:variance_test} shows the result of estimating the variance as the observationally-accessible---and commonly used---quantity $\hat{\sigma}^2_S=\mathrm{max}(S+\delta^2, \delta^2)$.  The mean $\chi_1^2$ asymptotes to unity only at $N_{\rm in}\ga 100$.   For $\delta\la S_{\rm in}\la 100$, the aforementioned bias toward $S<S_{\rm in}$ gives mean $\chi_1^2>1$ as the true variance is underestimated.  For the smallest signals, $S_{\rm in}\la \delta$,  $\chi^2_1<1$ as the $\mathrm{max}$ operation causes the true variance to be overestimated on average.

\smallskip
The red curve in Figure \ref{fig:variance_test} shows the result of taking the variance to be
\begin{equation}
    \hat{\sigma}^2_S=\mathrm{max}(S+2+\delta^2,0.6\delta^2),
    \label{eq:fudge}
\end{equation}
a formula that we found, via experiment, to bring mean values of $\chi^2_1$ closer to unity at all input signals specifically when $\delta\approx 2.6 e^-$; for other values of the Gaussian noise, the constants in Equation \ref{eq:fudge} would need to be re-determined.

\smallskip
Based on the above experiment, we use Equation \ref{eq:fudge} to estimate the variance at each pixel of the real M2FS images.  In our application to real data, we take the Poisson component to be $S=\hat{N}+t_{\rm exp}\hat{D}$, the sum of estimated source (including background) and dark counts, and the Gaussian component to be  $\delta^2=\hat{\sigma}^2_r+\sigma^2_{\hat{B}}+t_{\rm exp}^2\sigma^2_{\hat{D}}$, the sum of contributions from the estimated read noise and noise associated with empirical estimates of bias and dark count levels.  

\smallskip
The estimated M2FS read noise is typically $\hat{\sigma}_r\approx 2.6$ e$^-$, as calculated from the mean standard deviation over all pixels within individual images contributing to the master bias frames.  The master dark frame indicates a mean dark current rate of $\hat{D}\approx 2.0$ e$^-$ hour$^{-1}$.  The run-dependent master bias frames and the global master dark frame have typical uncertainties of $\sigma_{\hat{B}}\approx 0.15$ e$^-$ and $\sigma_{\hat{D}}\approx 0.25$ e$^-$ hour$^{-1}$, respectively, calculated as the standard deviations over the individual calibration frames divided by the square root of the number of calibration frames, and converted to units of electrons.  

\smallskip
We reiterate that our application of Equation \ref{eq:fudge} represents an ad hoc solution to the problem of estimating variances of pixel counts directly from the data.  It is tuned specifically to produce  $\chi^2_1\approx 1$ at $N_{\rm in}\la 100$ electrons, given M2FS-like read noise; at other levels of read noise the form of Equation \ref{eq:fudge} would need to be re-determined.  We note that there exist alternative solutions; e.g., \citet{guy22} develop a full model of the CCD image in order to estimate the variance at each pixel.  

\smallskip
Finally, for each channel we stitch together the four independently-processed  sections read by each amplifier in order to obtain a single image of size $2048$ (columns) $\times 2056$ (rows) square pixels.  Figure \ref{fig:m2fs_image} displays examples of the stitched frames obtained for four types of exposures, with illumination by: LED (top-left), twilight sky (top-right), Thorium-Argon-Neon lamp (bottom-left), and target stars  (bottom-right). 
 Single-order spectra appear as horizontal bands, each spanning 5130--5190~\AA\ over columns 300--1400. 
 Signals outside this range are contributed by light from adjacent orders, which we discard (see below).

\subsection{Identification and Tracing of Spectral Apertures}
\label{subsec:identify/trace}

\begin{figure*}
    \begin{tabular}{@{}llll@{}}
    \renewcommand{\arraystretch}{0}
%    \vspace{-1.5in}
    \includegraphics[width=4in,trim=0.8in 0.1in 0in 0in, clip]{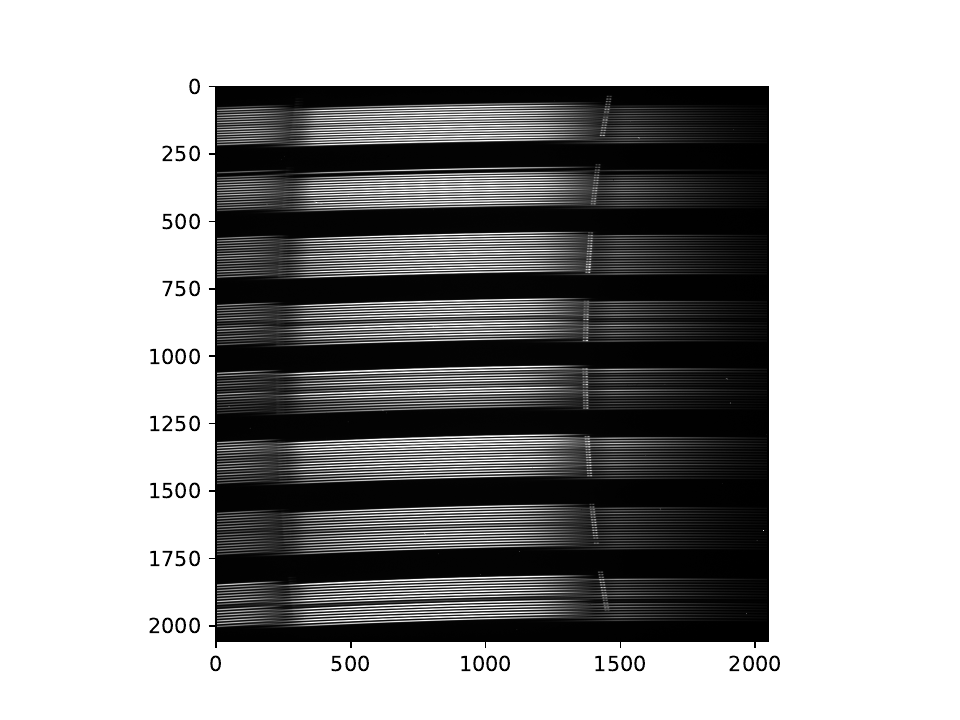}& \hspace{-1in}\includegraphics[width=4in,trim=0.8in 0.1in 0in 0in, clip]{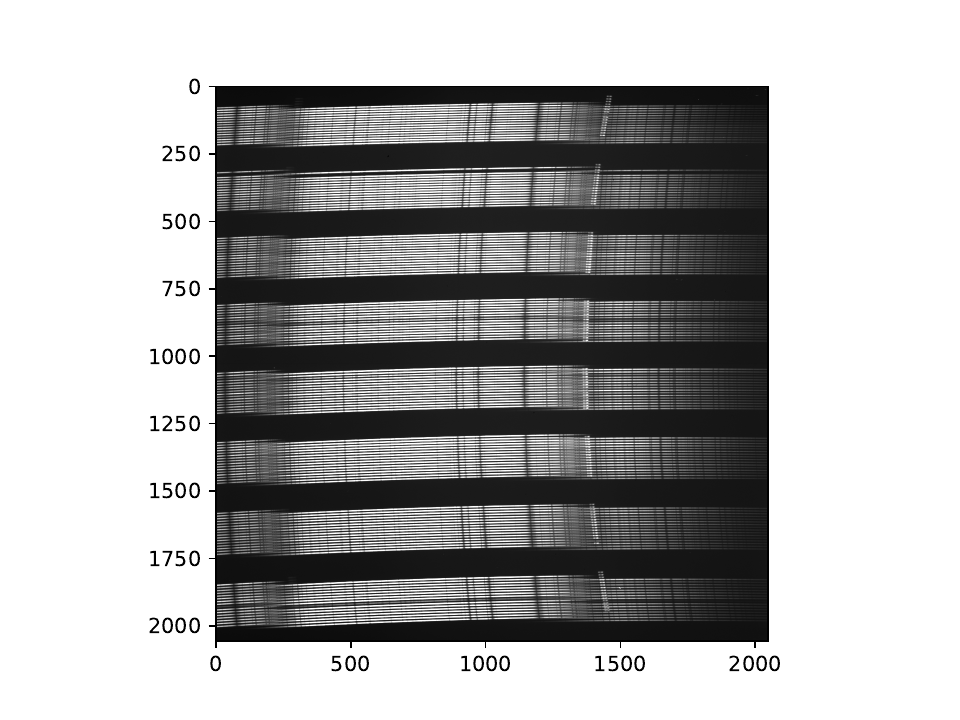}\\
%    \vspace{-1.5in}
    \includegraphics[width=4in,trim=0.8in 0.1in 0in 0in, clip]{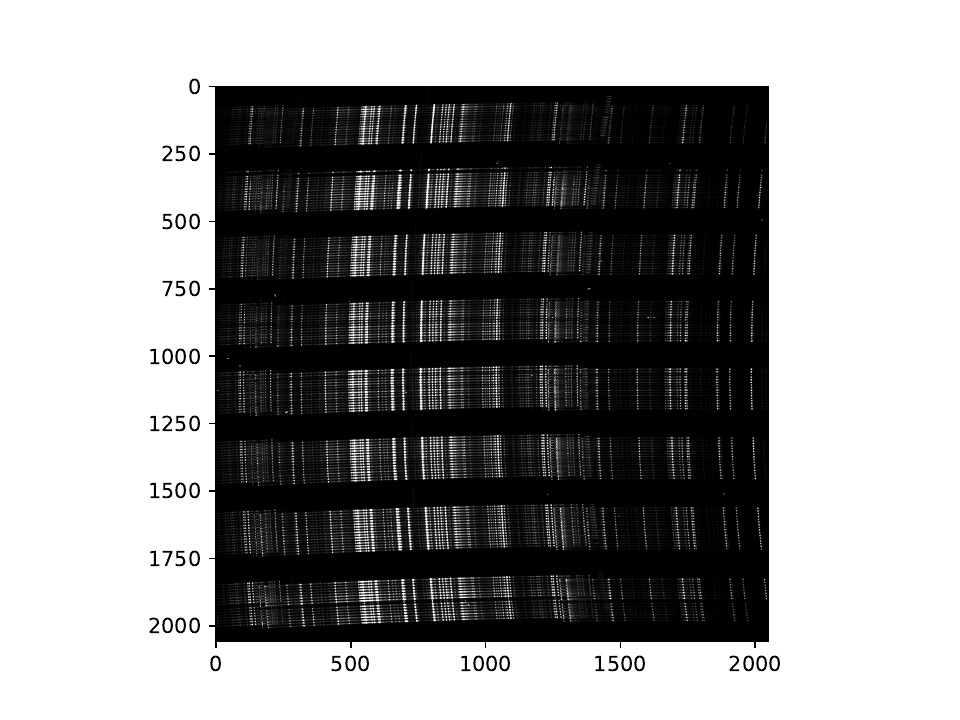}& \hspace{-1in}\includegraphics[width=4in,trim=0.8in 0.1in 0in 0in, clip]{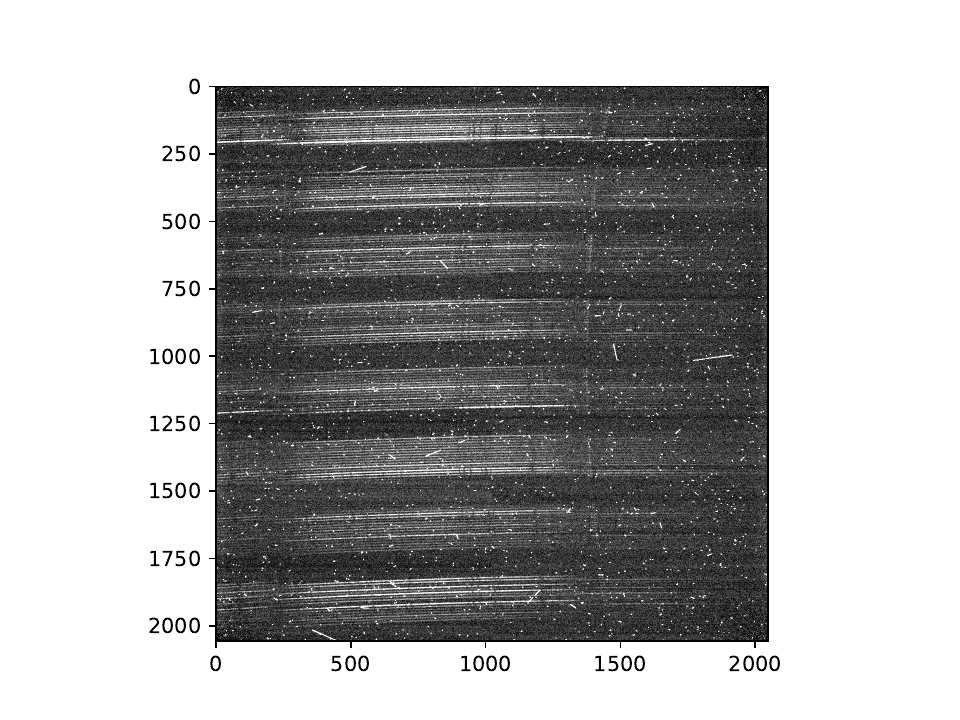}
    \end{tabular}
    \caption{Examples of raw M2FS (HiRes configuration)  images obtained during exposures of a calibration LED source (top left), evening twilight (top right), ThArNe arc lamp (bottom left) and a science field (bottom right).  Single-order spectra appear as horizontal bands, each  spanning 5130--5190~\AA\ over columns $\sim$~300--1400 (signal outside this column range is contributed by light from adjacent orders and is not used).  The separation into eight groups of 16 apertures reflects the physical bundling of fiber ends at the spectrograph. }
\label{fig:m2fs_image}
\end{figure*}

M2FS disperses light approximately along the direction parallel to rows in the stitched images, henceforth called the $x$ direction, where $x$ is a continuous variable along the discrete `column' axis (see Figure \ref{fig:m2fs_image}).  Adjacent spectra are offset approximately along the `row' axis, which we represent with continuous variable $y$.  In order to identify and trace spectral apertures, we follow procedures similar to those performed by IRAF's `apall' package.  For each science field, we operate on the  corresponding stitched LED frame (top-left panel of Figure \ref{fig:m2fs_image}), as it contains sufficient counts to identify and trace most spectral apertures.  For calibration exposures of standard stars or of twilight sky, counts are sufficiently high that we can operate directly on the stitched standard and twilight frames themselves.  The top-right panel of Figure \ref{fig:m2fs_image} displays the raw image obtained from an exposure taken during evening twilight.  

\begin{figure}
\includegraphics[width=3.5in, trim=0.0in 0in 0in 0.in, clip]{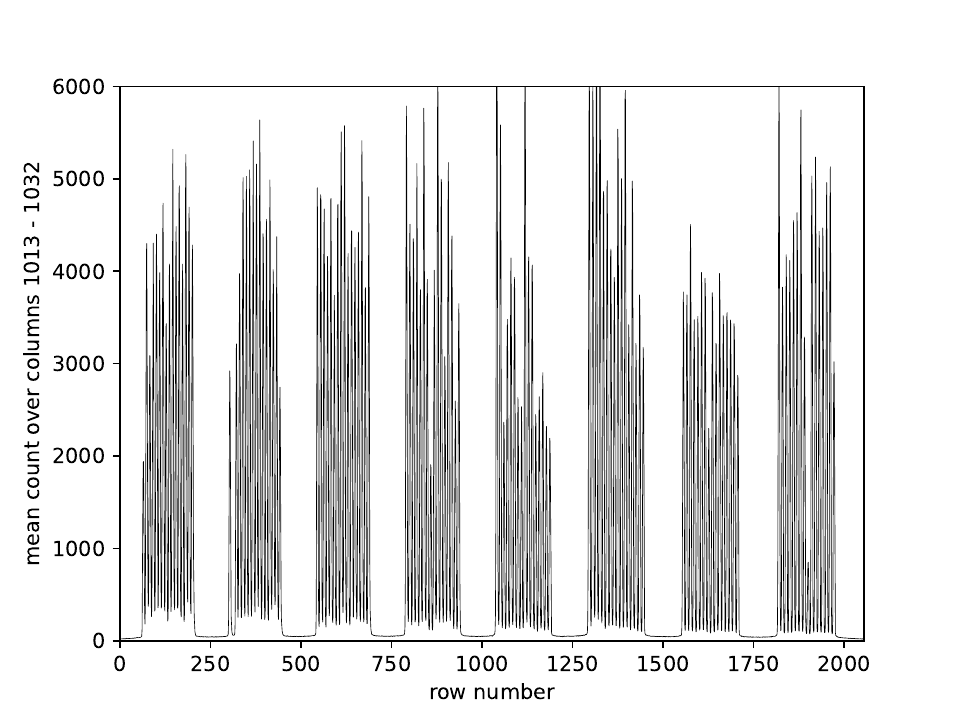}
    \caption{Identification of spectral apertures in an example M2FS frame.  Plotted for each row on the detector is the mean count recorded in columns $1013 - 1032$ (the middle 20 columns).  Local maxima signify the centers of spectral apertures.  
\label{fig:m2fs_find}}
\end{figure}

\smallskip
We begin by bundling the central 20 columns (columns 1013--1032), effectively combining them by storing their mean count as a function of row number ($y$ value).  Figure \ref{fig:m2fs_find} displays a characteristic example of this function, which resembles an emission-line spectrum; but of course here the `lines' are central aperture illumination profiles.  We use the astropy.modeling package to fit a Chebyshev polynomial that represents the `continuum' of this psuedo-spectrum.  After subtracting the best-fitting model continuum, we use the find\_lines\_derivative function from the astropy.specutils package to identify aperture centers as local maxima.  

\smallskip
We use these centers to initialize Gaussian fits (again via the astropy.modeling package) to the `continuum'-subtracted pseudo-spectrum, restricting our fits to the 10 rows around the centers returned by the  find\_lines\_derivative function, but re-fitting those centers under the Gaussian model.  The fitted Gaussian functions then represent the aperture illumination profiles across the center of the stitched image.  We repeat the this process for all 102 bundles of 20 (non-overlapping) consecutive columns, allowing us to quantify how the centers and widths of aperture profiles vary with column number across the stitched image.  

\smallskip
Next we inspect, by eye, the pseudo-spectrum along the central column bundle, as well as the Gaussian fit to each aperture profile along that central bundle.  Since the apertures in both `blue' and   `red' channels are known to follow a regular pattern of eight  approximately evenly-spaced groups, with each group  containing 16 approximately evenly-spaced apertures (see Figure \ref{fig:m2fs_image}), we can delete any obviously spurious aperture detections and insert artificial placeholders to represent (for book-keeping purposes) apertures corresponding to unassigned and/or broken fibers.  

\smallskip
Then, for each visually-confirmed aperture, we trace the full 2D  shape by `marching' from the center to each edge of the useful region (columns $\sim$~300 --1400; see Figure \ref{fig:m2fs_image}) in the stitched image.  We begin at the position whose $x$ coordinate is the median column number of the central column bundle, and whose $y$ coordinate is the fitted center of the aperture profile in that bundle.  We then find the fitted aperture center in the adjacent bundle that has the smallest deviation in its $y$ coordinate.  If the deviation has absolute value smaller than some threshold (we use 1.5 pixels), we step to a new position whose $x$ coordinate is the median column number of that adjacent bundle, and whose $y$ value is the fitted center of that bundle's aperture profile.  We proceed in this manner either to the edge of the useful region in the image, or until three consecutive column bundles have no aperture whose center deviates from the current $y$ coordinate by less than the specified threshold.  We then return to the center column and march, in the same manner, toward the opposite edge.  We thus obtain a list of $(x,y)$ positions that sample the aperture's 2D trace pattern.  To these data we fit and store  a $4^{\rm th}$-order polynomial function, iteratively rejecting $3\sigma$
outliers.  We also fit and store $4^{\rm th}$-order polynomials to the stored amplitudes and standard deviations; these two functions then characterize the aperture profile as a function of $x$.  

\subsection{Correction for Variations in Pixel Sensitivity}

M2FS does not have an internal lamp that uniformly illuminates the detectors; all incident light travels through the fibers.  In order to correct for random variations in pixel sensitivity within a given aperture, we use the previously-fit (Section \ref{subsec:identify/trace}) polynomials that represent center, amplitude and standard deviation of the LED aperture profile, all as functions of $x$, to generate a model 2D aperture image.  At a given column within the aperture, we evaluate the fitted polynomials to specify the parameters (center, amplitude, standard deviation) of the Gaussian aperture profile model.  We integrate that model to estimate the expected count within each pixel along the column, including all rows whose centers are within 3 aperture profile standard deviations of the aperture center.  Repeating this procedure at each column, we obtain a pixelated model of the  two-dimensional LED spectrum.   

\smallskip
Dividing the actual 2D LED spectrum by this model, we obtain the equivalent of a normalized `flat field' spectrum.  After repeating for each aperture, we divide the normalized `flat field' frame into each individual stitched image whose random variations in pixel sensitivity we wish to correct (these include science, twilight, and arc-lamp exposures). 
 
\subsection{Correction for Scattered Light} 

Having applied flat-field corrections to the stitched (science and calibration) images, next we estimate and remove scattered light.  We first use the corresponding LED exposures (or bright standard star and/or twilight exposures) to mask the regions corresponding to the identified and traced spectral apertures.  Specifically, we mask all pixels whose centers lie more than 3 standard deviations away from the center of the nearest aperture trace pattern, where the center and scale length are obtained by evaluating the polynomial functions fit to the aperture trace and aperture profile, respectively, at the pixel's $x$ coordinate (Section \ref{subsec:identify/trace}).  

\smallskip
Returning to the frame of interest, we then fit a 2D $4^{\rm th}$-order polynomial to the unmasked pixels, iteratively rejecting $3\sigma$ outliers, in order to estimate the contribution from scattered light.  We remove scattered light by subtracting this function from the frame of interest.

\subsection{Extraction of 1D spectra}
In order to extract 1D spectra from each aperture, we collapse each column within the aperture into a single pixel regardless of the aperture trace pattern, thereby preserving independence between adjacent columns.  This strategy would be optimal in the case that the spectral dispersion axis is exactly parallel to the detector's $x$ axis.  In reality the spectral apertures have nonzero curvature (Section \ref{subsec:identify/trace}); our procedure therefore results in some degradation of spectral resolution.

\smallskip
Let $\hat{N}(X,Y)$ and $\hat{\sigma}^2(X,Y)$ be the estimated count (in electrons) and estimated variance (in electrons$^2$), respectively, at discrete pixel $(X,Y)$.  Let $f(x,y)$ be the function that generates the 2D image of the spectrum---i.e., $f(x,y)\,dx\,dy$ is the expected count within area element $dx\,dy$ on the detector.  Physically, the function $f(x,y)$ is set by the intrinsic source (plus background) spectrum, the spectral resolution and the geometry of both aperture and detector.  We assume that, perpendicular to the spectral dispersion direction (i.e. the `spatial' direction, taken to be along the $y$ axis), the signal decays according to the Gaussian aperture profile whose parameters we evaluate from our polynomial fits described in Section \ref{subsec:identify/trace}, such that $f(y|x)=\mathcal{N}(y_0(x),\sigma^2(x))$, where $y_0(x)$ is the center of the aperture profile at dispersion coordinate $x$, and $\sigma(x)$ is the standard deviation.\footnote{Any functional dependence of $y_0$ on $x$ violates our starting assumption that the spectra are parallel to the $x$ axis; however, in practice the spectra are approximately aligned such that the dependence is weak.} 

\smallskip
Under this model, the predicted count at discrete pixel $(X,Y)$ is
\begin{eqnarray}
N_{\rm mod}(X,Y)=\displaystyle\int_{X_1}^{X_2} \mathrm{d}x \displaystyle\int_{Y_1}^{Y_2} \mathrm{d}y \, f(x,y)\nonumber\\
=\displaystyle\int_{X_1}^{X_2} \mathrm{d}x \displaystyle\int_{Y_1}^{Y_2} \mathrm{d}y \, f(x)\,f(y|x))\nonumber\\
=\displaystyle\int_{X_1}^{X_2}\mathrm{d}x\, f(x)\displaystyle\int_{Y_1}^{Y_2}\mathrm{d}y\,  \mathcal{N}(y_0(x),\sigma^2(x))\nonumber\\
=N_{\rm mod}(X)\displaystyle\int_{Y_1}^{Y_2}\mathcal{N}(y_0(X),\sigma^2(X)),
\label{eq:nxy}
\end{eqnarray}
where $(X_1,Y_1)$ and $(X_2,Y_2)$ are  corners across the diagonal of the pixel.  The count $N_{\rm mod}(X)$ is, by definition, the expectation value of $f(x)=\int p(x,y)\,dy$ at column $X$. 

\smallskip
Within a given aperture, we take each of the $N_{\rm row}$ pixels in column $X$ to be drawn independently from a normal distribution with mean predicted by Equation \ref{eq:nxy} and variance equal to the estimated value, $\hat{\sigma}^2(X,Y)$.  We define 
\begin{equation}
\chi^2\equiv \displaystyle\sum_{i=1}^{N_{\rm row}}\frac{\bigl [\hat{N}(X,Y_i)-N_{\rm mod}(X,Y_i)\bigr ]^2}{\hat{\sigma}^2(X,Y_i)}
\end{equation}
Minimizing $\chi^2$ with respect to $N_{\rm mod}(X)$, we recover the `optimal' estimator of \citet{horne86},
\begin{equation}
\hat{N}(X)=\frac{\sum_{i=1}^{N_{\rm pix}}\frac{\hat{N}(X,Y_i)I_i}{\hat{\sigma}^2(X,Y_i)}}{\sum_{i=1}^N\frac{I_i^2}{\hat{\sigma}^2(X,Y_i)}}
\label{eq:extract1d}
\end{equation}
where $I_i\equiv \displaystyle\int_{Y_{1_i}}^{Y_{2_i}}\mathrm{d}y\,\mathcal{N}\bigl (y0(X),\sigma^2(X)\bigr )$. 
 Given the data in the 2D image, and the Gaussian aperture profile parameters fit to spectral apertures in the LED frame (Section \ref{subsec:identify/trace}), the estimator in Equation \ref{eq:extract1d} is fully specified.  For all science and calibration frames, we use Equation \ref{eq:extract1d} to extract 1D spectra at every column of every aperture.  We propagate the estimated variance as
\begin{equation}
\hat{\sigma}^2[\hat{N}(X)]=\biggl (\displaystyle\sum_{i=1}^{N_{\rm pix}} \frac{I_i^2}{\hat{\sigma}^2(X_,Y_i)}\biggr ) ^{-1}.
\end{equation}

\subsection{Wavelength Calibration}
\label{sec:wavelength}

We calibrate wavelengths using the 1D spectra extracted from exposures of the illuminated arc lamp containing Thorium, Argon and Neon (`ThArNe') gases.  At the outset, for each individually-extracted 1D ThArNe spectrum, we use a $5^{\rm th}$-order polynomial to fit and subtract the continuum component, iteratively rejecting outliers at more than $5\sigma$ below the fit or more than $1\sigma$ above (the asymmetry effectively rejects pixels that sample emission features).  To the continuum-subtracted spectrum, we  then use the `find\_lines\_derivative' function from the astropy.specutils package to find emission features and estimate their centers in pixel space.  Within the ten pixels around the center of each identified emission line, we fit a Gaussian function and store the best-fitting center, standard deviation and amplitude.  The standard deviation quantifies the local spectral resolution.   

\smallskip
Next we manually identify emission lines in a single 1D extracted ThArNe spectrum (i.e., the spectrum obtained in a single aperture), which thereafter serves as a template for automatically identifying emission lines in all other ThArNe spectra in all apertures in all ThArNe exposures acquired using the same M2FS configuration.  Operating on the template spectrum only, we use  NOIRLab's thorium-argon spectral atlas \citep{palmer83} to visually identify individual emission lines interactively by eye.  We store the atlas wavelength and pixel coordinate (from the Gaussian fit described above) of each line center. 

\smallskip
Since we retain the pixelation native to the detector along the $x$ (column) axis, we expect that the wavelength/pixel relationship will be unique for each aperture and---given small temporal changes in the aperture trace pattern---unique for each exposure. The next task, then, is to transfer our mapping of emission lines in the template ThArNe spectrum automatically to all individual non-template ThArNe spectra.  For a given non-template spectrum, we begin by fitting a polynomial function that effectively distorts the template's pixel scale to bring the template's emission lines into alignment with those of the non-template spectrum.  That is, letting $T(X)$ and $F(X)$ denote continuum-subtracted template and non-template ThArNe counts as functions of pixel number $X$, we find the order-$m$ polynomial $P_m(x)=c_0+c_1\bigl(\frac{x-x_0}{x_s}\bigr )+c_2\bigl(\frac{x-x_0}{x_s}\bigr )^2+\ldots+c_m\bigl(\frac{x-x_0}{x_s}\bigr )^m$ that minimizes the sum of squared residuals $\sum_{i=1}^{N_{\rm pix}}(F(X_i)-A_1I(X_i))^2$,
where $x_0\equiv 0.5\bigl (X_{max}+X_{min}\bigr )$ is the midpoint of the template spectrum, $x_s\equiv 0.5\bigl (X_{max}-X_{min}\bigr )$ is half the range of the template spectrum, and $I(X)$ is the linear interpolation of $T\bigl (A_2+x(1+P_m(x))\bigr )$ at $X$.  We adopt $m=4$; free parameters include the five polynomial coefficients and constants $A_1$, $A_2$.  

\smallskip
We use the best-fitting model to transform the pixel coordinates of known emission lines in the template ThArNe spectrum to pixel coordinates in the non-template spectrum.  To each emission line in the non-template spectrum, we assign the atlas wavelength of the nearest line in the transformed template spectrum, tolerating coordinate mismatches of $\leq 2$  pixels.  
We then conduct the following iterative procedure: 1) Using only the matched features (which typically number between 25-40 per non-template spectrum), we fit a 5th-order polynomial to the atlas wavelength as a function of pixel coordinate at the line center; 2)  using this updated wavelength/pixel relation for the non-template spectrum, we assign atlas wavelengths to any as-yet unidentified emission lines in the non-template spectrum if their central wavelengths match those of as-yet unused template lines within a tolerance of $\leq 0.05$ \AA.  After iterating up to 10 times, we save for each non-template ThArNe spectrum the pixel coordinates at atlas wavelengths of the identified emission lines, coefficients of the final polynomial fit to the wavelength/pixel relation, the number of emission features used in the wavelength/pixel fit, and the rms of residuals to the fit.  For the HiRes (resp. MedRes) configuration, over \mtwofsHiReswavrmsn\ (\mtwofsMedReswavrmsn) non-template ThArNe spectra, the mean rms residual, after excluding those below the 1st percentile and those above the 99th, is \mtwofsHiReswavrmsmean\ \AA\ (\mtwofsMedReswavrmsmean\ \AA), with standard deviation \mtwofsHiReswavrmsstd\ \AA\ (\mtwofsMedReswavrmsstd\ \AA).   

\begin{figure*}
    \begin{tabular}{@{}llll@{}}
    \renewcommand{\arraystretch}{0}
%    \vspace{-1.5in}
    \includegraphics[width=4.in,trim=0.in 0.in 0.08in 0in, clip]{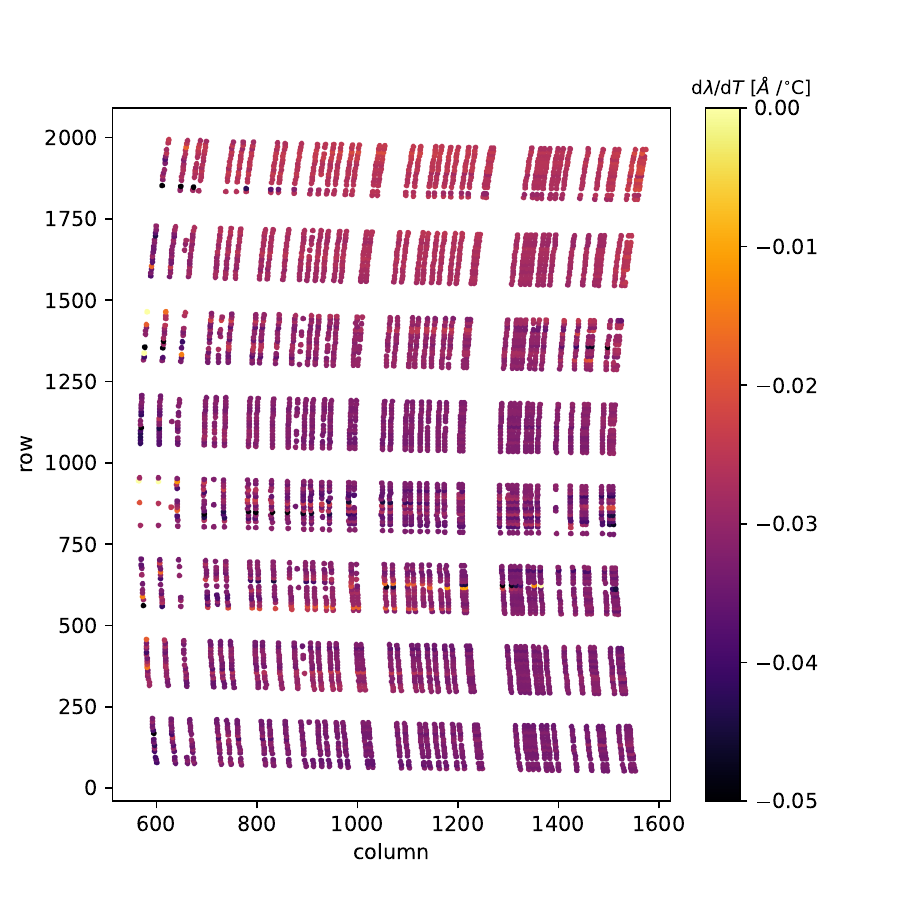}& \hspace{-1in}\includegraphics[width=4.in,trim=0.in 0.in 0in 0in, clip]{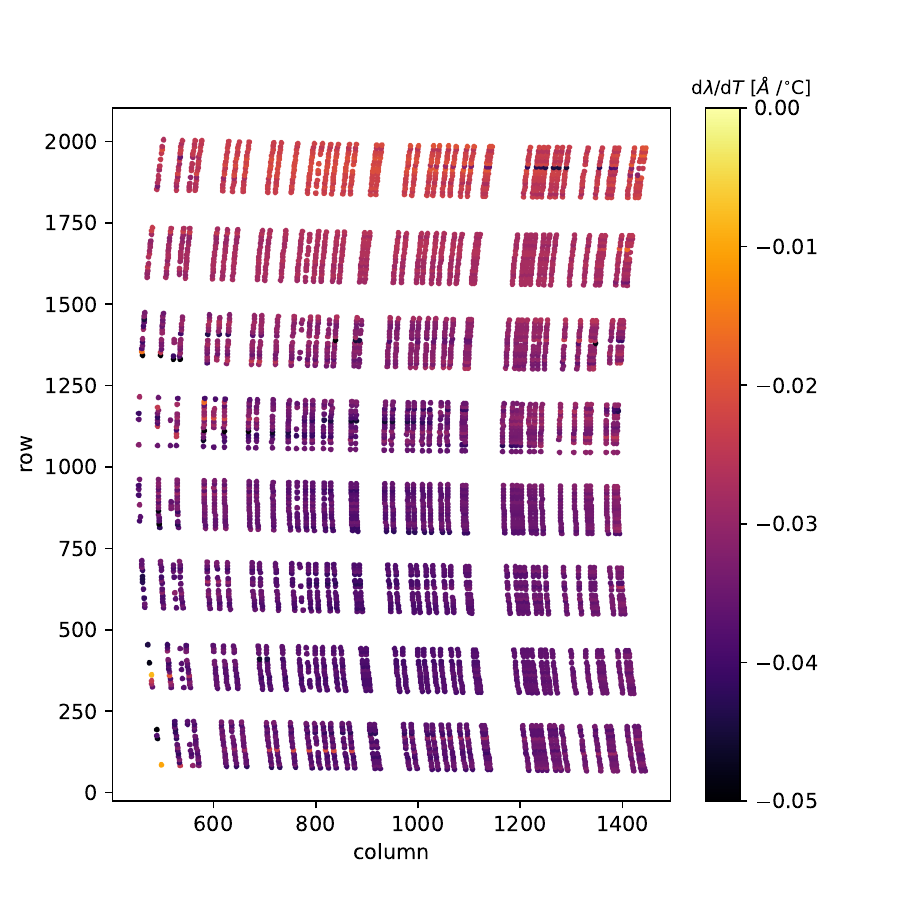}\\
%    \vspace{-1.5in}
    \end{tabular}
    \caption{Change in wavelength per change in temperature (as measured at the detector), from emission lines observed in calibration exposures acquired immediately before and immediately after M2FS (HiRes configuration) observations of one example science field.  Left/right panels show results for the blue and red channel, respectively.}
\label{fig:m2fs_dtemp}
\end{figure*}

\smallskip
The next step is to use the wavelength/pixel relations obtained for the ThArNe spectra to estimate wavelengths at all pixels of each individual science frame exposure.  Typically we obtain ThArNe calibration frames before and after each set of science exposures for a given target field, sometimes with an additional ThArNe exposure taken in between individual science exposures.  These sequences let us quantify systematic shifts in the wavelength/pixel relationship that we expect to be due to flexure of the detector hardware and its sensitivity to temperature (as measured within the spectrograph cell) changes.  Using one science field's set of ThArNe exposures as an example (observed with the HiRes grating), Figure \ref{fig:m2fs_dtemp} displays, across both detector arrays, the slopes $d\lambda/dT$ that we fit to the wavelength/temperature relation at the location of each identified ThArNe emission line.  We detect smooth variation across both detectors, with slope ranging from $\sim 0$ to $\sim 0.04$ \AA/K.  

\smallskip
In order to compensate for these systematic drifts of the wavelength/pixel relation, for every pixel in the set of ThArNe exposures corresponding to a given science field, we fit a linear model for pixel wavelength as a function of time.  Individual wavelengths are  weighted by the inverse-square of the rms residual with respect to the fitted wavelength/pixel relation.  For the time coordinate, we use the time at the exposure midpoint.  At every pixel of a given science exposure, we then assign the wavelength obtained by evaluating the linear wavelength/time function at the temporal midpoint of the science exposure.  In cases where multiple ThArNe exposures are not available for monitoring temperature and/or time dependence of the wavelength solution, we flag the corresponding catalog entries accordingly (see Section \ref{sec:dataset}).  The catalogs contain a column `n\_wav\_calibrations' that states the number of independent ThArNe exposures used in the wavelength calibration.  Two other columns, `temp\_min' and `temp\_max', give the minimum and maximum spectrograph temperature, across the science sub-exposures.  For the \mtwofswavflagn\ spectra where n\_wav\_cal=1 and temp\_max$-$temp\_min $>1$ K (\mtwofswavflaggoodn\ of which yield measurements passing our crude quality-control filter based on velocity uncertainty), we set flag wav\_cal\_flag=True in the M2FS catalogs.  

\smallskip
We do not apply heliocentric corrections to the calibrated wavelengths, which therefore include Doppler shifts due to the line-of-sight component of the observatory's velocity with respect to the barycentric rest frame.  Instead we apply heliocentric corrections directly to the line-of-sight velocities estimated using the observed wavelengths (Section \ref{sec:inference}).  

\subsection{Identification and masking of cosmic rays}
\label{sec:cr_rejection}
It is at this point that we identify and mask pixels in the extracted, wavelength-calibrated 1D science spectra that are affected by cosmic rays.  To each science spectrum, we first fit the continuum level using a 4th-order polynomial, iteratively rejecting outliers more than $2\sigma$ below or $3\sigma$ above the fit, where $\sigma$ is the root-mean-square value of residuals.  We then flag as a likely cosmic ray signal any pixel value that exceeds the fitted continuum level by more than $5\sigma$.  In subsequent analysis, we mask these as well as the four nearest pixels. 
 While this procedure will similarly mask bona fide emission features, we expect emission lines to be largely absent from the targeted stellar spectra over the observed spectral region. 

\subsection{Correction for variations in fiber throughput}

We use the  entire set of twilight exposures acquired during the observing run to estimate  relative throughput as functions of fiber and wavelength.  We begin by averaging, on a pixel-by-pixel basis within each aperture,  the (3--10) 1D spectra from individual exposures during a given twilight sequence (i.e., the set of exposures taken during a given evening/morning twilight).  When computing the mean, we weight the count in each pixel by its inverse variance.  We then combine these nightly weighted-mean twilight spectra across all twilight observations within a given run, taking a new  weighted mean count on a pixel-by-pixel basis within each aperture.  This second averaging is unique to each science exposure, as the count in each pixel is weighted by the inverse-squared difference in time between the midpoint of the nightly twilight sequence and the midpoint of the science exposure.

\smallskip
In order to estimate the relative fiber throughputs that pertain to a given science exposure, we operate on the corresponding run-averaged twilight frame, where the dominant spectral features are solar absorption lines.  Within each aperture, we fit a 4th-order polynomial to the mean twilight count as a function of wavelength, iteratively rejecting outliers more than 3$\sigma$ above and more than 1$ \sigma$ below the fit in order to isolate the continuum component.  For each pixel of a  given science spectrum, we evaluate the twilight-continuum polynomials from all apertures at the wavelength of the pixel in the science spectrum.  We then apply a wavelength-dependent throughput correction by dividing the count at each pixel by the ratio of the aperture's twilight continuum level to the median level across all apertures. 

\subsection{Sky subtraction}
A typical M2FS observation allocates 20--40 fibers to regions of blank sky, split approximately evenly among the two spectrographs.  For each field and spectrograph, we combine the throughput-corrected sky spectra to obtain a median sky spectrum, and then subtract the mean sky spectrum from the all throughput-corrected spectra for all targets observed with that spectrograph.  

\smallskip
When combining individual sky spectra to obtain a single  median spectrum, we must again contend with the fact that the wavelength/pixel relation is unique to each individual spectrum.  Following \citet{koposov11}, we interpolate all individual sky spectra onto a common wavelength grid that oversamples, with ten times the number of pixels, the original spectrum.  We then store the median sky spectrum in the oversampled space, and record the variance at each pixel as $2.198\pi \mathrm{MAD}^2/(2N_{\rm sky})$, where $N_{\rm sky}$ is the number of individual sky spectra and MAD is the median absolute deviation \citep{koposov11}.   From each individual science spectrum, we then interpolate the median sky (and variance) spectrum onto the pixel scale of the science spectrum, letting us then perform the sky subtraction directly on a pixel by pixel basis. 

\subsection{Stacking subexposures}
The final step of our M2FS image processing is to combine, on an aperture-by-aperture basis, the  spectra obtained in multiple exposures.  For a given aperture, we combine spectra from multiple exposures by taking the weighted mean (sky-subtracted) count at each pixel.  One drawback of this stacking on a pixel-by-pixel basis is that it can exacerbate the effect of temperature changes inside the spectrograph, which tend to cause the aperture trace pattern and wavelength/pixel relation to drift (Section \ref{sec:wavelength}.  In order to compensate for this effect and assign wavelengths to individual pixels in the stacked spectra, we follow the same procedure described in Section \ref{sec:wavelength}, where we evaluate for each pixel the linear wavelength vs. time relation determined from ThArNe exposures.  For each pixel in the stacked spectrum, we adopt as the time coordinate the mean midpoint of the individual science exposures, weighted by the inverse variance of the sky-subtracted count.

\subsection{Products}
 All processed M2FS spectra are available for download from the Zenodo database\footnote{DOI: 10.5281/zenodo.7837922}.  For each frame of (up to) 128 spectra obtained on one of the spectrograph channels, a fits file contains the pixel wavelengths (as calibrated to the observatory rest frame---i.e., not shifted to the heliocentric frame), the sky-subtracted counts and their variances, the sky spectrum that was subtracted, the pixel mask, and the best-fitting model spectrum (Section \ref{sec:analysis}), plus various observational details (e.g., date, time and spectrograph temperature of each individual exposure) and random samples from posterior probability distribution functions for model parameters inferred during analysis of the spectra (Section \ref{sec:analysis}). 

\begin{figure*}
    \includegraphics[width=7.5in, trim=0.0in 0in 0in 0.in, clip]{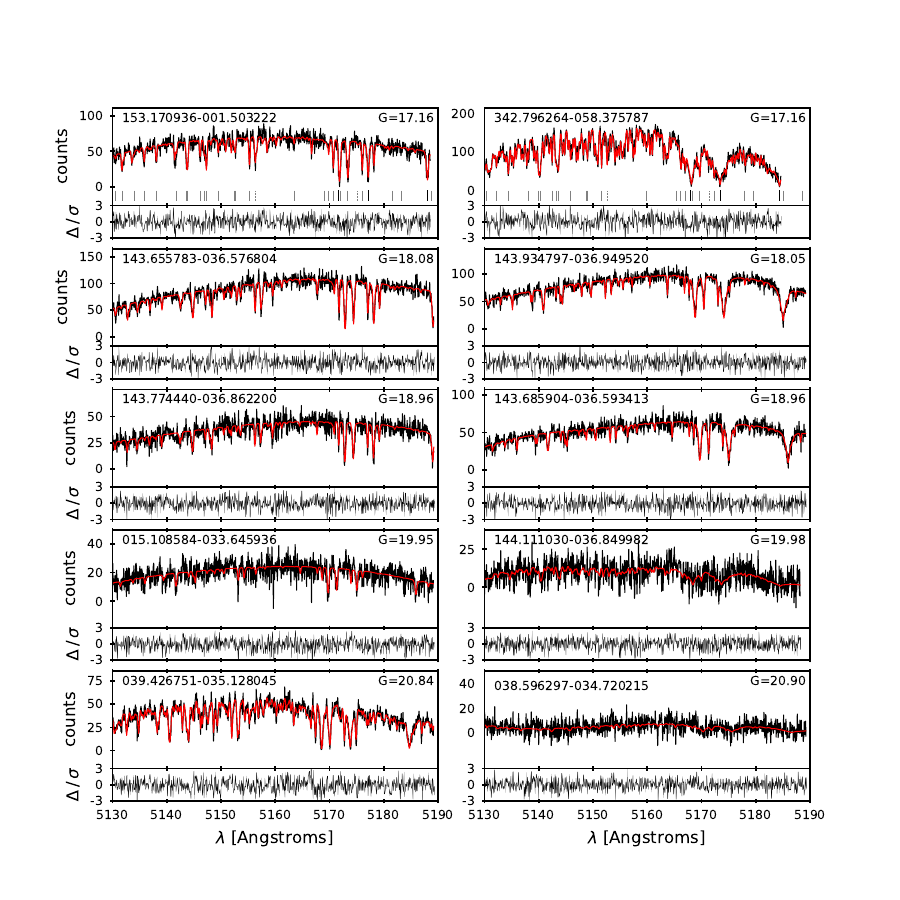}
    \caption{Examples of Magellan/M2FS HiRes spectra (black, main panels; sky-subtracted counts are scaled to the dimensions of the plotting window), which for our observing configuration cover 5125--5190~\AA\ at resolution $\mathcal{R} \approx 24,000$.  Text indicates Gaia ID and Gaia G-band magnitude.  Overplotted (red) are best-fitting model spectra.  Smaller panels display normalized (by the count error propagated through the processing pipeline) residual with respect to the best fit.  In the top panels, hash marks identify wavelengths (redshifted to match the observed spectrum) of known FeI (solid grey), FeII (broken grey) and MgI (solid black) lines.  Left-hand (resp. right-hand) panels depict spectra for likely red giant (dwarf) stars, with surface gravity measured to be \logg~$<1$ (\logg~$>4)$.  
\label{fig:m2fshires_spectra}}
\end{figure*}

\begin{figure*}
    \includegraphics[width=7.5in, trim=0.0in 2.9in 0in 0.in, clip]{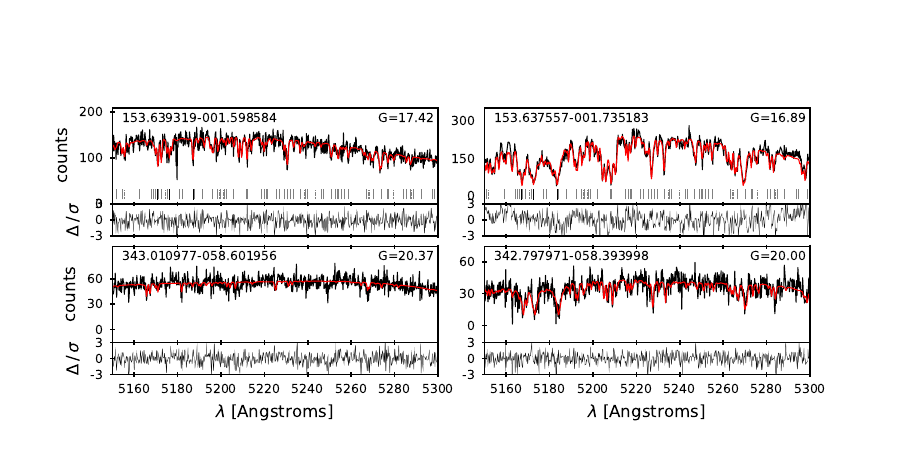}
    \caption{Same as Figure \ref{fig:m2fshires_spectra}, but for example Magellan/M2FS MedRes spectra, which span 5115--5300~\AA\ at $\mathcal{R}\approx 7000$.
\label{fig:m2fsmedres_spectra}}
\end{figure*}

\begin{figure*}
    \includegraphics[width=7.5in, trim=0.0in 0in 0in 0.in, clip]{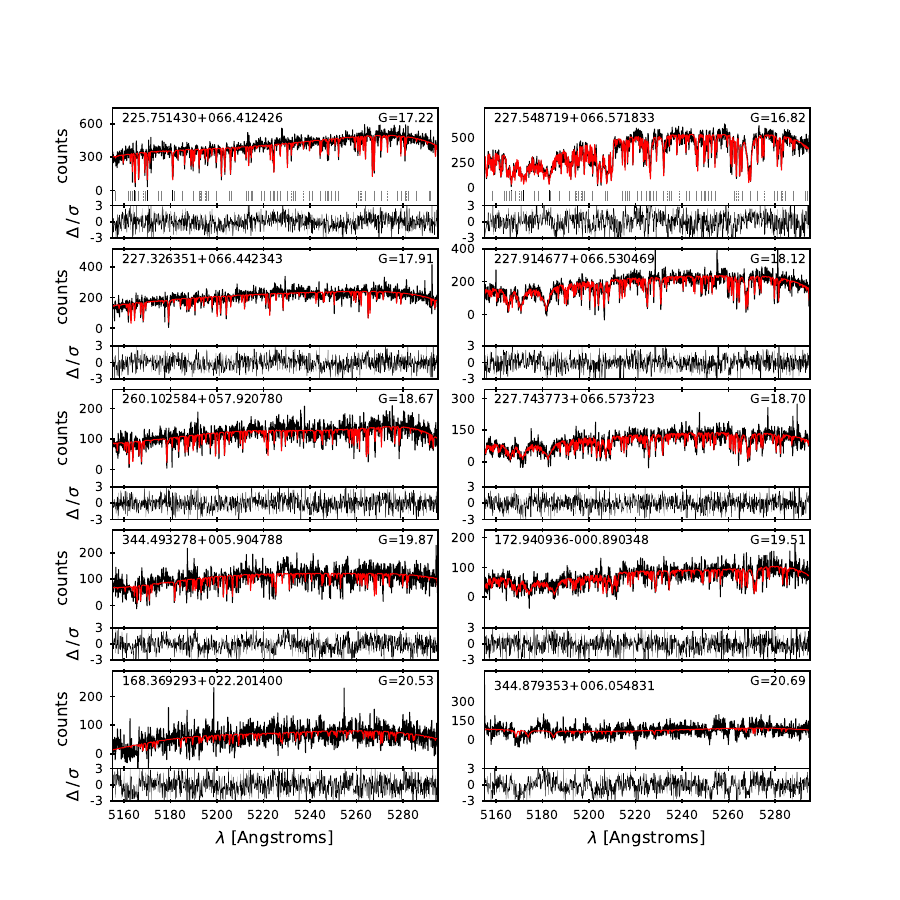}
    \caption{Same as Figure \ref{fig:m2fshires_spectra}, but for example MMT/Hectochelle spectra, which span 5150--5300~\AA\ at $\mathcal{R} \approx 32,000$.  The larger numbers of counts (cf. Figures \ref{fig:m2fshires_spectra} and \ref{fig:m2fsmedres_spectra}) reflect the fact that the Hectochelle pipeline calculates the sum of counts across sub-exposures, while the M2FS pipeline calculates the average.
\label{fig:hecto_spectra}}
\end{figure*}

\smallskip
Figures \ref{fig:m2fshires_spectra}, \ref{fig:m2fsmedres_spectra} and \ref{fig:hecto_spectra} display examples of fully processed spectra acquired with M2FS HiRes, M2FS MedRes and Hectochelle, respectively.  Source magnitude increases from top to bottom.  Left-hand and right-hand panels show spectra from stars measured to have weak and strong surface gravity, respectively, distinguishing the likely red giant stars within Galactic halo structures from dwarf stars in the Galactic foreground.  Sub-panels display residuals with respect to the best-fitting model spectra (Section \ref{sec:inference}), normalized by the propagated uncertainty in the observed count.   In the top two panels, hash marks identify wavelengths of known FeI, FeII and MgI absorption features that are listed in the database maintained by the Virtual Atomic and Molecular Data Centre (VALDC) Consortium, provided by the BASS2000 website\footnote{https://doi.org/10.25935/9TXJ-F095  }.

\section{Analysis of Magellan/M2FS and MMT/Hectochelle Spectra}
\label{sec:analysis}
We analyze each individual processed spectrum by fitting a model that is derived from a library of synthetic template spectra.  The procedure is similar to others previously deployed for modeling stellar spectra \citep[e.g.,][]{koleva09,koposov11,walker15,li19}.  Continuum-normalized synthetic spectra are computed over a grid of stellar-atmospheric parameters that has dimensions  \teff, \logg, \feh, \mgfe.  An additional grid dimension extends over a parameter, \lsf, that sets the spectral line spread function and thus the resolving power ($\mathcal{R}\approx\lambda/(2.355$\lsf$)$).  Given proposed values for these parameters, we generate a model spectrum by combining (via kernel smoothing) the surrounding templates within the multi-dimensional grid space, multiplying by a flexible continuum model and adjusting template wavelengths to account for source redshift as well as any low-order corrections to the wavelength/pixel relation.  We use this model spectrum to evaluate the likelihood of the observed spectrum.  We use the likelihood evaluations to perform Bayesian inference, ultimately obtaining a random sample from the posterior probability distribution function  (PDF) in model parameter space.  

\smallskip
We provide details of our analysis procedure below.  In most respects our procedure is identical to the one described by \citet{walker15} and subsequently followed by \citet{walker15b,spencer17,spencer18,buttry22,pace21}.  However, our current implementation differs in one significant way.  In previous work, we adopted synthetic template spectra originally used to analyze spectra from the Sloan Digital Sky Survey's SEGUE project \citep{lee08}, which implicitly assumed the abundance ratio of $\alpha$ elements to Fe to be a fixed function of \feh.   Now we use a new set of synthetic template spectra (Section \ref{sec:ian}) that we have computed over a range of \mgfe , with the value of \mgfe\ no longer dependent on \feh.

\subsection{Modeling}
Given a continnum-normalized, zero-redshift template spectrum, $T_{\mathbf{\theta}}\,(\lambda)$, corresponding to parameters $\mathbf{\theta}\equiv ($\teff,\logg,\feh,\mgfe,\lsf$)$, we compute a model stellar spectrum according to
 \begin{equation}
  M(\lambda)=P_l(\lambda)\,T_{\mathbf{\theta}}\biggl (\lambda\bigl [1+z+ Q_m(\lambda)\bigr ]\biggr ),
  \label{eq:model}
\end{equation}
where
\begin{eqnarray}
  P_l(\lambda)\equiv p_0+p_1\biggl [\frac{\lambda-\lambda_0}{\lambda_s}\biggr ]+p_2\biggl [\frac{(\lambda-\lambda_0)}{\lambda_s}\biggr ]^2\nonumber\\
+\ldots+p_l\biggl [\frac{(\lambda-\lambda_0)}{\lambda_s}\biggr ]^l
  \label{eq:polynomial1}
\end{eqnarray}
is an order-$l$ polynomial that represents a smooth continuum component.  In Equation \ref{eq:model}, rest wavelengths of the  template spectrum are modified according to source redshift (in the observatory rest frame), $z\approx V_{\rm LOS}/c,$ and an order-$m$ polynomial, 
\begin{eqnarray}
  Q_m(\lambda)\equiv \frac{q_1}{c}\biggl [\frac{\lambda-\lambda_0}{\lambda_s}\biggr ]+\frac{q_2}{c}\biggl [\frac{(\lambda-\lambda_0)}{\lambda_s}\biggr ]^2\nonumber\\
  +...+\frac{q_m}{c}\biggl [\frac{(\lambda-\lambda_0)}{\lambda_s}\biggr ]^m,
  \label{eq:polynomial2}
\end{eqnarray}
that can apply non-linear corrections to the wavelength/pixel relation.  Note that we omit from $Q_m(\lambda)$ a zero$^{\rm th}$-order term, as it would be entirely degenerate with source redshift in Equation \ref{eq:model}.  We examine zero-point redshift errors via direct comparison to external data sets (Section \ref{sec:external}).  

\smallskip
We choose $l=5$ and $m=2$, which provide sufficient flexibility to fit the continuum shape and to accommodate low-order corrections to the wavelength solution.  We adopt scale parameters $\lambda_0=\frac{1}{2}(\lambda_{\rm max}+\lambda_{\rm min})$ and $\lambda_s=\frac{1}{2}(\lambda_{\rm max}-\lambda_{\rm min})$ \AA, such that $-1 \leq (\lambda-\lambda_0)/\lambda_s \leq +1$ over the entire range of observed wavelengths.  For M2FS HiRes we use the range $\lambda_{\rm min}=5127$~\AA\ to  $\lambda_{\rm max}=5190$~\AA. 
For M2FS MedRes and Hectochelle we use the range $\lambda_{\rm min}=5155$~\AA\ to  $\lambda_{\rm max}=5295$~\AA. 

\subsubsection{Template Spectra}
\label{sec:ian}

We present a new high-resolution grid of template spectra
spanning 5050 $\leq \lambda \leq$ 5350~\AA\ 
around the Mg~\textsc{i} `b` triplet.
It is sampled at $\Delta\lambda$ = 
0.05~\AA\ intervals, 
yielding a resolving power of 
$\mathcal{R} \approx$ 104,000.
We generate these template spectra using a recent version
(2017) of the MOOG line analysis code \citep{sneden73,sobeck11}.
We interpolate model atmospheres from the
ATLAS9 grid \citep{castelli04}.

\smallskip
We generate line lists for the synthesis using the
LINEMAKE code\footnote{\url{https://github.com/vmplacco/linemake}} \citep{placco21linemake}.
LINEMAKE creates an initial list of lines
drawn from the \citet{kurucz11} line compendia.
It subsequently updates the transition probabilities,
hyperfine splitting structure, and isotope shifts
for lines with recent laboratory analysis
(e.g., \citealt{lawler09,lawler17review}).
LINEMAKE also incorporates recent laboratory work on
molecules, including CH, CN, C$_{2}$, and MgH
in this spectral range
\citep{hinkle13,masseron14,ram14,sneden14}.
The initial list includes more than 39,000 lines.
We remove the weakest lines, 
ones contributing less than 0.5\% to the 
line-to-continuum opacity ratio,
in a synthetic spectrum for a cool, metal-rich red giant
(\teff\ = 4000~K, \logg\ = 0.0, [Fe/H] = $+0.5$).
These lines contribute negligible absorption to
stars that are warmer and/or more metal poor.
The final line list contains 17,884 lines.

\begin{figure*}
    \includegraphics[width=3.45in]{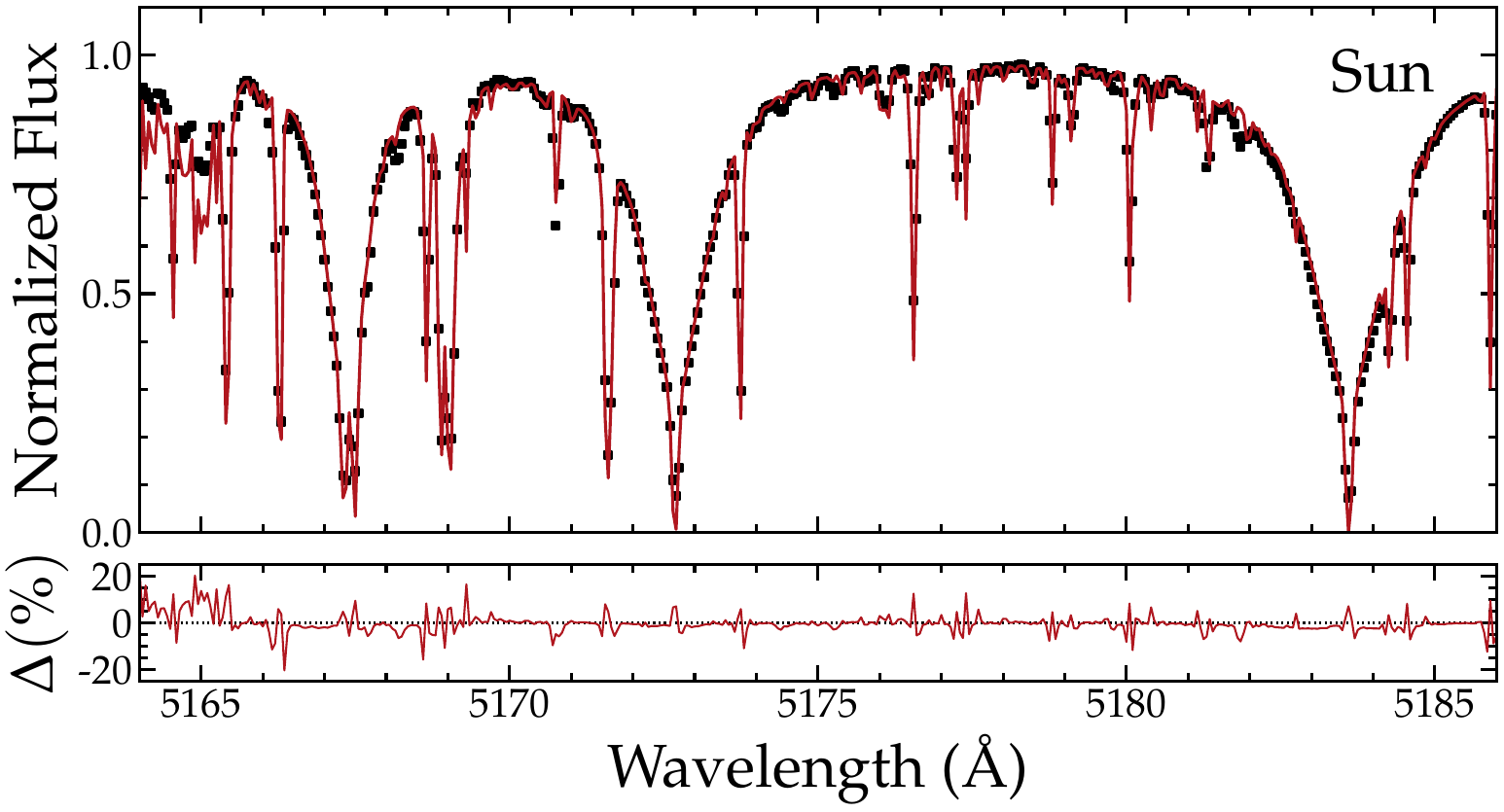}
    \hspace*{0.1in}
    \includegraphics[width=3.45in]{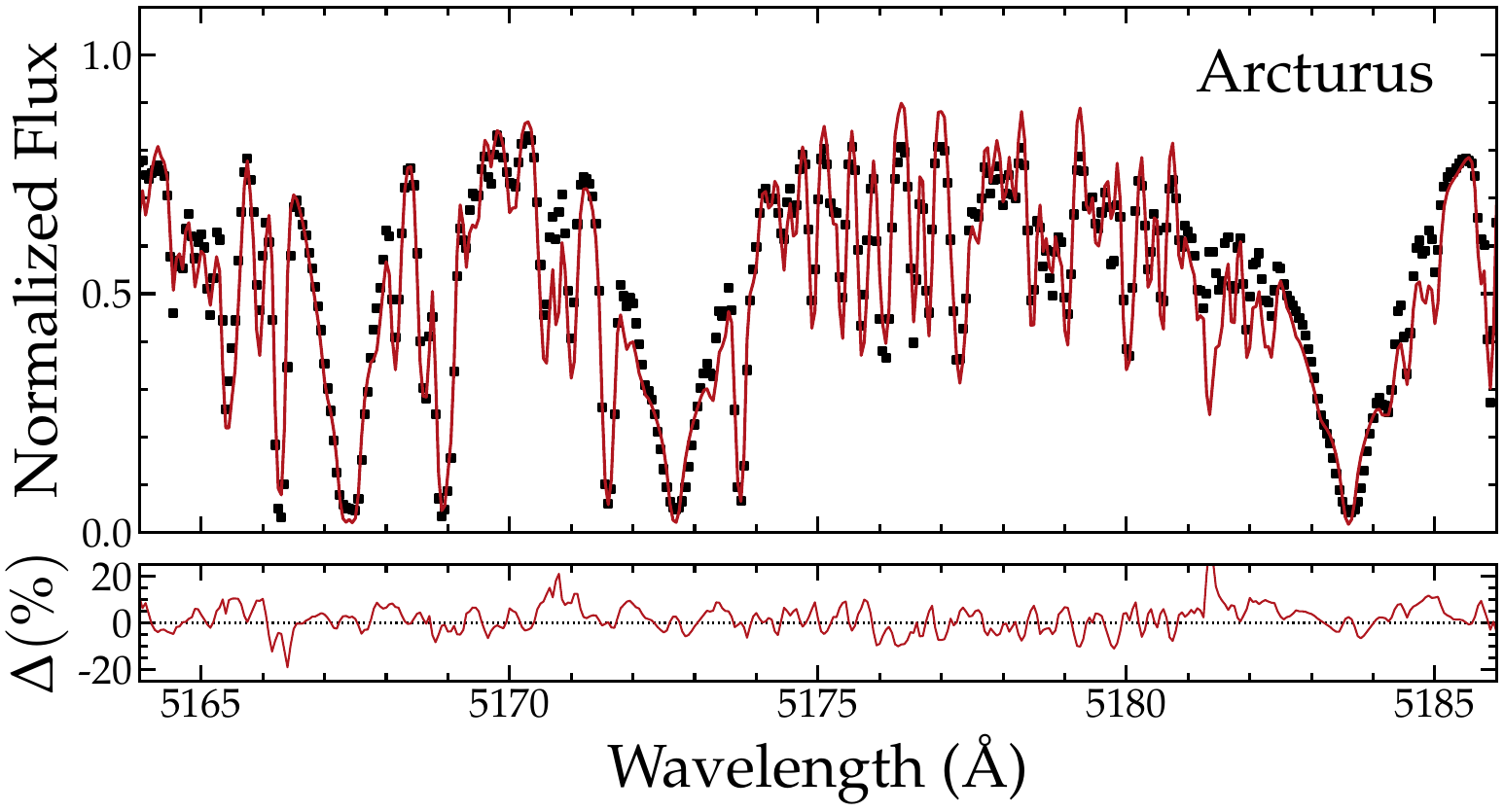}
    \caption{Validation of template spectra.  Top panels compare observed Solar (left) and Arcturus (right) 
    spectra with synthetic spectra generated 
    using the same tools as for our template grid.
    Data points represent the observed spectra, resampled to
    the resolution of our models, $\Delta\lambda$ = 0.05~\AA,
    which are shown by the red lines.
    Bottom panels illustrate the differences in percent.
\label{fig:speccompplot}}
\end{figure*}

\smallskip
As a proof of concept, we compare a small region of 
synthetic spectra generated using these tools
with the observed spectra of the Sun and Arcturus
\citep{kurucz84,hinkle00}
in Figure~\ref{fig:speccompplot}.
We adopt the \citet{holweger74} empirical model atmosphere for the Sun, and we adopt the
\citet{ramirez11} model atmosphere parameters for Arcturus
(\teff\ = 4286~K, 
\logg\ = 1.66, 
microturbulence velocity parameter (\vt) = 1.74~\kms, and
[Fe/H] = $-0.52$).
We also adopt
[Mg/Fe] = $+$0.37, 
[Si/Fe] = $+$0.33, and
[Ti/Fe] = $+$0.24 in our synthesis of the Arcturus spectrum.
We have empirically adjusted a small fraction
($\approx$0.4\%) of the \loggf\ values 
in our final linelist to better reproduce the 300~\AA\ 
region of interest for the
Solar and Arcturus spectra.
The overwhelming majority (75 of 77) of these changes are to lines without modern laboratory work,
and most are relatively weak and thus will have
negligible impact on the fitting of metal-poor stellar spectra.
The median absolute deviations for these regions of the
Solar and Arcturus spectra are 1.2\% and 3.8\%, respectively,
demonstrating the general reliability of our method.

\smallskip
We synthesize a grid spanning
3900 $\leq$ \teff\ $\leq$ 7500~K in intervals of 100~K,
0.0 $\leq$ \logg\ $\leq$ 5.0 [cgs] in intervals of 0.25~dex,
$-4.0 \leq$ [Fe/H] $\leq +1.0$ in intervals of 0.25~dex, and
$-1.0 \leq$ [Mg/Fe] $\leq +1.4$ in intervals of 0.20~dex.
A few regions near the edge of this grid are excluded because
they represent non-physical combinations of parameters or 
they extend beyond the ATLAS9 grid.
ATLAS9 models with $\alpha$ enhancement are adopted 
when [Mg/Fe] $\geq +0.1$.
The microturbulence velocity parameter is
adopted as a function of \logg:\ 
\vt\ = 1.0~\kms\ for dwarfs (\logg\ $\geq$ 4.0),
\vt\ = 2.0~\kms\ for giants (\logg\ $\leq$ 1.0), and
varying linearly between these two points.
The macroturbulence velocity is assumed to be
3.0~\kms\ for dwarfs and subgiants
(\logg\ $\geq$ 3.0),
8.0~\kms\ at \logg\ = 0.0, and
varying linearly between these two points.
We adopt the Solar values for 
carbon ($^{12}$C/$^{13}$C = 89/1) and
magnesium ($^{24}$Mg/$^{25}$Mg/$^{26}$Mg = 79/10/11)
isotope ratios.
Our final grid contains a total of 186071 model spectra, all of which we make publicly available at the Zenodo database (DOI: 10.5281/zenodo.7837922).  

\smallskip
We account for the finite spectral resolution of M2FS (resolving power $\mathcal{R}\approx 24,000$ in our chosen configuration) and Hectochelle ($\mathcal{R}\approx 32,000$) by broadening each template spectrum via Gaussian kernel smoothing.  We repeat for six different values of smoothing bandwidths: for modeling M2FS `HiRes' and Hectochelle spectra we use \lsf$=0.06$~\AA\ , $0.09$~\AA, and $0.12$~\AA\ (resolving power $\mathcal{R}\approx 37,000$, $24,000$, and $18,000$, respectively, at $\lambda=5200$~\AA).  For modeling M2FS `MedRes' spectra we use \lsf$=0.20$~\AA\ , $0.30$~\AA, and $0.40$~\AA\ (resolving power $\mathcal{R}\approx 11,000$, $7,400$, and $5,500$, respectively.

\smallskip
Thus we obtain a library of `raw' synthetic stellar template spectra that discretely samples over a regular grid spanning a finite, 5-dimensional volume.  We denote as $T_{{\mathbf{\theta}_0}}(\lambda)$ the raw template corresponding to grid point  $\mathbf{\theta}_0\equiv ($\teff, \logg, \feh, \mgfe, and \lsf$)$.  In order to evaluate models at arbitrary location (i.e., not necessarily at grid points), we combine the $2^5=32$ surrounding raw templates via five-dimensional Gaussian kernel smoothing:
\begin{equation}
    T_{\mathbf{\theta}}\,(\lambda)=\frac{\displaystyle\sum_{i=1}^{32}T_{\mathbf{\theta}_{0},i}(\lambda)\,K_H\bigl (\mathbf{\theta}_{0,i}-\mathbf{\theta}\bigr )}{\displaystyle\sum_{i=1}^{32}K_H\bigl (\mathbf{\theta}_{0,i}-\mathbf{\theta}\bigr )}, 
\end{equation}
where $K_H(\mathbf{x})\equiv \exp\bigl [-\frac{1}{2}\mathbf{x}^T\mathbf{H}^{-1} \mathbf{x}\bigr ]$, and we adopt diagonal bandwidth matrix $\mathbf{H}=\mathrm{diag}(\mathbf{h}\circ\mathbf{h})$, with $\mathbf{h}=(300\,\mathrm{K},0.5, 0.25, 0.2, 0.03\AA)$ so that the smoothing bandwidth in each dimension equals the grid spacing.  We note that, as a result of this nearest-neighbor smoothing, $T_{\mathbf{\theta}}(\lambda)$ is not strictly a continuous function of $\mathbf{\theta}$ and does not necessarily equal $T_{\mathbf{\theta}_0}(\lambda)$ when evaluated at grid points.  Nevertheless, tests with mock spectra generated directly from templates indicate reliable recovery of input parameters \citep{walker15}.  

\subsubsection{Inference}
\label{sec:inference}

We estimate model parameters via Bayesian inference.  Given observed spectrum $S$, the model specified by free parameter vector $\mathbf{\theta}$ has posterior probability distribution
\begin{equation}
    P(\mathbf{\theta}|S)=\frac{P(S|\mathbf{\theta})\,P(\mathbf{\theta})}{P(S)},
    \label{eq:bayes}
\end{equation}
where $P(S|\mathbf{\theta})$ is the conditional probability, given the model (or `likelihood'), of obtaining the observed spectrum, $P(\mathbf{\theta})$ is the prior probability distribution function for model parameters, and 
\begin{equation}
    P(S)\equiv\int P(S|\mathbf{\theta})\,P(\mathbf{\theta})\,d\mathbf{\theta}
    \label{eq:evidence}
\end{equation}
is the marginal likelihood.  Assuming independence among the counts at all $N_{\rm pix}$ pixels, the spectrum has likelihood
\begin{equation}
    P(S|\mathbf{\theta})=\displaystyle\prod_{i=1}^{N_{\rm pix}}\mathcal{N}_{S_i}\bigl (M(\lambda_i),\sigma_i^2\bigr ),
    \label{eq:likelihood}
\end{equation}
where
\begin{equation}
    \mathcal{N}_{S_i}(M(\lambda_i),\sigma_i^2)\equiv \frac{1}{\sqrt{2\pi \sigma_i^2}}\exp\biggl [-\frac{1}{2}\frac{\bigl (S_i-M(\lambda_i)\bigr )^2}{\sigma_i^2}  \biggr]
    \label{eq:gaussian}
\end{equation}
is the normal distribution, with mean $M(\lambda_i)$ equal to the model prediction (Equation \ref{eq:model}) for the count at the wavelength assigned to pixel $i$ in the observed spectrum, and variance
\begin{equation}
    \sigma_i^2\equiv s_1\hat{\sigma}^2_{S_i}+s_2^2
    \label{eq:rescale}
\end{equation}
allows for a linear correction to the variance originally estimated for the observed count.  In practice, given the fixed and discrete wavelength sampling of our template spectra, we evaluate (the logarithm of) Equation \ref{eq:likelihood} after performing a linear interpolation of $M(\lambda)$ onto the wavelengths assigned to pixels in the  observed spectrum. 

\smallskip
Our model contains 16 free parameters.  Table \ref{tab:priors} lists each parameter, along with the range over which the priors that we adopt are uniform and nonzero. 

\begin{table*}
  \scriptsize
  \centering
%  \begin{minipage}{240mm}
    \caption{Free parameters and priors of Spectral Model   \label{tab:priors}}
    \begin{tabular}{@{}lllllll@{}}
      \hline
      parameter& prior& description\\
      \hline
    \vlos/(\kms)&uniform between $-500,+500$&line-of-sight velocity\\
      \teff/K&uniform between $3900,7500$&effective temperature\\
      \logg &uniform between $0,5$& base-10 logarithm of surface gravity, cgs units\\
      \feh &uniform between $-4.0,+0.5$& iron abundance\\
      \mgfe &uniform between $-0.8,+1.0$& magnesium abundance\\
      $p_0$&uniform between$^{a}$ $-\max[S(\lambda)],+\max[S(\lambda)]$&polynomial coefficient (continuum; eq \ref{eq:polynomial1})\\
      $p_1$&uniform between $-\max[S(\lambda)],+\max[S(\lambda)]$&polynomial coefficient (continuum; eq \ref{eq:polynomial1})\\
      $p_2$&uniform between $-\max[S(\lambda)],+\max[S(\lambda)]$&polynomial coefficient (continuum; eq \ref{eq:polynomial1})\\
      $p_3$&uniform between $-\max[S(\lambda)],+\max[S(\lambda)]$&polynomial coefficient (continuum; eq \ref{eq:polynomial1})\\
      $p_4$&uniform between $-\max[S(\lambda)],+\max[S(\lambda)]$&polynomial coefficient (continuum; eq \ref{eq:polynomial1})\\
      $p_5$&uniform between $-\max[S(\lambda)],+\max[S(\lambda)]$&polynomial coefficient (continuum; eq \ref{eq:polynomial1})\\
      $q_1/$(\kms)&uniform between $-10,+10$&polynomial coefficient (wavelength solution; eq. \ref{eq:polynomial2})\\
      $q_2$/(\kms)&uniform between $-10,+10$&polynomial coefficient (wavelength solution; eq. \ref{eq:polynomial2})\\
      $\sigma_{\rm LSF}/$\AA &uniform between $0.06,0.12$ (M2FS HiRes, Hectochelle)& bandwidth of Gaussian kernel to broaden line spread function\\
    $\sigma_{\rm LSF}/$\AA &uniform between $0.2,0.4$ (M2FS MedRes)& bandwidth of Gaussian kernel to broaden line spread function\\
      $\log_{10}s_1$&uniform between $-1,+6$&rescales observational errors (eq. \ref{eq:rescale}) \\
      $\log_{10}s_2$&uniform between $-2,+2$&adds to
      observational errors (eq. \ref{eq:rescale}) \\
      \hline
    \end{tabular}
    \\
    $^{a}$ $\max[S(\lambda)]$ is the maximum value (discounting
    pixels flagged as cosmic rays) of the sky-subtracted spectrum.
%  \end{minipage}
\end{table*}

\smallskip
We use the software package MultiNest \citep{feroz08,feroz09} to perform the inference.  MultiNest implements a nested sampling algorithm  \citep{skilling04} explicitly to compute the integral in Equation \ref{eq:evidence}.  As part of this procedure it obtains a random sample from the posterior PDF (Equation \ref{eq:bayes}).  These samples, for all of our M2FS and Hectochelle spectra, are provided along with the spectra at https://cmu.box.com/v/m2fs-hectochelle.  

\smallskip
For convenience and simplicity of downstream analysis, we use simple statistics to summarize the full posterior PDFs.  Specifically, we use MultiNest's random sampling of the PDF to estimate the mean, standard deviation, skew and kurtosis of the marginal (1D) posterior PDF for each model parameter.    

\subsection{Internal Validation}
\label{sec:internalvalidation}
In previous work we have used the  summary statistics for posterior PDFs to define quality control filters.  For example, \citet{walker15} discard any observation for which the sampled marginal PDF for \vlos\ has standard deviation $>5$ \kms, and/or skew and/or kurtosis deviating by more than one from the Gaussian value of zero.  However, we find that the skew/kurtosis filters are approximately redundant with the cut on standard deviation alone.  Therefore, in the analysis that follows, by default we discard only those observations for which the sampled marginal PDF for \vlos\ has standard deviation $>5$ \kms.  Of course, other users might reasonably choose other quality control criteria, depending on scientific goals.

\begin{figure}
    \includegraphics[width=3.5in, trim=0.0in 2.7in 3.in 0.5in, clip]{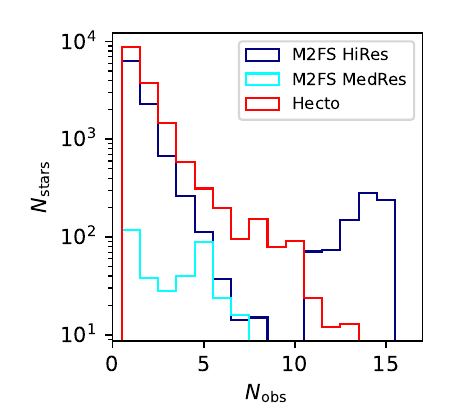}
    \caption{Distribution of number of independent measurements (having line-of-sight velocity error $<5$ \kms), for M2FS HiRes (blue), M2FS MedRes (cyan) and Hectochelle (red) samples.  The bump in the M2FS HiRes sample at $N_{\rm obs}>10$ is contributed entirely by repeated observations of the Tucana II dwarf galaxy.    
\label{fig:nobs}}
\end{figure}

\smallskip
In our M2FS HiRes (M2FS MedRes, Hectochelle) sample, \mtwofshiresgoodobs\  (\mtwofsmedresgoodobs, \hectogoodobs) spectra yield measurements that pass our simple quality-control filter.  These spectra come from \mtwofshiresgoodstar\  (\mtwofsmedresgoodstar, \hectogoodstar) unique sources, with \mtwofshiresrepeated\  (\mtwofsmedresrepeated, \hectorepeated) sources having multiple independent measurements.  Figure \ref{fig:nobs} displays the distribution of number of independent measurements per star.  As the number of independent measurements increases, the number of stars having that number of measurements declines approximately as a power law, with the the M2FS sample containing stars having as many as 16 measurements, and the Hectochelle sample containing stars having as many as 14 measurements.  In the M2FS sample, all stars having more than 10 measurements come from repeated observations of the Tucana II dwarf galaxy.  

\smallskip
We use the stars with repeat observations to fit models that specify the observational error associated with each measurement of each physical model parameter (\vlos,\teff,\logg,\feh,\mgfe).  For a given physical parameter, denoted here generically as $X$, we consider all pairs of independent measurements, $X_1$ and $X_2$, of the same sources.  Following \citet{li19}, we assume that deviations $\Delta X\equiv X_1-X_2$ are distributed as a mixture of two Gaussian distributions.  The first has variance set by formal observational errors; the second, which allows for `outlier' measurements---including spurious measurements and/or cases of true variability, as with velocities measured for stars in binary systems---has constant variance $\sigma^2_{\rm out}$ that is unrelated to formal observational errors.  That is, given zero-point offset $\mu_{\Delta X}$, variance $\sigma^2_{\Delta X}\equiv \sigma_{X_1}^2+\sigma_{X_2}^2$ that is set by formal observational errors $\sigma_{X_1}$ and $\sigma_{X_2}$, outlier variance $\sigma^2_{\rm out}$ and outlier fraction $f_{\rm out}$, the deviation between measurements 1 and 2 of a common source has probability
\begin{eqnarray}
    P(\Delta X|\,\mu_{\Delta X},\sigma^2_{\Delta X},f_{\rm out},\sigma_{\rm out})\hspace{1.2in}\nonumber\\
    =(1-f_{\rm out})\mathcal{N}(\mu_{\Delta X},\sigma_{\Delta X}^2)   +f_{\rm out}\,\mathcal{N}(0,\sigma_{\rm out}^2).\hspace{0.2in}
\label{eq:error_mixture}    
\end{eqnarray}
where $\mathcal{N}(\mu,\sigma^2)$ denotes the normal distribution with mean $\mu$ and variance $\sigma^2$.  We assume $\mu_{\Delta X}=0$ when comparing measurements from the same instrument, as in this section, but not when comparing measurements from different instruments, as in Section \ref{sec:external}.  We model the formal random errors as linear (in quadrature) functions of the standard deviations, denoted $\sigma_{X_{\rm 1,MN}}$ and $\sigma_{X_{\rm 2,MN}}$, obtained directly from MultiNest's random sampling of the marginal (1D) posterior PDF for parameter $X$.  That is, we assume  
\begin{eqnarray}
    \sigma^2_{X_1}=s^2+k^2\sigma_{X_{\rm 1,MN}}^2,\nonumber\\
    \sigma^2_{X_2}=s^2+k^2\sigma_{X_{\rm 2,MN}}^2,
    \label{eq:floorscale}
\end{eqnarray}
and similar for all pairs of measurements obtained for common sources that deviate by amounts smaller than a threshold, $|\Delta X|_{\rm out}$.  We assume that deviations larger than $|\Delta X|_{\rm out}$ are are contributed by spurious measurements, which we then exclude from our analysis (but not from the catalogs presented below).  We take $|\Delta V_{\rm LOS}|_{\rm out}= 100$ \kms, $|\Delta T_{\rm eff}|_{\rm out}=2000$ K, $|\Delta \log\mathrm{g}|_{\rm out}=2.5$ dex, $|\Delta [\mathrm{Fe/H}]|_{\rm out}=2.5$ dex and $|\Delta [\mathrm{Mg/Fe}]|_{\rm out}=1.0$ dex.  We assume that a single value of the error `floor', $s$, and a single value of scaling parameter, $k$, hold across the entire sample obtained with a given telescope/instrument.  The total set of deviations, over all $N_{\rm pair}$ pairs of measurements, has likelihood
\begin{equation}
\prod_{i=1}^{N_{\rm pair}}\frac{P(\Delta X_i|\mu_{\Delta X},\sigma_{X_1},\sigma_{X_2},f_{\rm out},\sigma_{\rm out})}{\displaystyle\int_{-|\Delta X|_{\rm out}}^{+|\Delta X|_{\rm out}} P(\Delta X_i|\mu_{\Delta X}\sigma_{X_1},\sigma_{X_2},f_{\rm out},\sigma_{\rm out}) \,d(\Delta X)}.
\label{eq:error_likelihood}
\end{equation}
We consider all pairs of measurements that both satisfy our crude quality-control criterion (velocity error $\leq 5$ \kms) for common sources, excluding measurements from sources listed in Gaia's (DR3) catalog of RR Lyrae variables (see Section \ref{sec:anomalies}).  This selection gives $N_{\rm pair}=$ \mtwofsvnpairsHiRes\ for M2FS HiRes, $N_{\rm pair}=$ \mtwofsvnpairsMedRes\ for M2FS MedRes, and $N_{\rm pair}=$ \hectovnpairs\ for Hectochelle.  We use MultiNest to perform the inference.  For each of the five physical parameters we infer from spectra, Table \ref{tab:error_adjust} lists the prior for each of the four parameters of our error model, as well as the mean and standard deviation of the marginal posterior PDF.  
\begin{figure*}
    \includegraphics[width=7.5in,trim=0.3in 2.0in 0.3in 0.5in, clip]{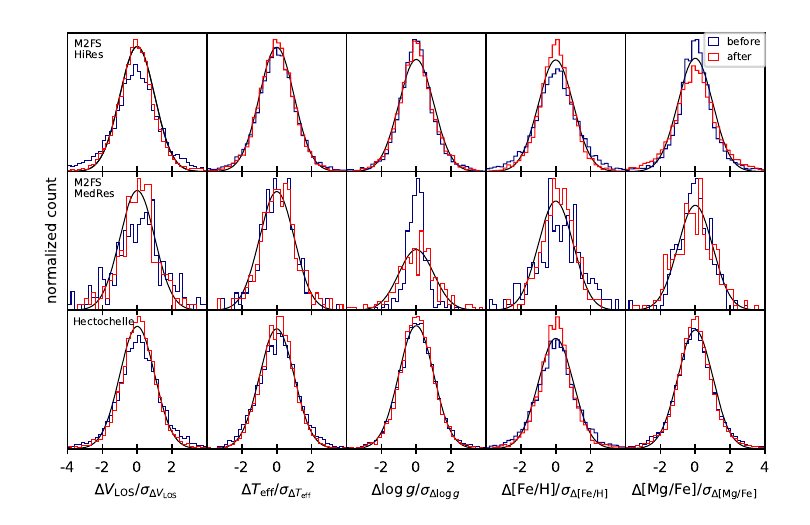}
    \caption{Internal validation of formal uncertainties. For each of the spectroscopic observables \vlos, \teff, \logg, \feh, \mgfe, panels indicate distributions of pair-wise deviations between independent measurements of the same target, normalized by the combined error in both measurements.  Errors are calculated using the standard deviation taken directly from the marginal posterior PDFs returned by MultiNest (blue `before' histograms), and using the formal errors obtained by fitting the error model described in Section \ref{sec:internalvalidation} (red `after' histograms).  Individual panels in the top, middle and bottom rows show results from \mtwofsvnpairsHiRes\ pairs of M2FS  HiRes observations, \mtwofsvnpairsMedRes\ pairs of M2FS MedRes observations, and \hectovnpairs\  pairs of Hectochelle observations, respectively.  In all panels, the solid black curve is the standard normal distribution.  
\label{fig:error_adjust}}
\end{figure*} 

\smallskip
For M2FS HiRes (MedRes), we infer error `floors' of $s_{V_{\rm LOS}}=$ \mtwofssigvfloorHiRes\ km s$^{-1}$ (\mtwofssigvfloorMedRes\ \kms), $s_{T_{\rm eff}}=$ \mtwofssigtefffloorHiRes\ K (\mtwofssigtefffloorMedRes\ K), $s_{\rm logg}=$ \mtwofssigloggfloorHiRes\ (\mtwofssigloggfloorMedRes), $s_{\rm [Fe/H]}=$ \mtwofssigfehfloorHiRes\ (\mtwofssigfehfloorMedRes), $s_{\rm [Mg/Fe]}=$ \mtwofssigmgfefloorHiRes\ (\mtwofssigmgfefloorMedRes).  For Hectochelle, the floors are all lower, presumably as a  benefit of wider spectral coverage, with $s_{V_{\rm LOS}}=$ \hectosigvfloor\ \kms, $s_{T_{\rm eff}}=$ \hectosigtefffloor\ K, $s_{\rm logg}=$ \hectosigloggfloor, $s_{\rm [Fe/H]}=$ \hectosigfehfloor, $s_{\rm [Mg/Fe]}=$ \hectosigmgfefloor.  The inferred scaling parameters are scattered around unity, in several cases (including the velocity measurements for M2FS HiRes and Hectochelle) consistent with a value of unity within the 99\% credible interval.  The outlier fraction tends to comprise $\la 10\%$ of the samples, except for the measurements of \feh\ and \mgfe, where the outlier fractions reach $\sim 20-50\%$.  Analyses of chemical abundance distributions may therefore benefit from stricter sample selection criteria than our fiducial one that is based solely on the formal error in \vlos.

\begin{deluxetable*}{lllllllll}
\tablewidth{0pt}
\tablecaption{Summary of posterior PDFs for parameters of ther model used to adjust observational errors (Section \ref{sec:internalvalidation}). \label{tab:error_adjust}
}
\tablehead{\colhead{quantity}&\colhead{$s$}&\colhead{$k$}&\colhead{$f_{\rm out}$}&\colhead{$\sigma_{\rm out}$}\\
&\colhead{(floor)}&\colhead{(multiplier)}&\colhead{(outlier fraction)}&\colhead{(outlier std. dev.)}
}
\startdata
\underline{M2FS HiRes}\\
$V_{\rm los}$ &\mtwofssigvfloorHiRes $\,$\kms&\mtwofssigvscaleHiRes&\mtwofssigvmixfracHiRes&\mtwofssigvsigoutHiRes$\,$\kms\\
$T_{\rm eff}$ &\mtwofssigtefffloorHiRes$\,$ K&\mtwofssigteffscaleHiRes&\mtwofssigteffmixfracHiRes&\mtwofssigteffsigoutHiRes$\,$ K\\
$\log_{10}[g]$ &\mtwofssigloggfloorHiRes&\mtwofssigloggscaleHiRes&\mtwofssigloggmixfracHiRes&\mtwofssigloggsigoutHiRes\\
$[\mathrm{Fe/H]}$ &\mtwofssigfehfloorHiRes&\mtwofssigfehscaleHiRes&\mtwofssigfehmixfracHiRes&\mtwofssigfehsigoutHiRes\\
$[\mathrm{Mg/Fe}]$ &\mtwofssigmgfefloorHiRes&\mtwofssigmgfescaleHiRes&\mtwofssigmgfemixfracHiRes&\mtwofssigmgfesigoutHiRes\\
\\
\underline{M2FS MedRes}\\
$V_{\rm los}$ &\mtwofssigvfloorMedRes $\,$\kms&\mtwofssigvscaleMedRes&\mtwofssigvmixfracMedRes&\mtwofssigvsigoutMedRes$\,$\kms\\
$T_{\rm eff}$ &\mtwofssigtefffloorMedRes$\,$ K&\mtwofssigteffscaleMedRes&\mtwofssigteffmixfracMedRes&\mtwofssigteffsigoutMedRes$\,$ K\\
$\log_{10}[g]$ &\mtwofssigloggfloorMedRes&\mtwofssigloggscaleMedRes&\mtwofssigloggmixfracMedRes&\mtwofssigloggsigoutMedRes\\
$[\mathrm{Fe/H]}$ &\mtwofssigfehfloorMedRes&\mtwofssigfehscaleMedRes&\mtwofssigfehmixfracMedRes&\mtwofssigfehsigoutMedRes\\
$[\mathrm{Mg/Fe}]$ &\mtwofssigmgfefloorMedRes&\mtwofssigmgfescaleMedRes&\mtwofssigmgfemixfracMedRes&\mtwofssigmgfesigoutMedRes\\
\\
\underline{MMT/Hectochelle}\\
$V_{\rm los}$ &\hectosigvfloor $\,$\kms &\hectosigvscale&\hectosigvmixfrac&\hectosigvsigout $\,$ \kms\\
$T_{\rm eff}$ &\hectosigtefffloor $\,$K&\hectosigteffscale&\hectosigteffmixfrac&\hectosigteffsigout $\,$K\\
$\log_{10}[g]$ &\hectosigloggfloor&\hectosigloggscale&\hectosigloggmixfrac&\hectosigloggsigout\\
$[\mathrm{Fe/H]}$ &\hectosigfehfloor&\hectosigfehscale&\hectosigfehmixfrac&\hectosigfehsigout\\
$[\mathrm{Mg/Fe}]$ &\hectosigmgfefloor&\hectosigmgfescale&\hectosigmgfemixfrac&\hectosigmgfesigout\\
\enddata
\end{deluxetable*}

\smallskip
Figure \ref{fig:error_adjust} shows distributions of pair-wise measurement deviations normalized by combined measurement errors, with the combined measurement error calculated from the standard deviations of the posterior PDF originally sampled by MultiNest, $\sigma_{\Delta X_{\rm MN}}=\sqrt{\sigma_{X_{\rm 1,MN}}^2+\sigma_{X_{\rm 2,MN}}^2}$ (black histograms), and from the formal errors returned by the best-fitting error model, $\sigma_{\Delta X}=\sqrt{\sigma_{X_1}^2+\sigma_{X_2}^2}$ (red histograms).  By design, the latter are generally closer to the standard normal distribution (solid black curves).  In our data catalogs, the columns `X\_error' list the errors for observable `X' after performing the adjustment of Equation \ref{eq:floorscale}, with mean values of error model parameters listed in Table \ref{tab:error_adjust}.  The columns `X\_error\_raw' list the pre-adjusted values obtained directly from the posterior sampled by MultiNest.

\subsection{External Comparisons}
\label{sec:external}

\begin{figure}    \includegraphics[width=3.75in,trim=0in 1.in 3in 0.5in, clip]{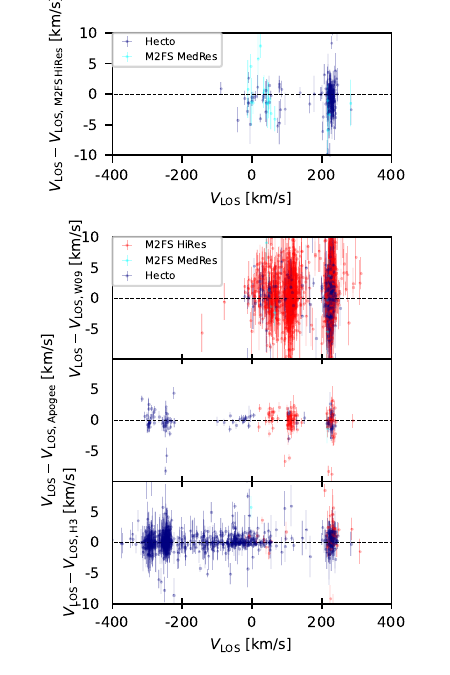}
  \caption{\textit{Top row:} Difference between line-of-sight velocities measured by M2FS HiRes and either M2FS MedRes (cyan) or Hectochelle (blue), for stars common to both samples.  \textit{Bottom three rows:} Differences between line-of-sight velocities we measure using M2FS HiRes (red), M2FS MedRes (cyan) or Hectochelle (blue) and those measured in external surveys by \citet[][second row]{walker09a}, APOGEE \citep[DR17;][third row]{apogee_dr17}, the H3 survey \citep[][fourth row]{conroy19}. 
    \label{fig:compare_vlos}}
\end{figure}

\begin{figure*}
    \includegraphics[width=7.5in, trim=0.in 1.35in 0in 0in, clip]{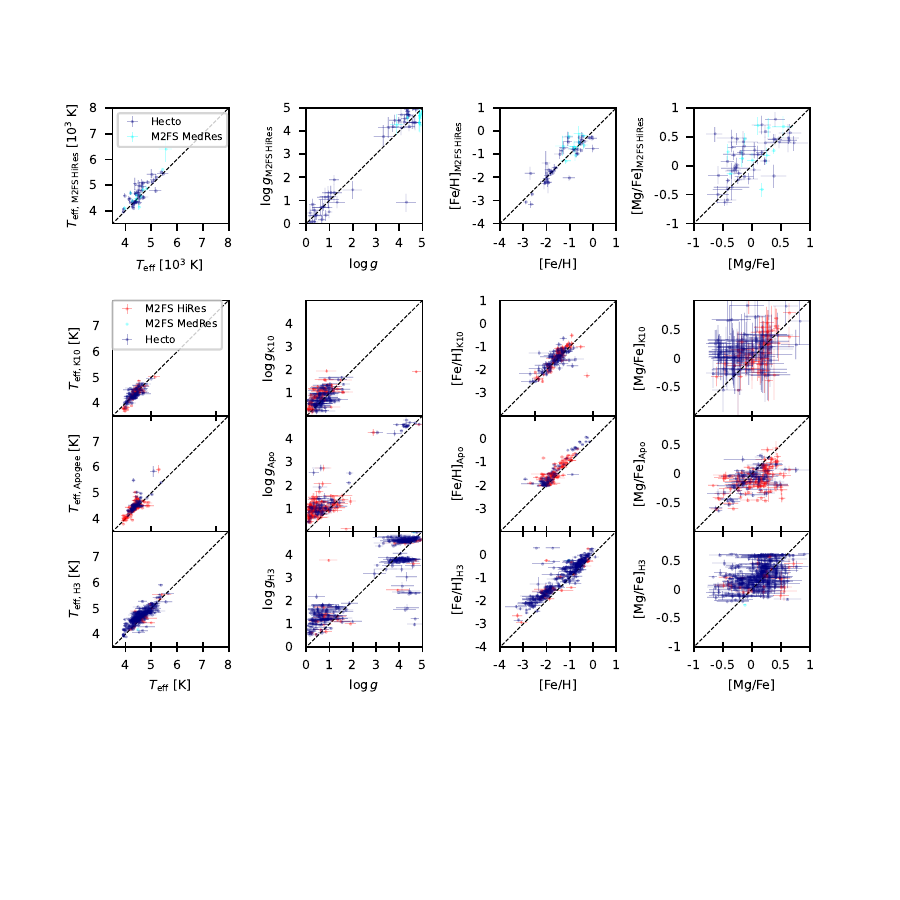}\\
    \caption{\textit{Top row:} Comparison of stellar-atmospheric parameters measured (before applying zero-point adjustments) by M2FS HiRes and either M2FS MedRes (cyan) or  Hectochelle (blue), for stars common to both samples.  \textit{Bottom three rows:} Comparison of parameters that we measure (before applying zero-point adjustments) using M2FS (red) and Hectochelle (blue) to those measured in external surveys by \citet[][second row]{kirby10}, APOGEE \citep[DR17;][third row]{apogee_dr17}, and the H3 survey \citep[][fourth row]{conroy19}.  In all panels of the bottom three rows, the quantity plotted along the horizontal axis is the measurement from M2FS (red) and/or Hectochelle (blue).
\label{fig:compare_atm}}
\end{figure*}

We compare our M2FS and Hectochelle catalogs directly to each other and to large spectroscopic data sets that are previously published and/or in progress.  Our primary goal is to detect and quantify systematic differences, e.g.,  zero-point offsets.  The top panels of Figures \ref{fig:compare_vlos} and \ref{fig:compare_atm} compare velocities and stellar-atmospheric parameters, respectively, that we measure with M2FS HiRes, M2FS MedRes and Hectochelle, for all stars that appear in at least two instrument-specific samples.  In both figures, the bottom three rows of panels compare our M2FS and Hectochelle measurements to those from external catalogs by \citet[][`W09' hereafter]{walker09a}, \citet[][`K10' hereafter]{kirby10}, the Sloan Digital Sky Survey's APOGEE project  \citep[DR17]{apogee_dr17}, and the  Hectochelle in the Halo at High Resolution Survey \citep[][`H3'  hereafter]{conroy19}.  

\smallskip
W09's catalog includes 8855 line-of-sight velocities measured for 7103 unique sources toward the dwarf spheroidal galaxies Carina, Fornax, Sculptor and Sextans.  The W09 spectra were acquired using the Michigan-MIKE Fiber System \citep{walker07c}, a precursor to M2FS at Magellan that operated at similar spectral resolution over a similar spectral range.  The W09 catalog has \mtwofshireswalkervn\ sources in common with our current M2FS HiRes sample, \walkermtwofsmedresvn\ sources in common with our M2FS MedRes sample, and \hectowalkervn\ sources  in common with our Hectochelle sample.  While W09 measure spectroscopic indices for iron and magnesium absorption features, they do not measure the set of stellar-atmospheric parameters that we have in our current samples.  Our comparisons to W09's catalog are therefore limited to line-of-sight velocities.  

\smallskip
K10 measure \teff, \logg, \feh\ and \mgfe\ for $\sim 3000$ stars in eight of the Milky Way's dSph satellites.  The K10 catalog has \mtwofshireskirbyteffn\ (all HiRes mode) and \hectokirbyteffn\ stars in common with our M2FS and Hectochelle samples, respectively.  The K10 spectra have resolving power $\mathcal{R} \sim 6500$ near the calcium triplet at $\lambda\sim 8500$ \AA, probing a different wavelength range at lower resolution than the other catalogs considered here.  In contrast to our estimates of \teff\ and \logg, which rely entirely on information contained in the spectrum, K10  incorporate stellar photometry into their estimate of \teff\, and use photometry alone to estimate \logg.  The K10 catalog does not list measurements of \vlos; therefore, our comparisons to K10's catalog are limited to stellar-atmospheric parameters.  

\smallskip
The APOGEE catalog, from the 17th data release of the Sloan Digital Sky Survey (DR 17; \citealt{apogee_dr17}), includes  line-of-sight velocities and stellar-atmospheric parameters measured from high-resolution ($\mathcal{R}\sim 22,500$ over $\sim 1.5-1.7$~microns in wavelength) spectra obtained for $\sim 650,000$ stars in the Milky Way and a few of its dwarf galaxy satellites.  We select all sources from the APOGEE DR17 `allstar' catalog for which the APOGEE Stellar Parameters and Abundances Pipeline (ASPCAP) returns measurements for all of \teff, \logg, \feh\ and \mgfe\ (ASPCAP lists separate measurements of  \mgfe\ and [$\alpha$/Fe]; we use only the former for purposes of direct comparison).  We then discard any sources for which the `RV\_FLAG' bitmask has the `RV\_SUSPECT' bit set, and/or the `ASPCAPFLAG' bitmask has the `STAR\_WARN' bit set.  After applying these filters and then removing stars for which we measure \feh$<-2.5$ (i.e., below the minimum metallicity of APOGEE's template spectra), there are \mtwofshiresapogeevn\ APOGEE stars in common with our M2FS HiRes sample, \mtwofsmedresapogeevn\ stars in common with our M2FS MedRes sample, and \hectoapogeevn\ in common with our Hectochelle sample.  For a given star, we take the mean APOGEE velocity as given by the `VHELIO\_AVG' parameter, with observational error given by `VERR'.  

\smallskip
Finally, the H3 Survey \citep{conroy19} is ongoing, using the same MMT/Hectochelle configuration that we do.  H3 is designed to map the Galactic stellar halo, targeting $\sim 2\times 10^5$ halo stars down to a magnitude limit of $r\lesssim 18$.  H3's and our spectra are acquired and modeled independently, but processed using the same CfA pipeline discussed at the beginning of Section \ref{sec:processing}.  The H3 team models individual spectra using the software package MINESweeper \citep{cargile20}, which simultaneously fits isochrone models to stellar magnitudes measured from broad-band photometry.  H3's incorporation of photometric information provides additional power to constrain stellar-atmospheric parameters, while also giving capability to infer spectro-photometric distances to individual sources.  

\smallskip 
The faint end of H3's sample overlaps only slightly with the bright end of ours, leaving relatively few stars common to both surveys.  In order to provide a more meaningful basis for comparison, the H3 team applied their MINESweeper analysis directly to our Hectochelle spectra from four different fields in the Sextans dSph galaxy (P. Cargile, private communication).  While not part of the actual H3 survey, this comparison `H3' sample contains  \mtwofshireshthreevn\ sources common to our M2FS HiRes sample, \hthreemtwofsmedresvn\ sources common to our M2FS MedRes sample, and \hectohthreevn\ sources common to our Hectochelle sample. 

\smallskip
For a given observable quantity, $X$, we infer zero-point offsets for each of the above catalogs simultaneously.  We begin by constructing vectors of deviations, $\Delta X\equiv X_1-X_2$, and corresponding errors, $\sigma_{\Delta X}=\sqrt{\sigma_{X,1}^2+\sigma_{X_2}^2}$, for all pairs of sources common to different catalogs `1' and `2'.  We loop over all possible combinations of catalogs, such that a star appearing at least once in all six catalogs will have 10 pairs of measurements\footnote{Recall that the W09 catalog lacks stellar-atmospheric parameters and the K10 catalog lacks line-of-sight velocities, so measurements of a given quantity for a given star can appear in up to five different catalogs.}; within a given catalog, multiple measurements of the same source are replaced by the inverse-variance-weighted mean value.

\smallskip
We assume that, for a given observable, $X$, the pair-wise deviations, $\Delta X,$ follow a Gaussian distribution, with standard deviation $\sigma_{\Delta X}$ and  pair-dependent mean, $\mu_{\Delta X}=\overline{\Delta X_1}-\overline{\Delta X_2}$, that is specified by the difference in mean offsets (from some standard zero point) of catalogs 1 and 2.  This model and the corresponding likelihood function can be specified by Equations \ref{eq:error_mixture} and \ref{eq:error_likelihood}, respectively, only now with the outlier fraction assumed to be $f_{\rm out}=0$ and the catalogued observational errors taken at face value ($s=0,\, k=1$).  In order to guard against catastrophic outliers, as described in Section \ref{sec:internalvalidation}, we discard pairs with deviations in excess of $|\Delta V_{\rm LOS}|_{\rm out}= 100$ \kms, $|\Delta T_{\rm eff}|_{\rm out}=2000$ K, $|\Delta \log\mathrm{g}|_{\rm out}=2.5$ dex, $|\Delta [\mathrm{Fe/H}]|_{\rm out}=2.5$ dex and $|\Delta [\mathrm{Mg/Fe}]|_{\rm out}=1.0$ dex.  Four free parameters specify zero-point offsets: $\overline{\Delta X}_{\rm M2FS}$, $\overline{\Delta X}_{\rm Hecto}$, $\overline{\Delta X}_{\rm H3}$, and $\overline{\Delta X}_{\rm  W09}$ (if $X=$\vlos), $\overline{\Delta X}_{\rm K10}$ (if $X=$\teff, \logg, \feh\ or \mgfe).  

\smallskip
Given the APOGEE catalog's size and widespread use across different sub-fields, we choose that catalog to define the absolute zero point, assuming $\overline{\Delta X}_{\rm Apo}=0$ for all $X$.  Table \ref{tab:compare_previous} lists offsets, relative to the APOGEE zero point, that we infer (again via MultiNest, as in Section \ref{sec:internalvalidation}) for each observable and each catalog.  Positive offsets, $\overline{\Delta X}>0$,  imply that a catalog's zero point is more positive than APOGEE's.  For each catalog named in Column 1, Columns 2--7 identify the number of pairs of sources in common with each of the other individual catalogs. 

\begin{deluxetable*}{llllllllllllll}
\tablewidth{0pt}
\tablecaption{Zero-point offsets (with respect to APOGEE DR17) inferred for M2FS, Hectochelle and external data sets (Section \ref{sec:external}).
\label{tab:compare_previous}
}
\tablehead{\colhead{Sample\tablenotemark{a}}&\colhead{$N_{\rm 1}$}&\colhead{$N_{\rm 2}$}&\colhead{$N_{\rm 3}$}&\colhead{$N_{\rm 4}$}&\colhead{$N_{\rm 5}$}&\colhead{$N_{\rm 6}$}&\colhead{$N_{\rm 7}$}&\colhead{$\overline{\Delta V_{\rm LOS}}$\tablenotemark{b}}&\colhead{$\overline{\Delta T_{\rm eff}}$}&\colhead{$\overline{\Delta \log \mathrm{g}}$}&\colhead{$\overline{\Delta \mathrm{[Fe/H]}}$}&\colhead{$\overline{\Delta \mathrm{[Mg/Fe]}}$}}
\startdata 
M2FS HiRes &--- & \mtwofshireshectovn & \mtwofshiresmtwofsmedresvn &
\mtwofshireswalkervn & \mtwofshireskirbyteffn & \mtwofshireshthreevn & \mtwofshiresapogeevn & \mtwofshiresvoffset & \mtwofshiresteffoffset & \mtwofshiresloggoffset & \mtwofshiresfehoffset & \mtwofshiresalphaoffset\\
M2FS MedRes && --- & \hectomtwofsmedresvn & \walkermtwofsmedresvn & 0 & \hthreemtwofsmedresvn & \mtwofsmedresapogeevn & \mtwofsmedresvoffset & \mtwofsmedresteffoffset & \mtwofsmedresloggoffset & \mtwofsmedresfehoffset & \mtwofsmedresalphaoffset\\
Hectochelle &&& --- & \hectowalkervn & \hectokirbyteffn & \hectohthreevn & \hectoapogeevn & \hectovoffset & \hectoteffoffset & \hectologgoffset & \hectofehoffset & \hectoalphaoffset\\
W09 &&&& ---& \nodata & \hthreewalkervn & \walkerapogeevn & \walkervoffset & \nodata & \nodata &\nodata &\nodata\\
K10 &&&&& --- & \kirbyhthreeteffn & \kirbyapogeeteffn & \nodata & \kirbyteffoffset & \kirbyloggoffset & \kirbyfehoffset & \kirbyalphaoffset\\
H3\tablenotemark{c} &&&&&& --- &\hthreeapogeevn & \hthreevoffset & \hthreeteffoffset & \hthreeloggoffset & \hthreefehoffset & \hthreealphaoffset
\enddata
\tablenotetext{a}{Samples: 1=M2FS HiRes; 2=M2FS MedRes; 3=Hectochelle; 4=\citet{walker09a}; 5=\citet{kirby10}; 6=H3; 7=APOGEE DR17}
\tablenotetext{b}{A value $\overline{\Delta X}\equiv \overline{X-X_{\rm 7}}>0$ implies a zero point that is more positive than that of the APOGEE catalog.}
\tablenotetext{c}{The `H3' sample  that we use here is from the H3 team's analysis of a subset of $\sim 750$ spectra from our program.}
\end{deluxetable*}

\smallskip
Examining the results for our M2FS and Hectochelle samples, we find that, whereas the Hectochelle sample shows little velocity offset with respect to APOGEE ($\overline{\Delta V_{\rm LOS}}_{\rm Hecto}=$ \hectovoffset\ \kms), the M2FS HiRes sample is systematically offset by $\overline{\Delta V_{\rm LOS}}_{\rm M2FS,HiRes}=$ \mtwofshiresvoffset\ \kms.  The M2FS MedRes sample shows no significant offset, with $\overline{\Delta V_{\rm LOS}}_{\rm M2FS,MedRes}=$ \mtwofsmedresvoffset\ \kms, but with a large uncertainty reflecting the fact that the M2FS MedRes sample has relatively few stars in common with the other samples.  For most stellar-atmospheric parameters, both M2FS HiRes and Hectochelle samples show statistically significant offsets from APOGEE.  The offsets in surface gravity ($\overline{\Delta  \log g}_{\rm M2FS\,HiRes}=$ \mtwofshiresloggoffset\   and $\overline{\Delta  \log g}_{\rm Hecto}=$ \hectologgoffset) and metallicity ($\overline{\Delta \mathrm{[Fe/H]}}_{\rm M2FS\,HiRes}=$ \mtwofshiresfehoffset\ and $\overline{\Delta \mathrm{[Fe/H]}}_{\rm Hecto}=$ \hectofehoffset) are similar for both samples, while the difference in temperature offsets ($\overline{\Delta T_{\rm eff}}_{\rm M2FS\,HiRes}=$ \mtwofshiresteffoffset\ K and $\overline{\Delta T_{\rm eff}}_{\rm Hecto}=$ \hectoteffoffset\ K) likely reflects the different wavelength coverage of the different instruments/configurations.  However, the smaller temperature offset of the H3 sample ($\overline{\Delta T_{\rm eff}}_{\rm H3}=$ \hthreeteffoffset\ K), which uses the same Hectochelle configuration that we do, also implicates differences in analysis procedure as a source of systematic error.  Finally, while our Hectochelle sample shows good agreement with APOGEE in terms of the magnesium abundance ($\overline{\Delta \mathrm{[Mg/Fe]}}_{\rm Hecto}=$ \hectoalphaoffset), the M2FS sample is offset by $\overline{\Delta \mathrm{[Mg/Fe]}}_{\rm M2FS}=$ \mtwofshiresalphaoffset.

\smallskip
Perhaps most eye-catching among the external comparisons are those involving surface gravity in the H3 catalog (bottom row, second column of Figure \ref{fig:compare_atm}).  The H3 surface gravities are multi-modal at \logg\ $\gtrsim 2$.  This feature is likely real (i.e., reflecting a true multi-modality among the observed high-gravity stars) and detectable here because of H3's simultaneous fitting of isochrone and spectral models.  The modes at \logg\ $\sim 4.5$, \logg\ $\sim 3.7$ and \logg\ $\sim 2.5$  correspond to the main sequence, sub-giant and horizontal branches,  respectively, all of which are confined to distinct ranges of surface gravity in isochrone space.  H3's fitting of isochrone models to broad-band photometry effectively requires these evolutionary stage to be separated, giving rise to the observed multi-modality in \logg\ space. 

\smallskip
The primary lesson we take from all of these  external comparisons is that zero-point offsets among \textit{all} of the independent datasets are common at the level of a few $\times 0.1$ \kms\ in line-of-sight velocity, $\sim 100$ K in effective temperature, and a few $\times 0.1$ dex in surface gravity, metallicity and magnesium abundance.  Offsets of these magnitudes are perhaps not surprising, given the variety of spectral resolutions, wavelength ranges and analysis techniques employed.  We acknowledge that our M2FS+Hectochelle results for individual stars are susceptible to systematic errors at these levels.

\smallskip
In the M2FS (HiRes and MedRes) and Hectochelle catalogs presented below, we subtract from each individual measurement of \vlos, \teff, \logg, \feh\ and \mgfe\ the zero-point offset listed in Table \ref{tab:compare_previous}, such that the catalogs are effectively shifted to the APOGEE zero point.  Table columns labeled `X' list values of observable `X' after shifting to the Apogee zero point.  Columns labeled `X\_raw' list the original values---i.e., before applying the zero-point correction.

\smallskip
After applying the zero-point corrections, we compare our current M2FS and Hectochelle catalogs to measurements that we have previously published for subsets of the current samples---including stellar targets in the dwarf galaxies Draco, Reticulum~II, Tucana~II, Grus~I, Crater~II, Leo~II, Ursa Minor,  Hydrus~I and Fornax \citep{walker15,walker15b,walker16,caldwell17,spencer17,spencer18,koposov18,pace21}.  Despite using the same raw M2FS+Hectochelle spectra, the previously-published measurements can differ systematically from current ones even before applying zero-point corrections, as they are derived using an entirely different library of synthetic template spectra.  Specifically, the previously-published measurements are based not on the library we introduce in Section \ref{sec:ian}, but instead on a library that was designed originally for use with the SDSS Segue Stellar Parameter Pipeline \citep[`SSPP'][]{lee08}.  The SSPP library is computed over a fixed grid in \teff\, \logg\ and \feh, and assumes a monotonic relationship between $\alpha$-element abundance and \feh.  Experimenting with three independent libraries of synthetic template spectra, \citet{walker15} observed library-dependent zero-point offsets as large as $\overline{\Delta V_{\rm LOS}}\sim 0.5$ \kms,  $\overline{\Delta T_{\rm eff}}\sim 300$ K, $\overline{\Delta\log\mathrm{g}} \sim 0.7$ dex and $\overline{\Delta\mathrm{[Fe/H]}}\sim 0.5$ dex.

\smallskip
The previously published M2FS HiRes, M2FS MedRes and Hectochelle data sets contain \mtwofshiresssppvn, \mtwofsmedresssppvn\ and \hectossppvn\ sources, respectively, from our current samples.  Comparing these measurements directly to the current ones, we find that the previously-published M2FS HiRes (M2FS MedRes) measurements are offset from current (raw, i.e., before applying an offset to the APOGEE zero point) values by  $\overline{\Delta V_{\rm LOS}}=$ \mtwofshiresssppvoffset\ \kms\ (\mtwofsmedresssppvoffset\ \kms), $\overline{\Delta T_{\rm eff}}=$ \mtwofshiresssppteffoffset\ K (\mtwofsmedresssppteffoffset\ K), $\overline{\Delta\log\mathrm{g}}=$ \mtwofshiressspploggoffset\ dex (\mtwofsmedressspploggoffset\ dex) and $\overline{\Delta \mathrm{[Fe/H]}}=$ \mtwofshiresssppfehoffset\ dex (\mtwofsmedresssppfehoffset\ dex), where positive values imply that the current measurements are, on average, larger than the previously-published ones.  The previously-published Hectochelle measurements show offsets of similar magnitude, with $\overline{\Delta V_{\rm LOS}}=$ \hectossppvoffset\ \kms, $\overline{\Delta T_{\rm eff}}=$ \hectossppteffoffset\ K, $\overline{\Delta\log\mathrm{g}}=$ \hectosspploggoffset\ dex and $\overline{\Delta \mathrm{[Fe/H]}}=$ \hectossppfehoffset\ dex.  We notice that these offsets with respect to current values are similar to, or smaller than, the zero-point shifts that were applied to raw measurements in the previously-published work (see \citealt{walker15,walker15b} for details).  Those shifts were determined empirically, based on observed offsets between known solar values and values measured from high-S/N spectra acquired during twilight exposures.  Specifically, the previously-published M2FS measurements include zero-point shifts (i.e., quantities that were added to raw measurements) of $\Delta V_{\rm LOS}=0$ \kms, $\Delta T_{\rm eff}=-69$ K, $\Delta\log g=-0.09$ dex, $\Delta \mathrm{[Fe/H]}=+0.20$ dex, while the previously-published  Hectochelle measurements include shifts of $\Delta V_{\rm LOS}=-0.81$ \kms, $\Delta T_{\rm eff}=+303$ K, $\Delta\log g=+0.63$ dex, $\Delta \mathrm{[Fe/H]}=+0.48$ dex.  Based on these direct comparisons, then, we find that our switch to the new template library (described in Section \ref{sec:ian}), followed by our new zero-point calibration based on external comparisons, results in relatively small offsets from previous values.  

\begin{figure*}
    \begin{tabular}{@{}llll@{}}
    \renewcommand{\arraystretch}{0}
%    \vspace{-1.5in}
    \includegraphics[width=3.5in,trim=0.in 3.in 2.75in 0in, clip]{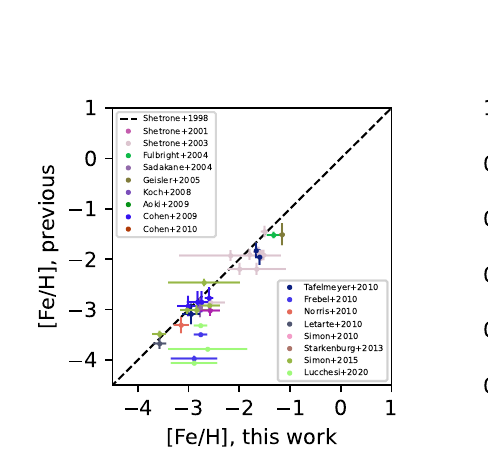}& \hspace{-0.75in}\includegraphics[width=3.5in,trim=2.75in 3.in 0in 0in, clip]{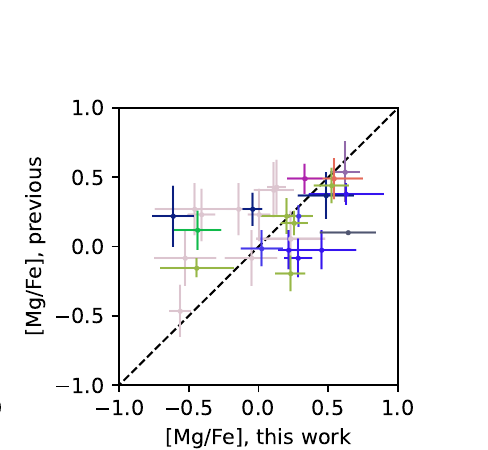}\\
%    \vspace{-1.5in}
    \end{tabular}
    \caption{Comparison of current M2FS+Hectochelle measurements of \feh\  (left) and \mgfe\ (right) to previously-published values derived from high-resolution spectra.   \label{fig:compare_lowmetallicity}}
%\label{fig:compare_lowmetallicity}
\end{figure*}

\smallskip
Finally, after having applied the zero-point calibration as discussed above, we compare our measurements of \feh\ and \mgfe\ directly to previously-published abundance measurements derived from high-resolution spectra acquired for relatively small samples of individual stars in dSph galaxies.  The external samples come from observations with the HIRES spectrograph at the Keck Telescopes \citep{shetrone01,fulbright04,cohen09,cohen10,frebel10b}, the High Dispersion Spectrograph at the Subaru Telescope \citep{sadakane04,aoki09}, the UVES \citep{shetrone03,norris10,tafelmeyer10,lucchesi20} and X-Shooter \citep{starkenburg13} spectrographs at the Very Large Telescope, and the MIKE spectrograph at Magellan \citep{simon15}.  Figure \ref{fig:compare_lowmetallicity} displays the comparisons.  

\smallskip
Comparing \feh\ metallicities (left panel of Figure \ref{fig:compare_lowmetallicity}) we find generally good agreement with the high-resolution studies.  The bulk of measurements are consistent with a small offset such that our values may be systematically metal-rich by $\sim 0.1$ dex, with no significant dependence on additional stellar-atmospheric parameters like \teff\ or \logg.  At the very metal-poor end, however, our measurements for two stars (both in the Sculptor dSph galaxy) with previously-published values \feh\ $\lesssim -4$ \citep{tafelmeyer10,simon15} both come in at \feh\ $\gtrsim -3$ in our work, disagreeing with the previous measurements at the $\sim 2\sigma$ level.  One of these stars, Scl07-50, has been identified (based on the previous measurements) as the most metal-poor star known in an external galaxy \citep{tafelmeyer10}.  It is potentially concerning that our measurements do not reproduce this result.  However, we note that our library of template spectra includes only metallicities \feh~$\geq -4$, and that our applied offsets of $\overline{\Delta\mathrm{[Fe/H]}}$ imply that the minimum metallicity that we can in principle measure is \feh\ $\sim -3.75$.  The relatively large uncertainties on our measurements of these two stars imply that the mean values will be correspondingly larger than this minimum.  We expect, therefore, that users may choose to apply stricter quality-control filters (e.g., a threshold in formal uncertainty) when analyzing chemical abundances, especially when working near the limits of our metallicity scale.  

\smallskip
Comparing \mgfe\ abundances (right panel of Figure \ref{fig:compare_lowmetallicity}, we see what is perhaps the opposite problem, as our template library extends to lower \mgfe\ than is allowed in some previous studies.  At solar and higher values of \mgfe\, we find generally good agreement with the results of previous high-resolution studies.  At sub-solar abundance, our measurements of \mgfe\ tend to be lower than those previously reported.  Again, we expect that users may want to tighten quality-control filters when analyzing chemical abundances; we note that requiring our measurement of \feh\ to have uncertainty smaller than 0.5 dex would remove from the comparison sample all but one of the stars for which we measure \mgfe\ to be sub-solar.  

\begin{figure}    \includegraphics[width=3.75in,trim=0.5in 0.2in 2.5in 0.5in, clip]{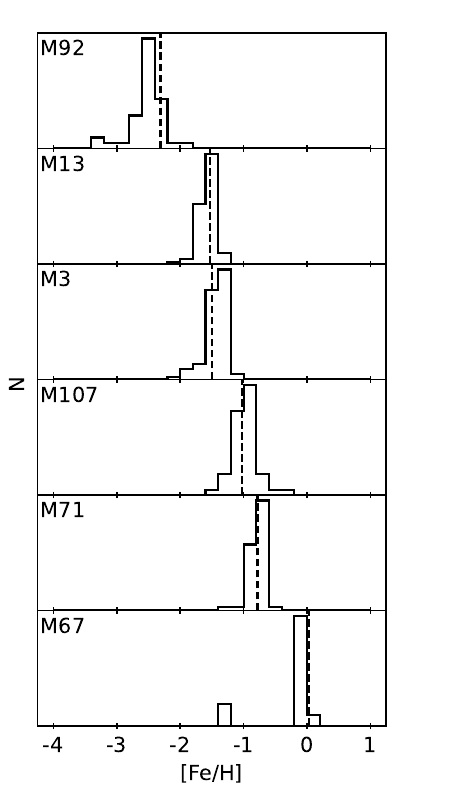}
  \caption{Histograms of metallicities we infer from archival Hectochelle observations of targets in the star clusters (top to bottom) M92, M13, M3, M107, M71 and M67.  In each panel, the the dashed vertical line indicates the metallicity tabulated by \citet{harris96}, except for the metallicity of M67, which we adopt from \citet{pace08}.  
    \label{fig:compare_gcs}}
\end{figure}

\smallskip
We perform one additional external cross-check on our metallicity measurements, fitting our spectral models (Section \ref{sec:analysis}) to archival Hectochelle spectra acquired during observations of globular and open star clusters.  These observations, performed by other investigators (including the H3 team), used the same spectrograph configuration  and processing pipeline that we employ for our own Hectochelle spectra.  Figure \ref{fig:compare_gcs} displays histograms of \feh\ that we obtain for each of the clusters M3, M13, M67, M71, M92, M107, which span a range of $-2.2\la $\feh\ $\la 0$ in metallicity. 
 For each cluster we keep only stars for which our measurements have velocity error $< 5$ km s$^{-1}$, metallicity error $< 0.5$ dex, and --- in order to reduce contamination from non-member sources --- \logg$<3$ and \vlos\ within 10 km s$^{-1}$ of the systemic mean tabulated by \citet{harris96}, except for M67, for which we adopt the spectroscopic mean velocity and metallicity measured by \citet{pace08}.  Figure \ref{fig:compare_gcs} shows the resulting distributions of \feh\ observed toward each cluster, with clear peaks associated with cluster members.  We find good agreement with the previously-published mean metallicities, giving confidence that our calibrated zero-point is accurate.

\subsection{Anomalous Sources}
\label{sec:anomalies}

Our target selection filters (Section \ref{sec:observations}) are designed to isolate primarily red giant stars in the Galactic halo substructures of interest, with contamination contributed mainly by dwarf stars in the Galactic foreground.  Our spectral templates are designed to fit individual stars within the limited range of stellar-atmospheric parameters identified in Section \ref{sec:ian}, which can accommodate the vast majority of selected targets.  Nevertheless, we expect our target selection filters to admit various kinds of anomalous sources for which our templates may provide relatively poor fits---e.g., carbon-enhanced stars, unresolved galaxies and quasars.  

\begin{figure}
\includegraphics[width=3.75in,trim=0.2in 0.1in 3.in 0.5in, clip]{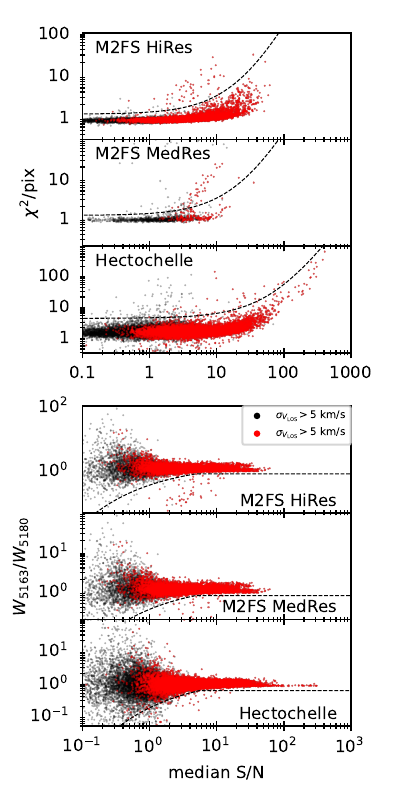}
\caption{\textit{Top three panels:} $\chi^2$ per pixel vs median S/N ratio per pixel, from the best-fitting model for each individual spectrum obtained with M2FS HiRes, M2FS MedRes
 and Hectochelle.  Red points identify observations that pass our crude quality-control filter, with raw velocity error $\leq 5$ km s$^{-1}$.  Outliers having $\chi^2$/pix above the dashed curves tend to correspond to anomalous sources, primarily carbon stars, background galaxies and quasars. 
  \textit{Bottom three panels:} Ratio of median flux in the $5160-5167$ \AA\ bandpass to the median flux in the $5176-5183$ \AA\ bandpass, vs. median S/N ratio.  Outliers having flux ratios below the dashed curves are flagged in our data catalogs as likely carbon stars. 
 \label{fig:m2fs_hecto_weird}}
\end{figure}

\smallskip
In order to identify anomalous sources systematically, first we look for cases where the observed spectrum, $S(\lambda)$, exhibits relatively large residuals with respect to the best-fitting model spectrum, $M(\lambda)$.  For each individual spectrum in our M2FS (top) and Hectochelle (bottom) samples, the top two panels of Figure \ref{fig:m2fs_hecto_weird} plot the mean value of $\chi^2\equiv \sum_{i=1}^{N_{\rm pix}}(S_i-M(\lambda_i))^2/\mathrm{Var}[S_i]$ as a function of the median S/N ratio, where the mean and median are evaluated over all $N_{\rm pix}$ unmasked pixels.  The variance spectrum, $\mathrm{Var}[S]$, is the original one, uncorrected by the linear re-scaling parameters inferred as part of the spectral fit (see Equation \ref{eq:rescale}), as the re-scaled variance will be inflated to compensate for template mismatch.  For both M2FS and Hectochelle, we find that the mean value of $\chi^2$ is approximately constant at median S/N$\lesssim 10$, with characteristic values of $\chi^2/\mathrm{pix}\sim 1.0$ for M2FS and $\chi^2/\mathrm{pix}\sim 1.5$ for Hectochelle, suggesting that the uncertainties in pixel counts estimated by the Hectochelle pipeline tend to be under-estimated by $\sim 20\%$.  We reiterate that, by design, our linear re-scaling of the raw variances (Equation \ref{eq:rescale}) brings the typical values to $\chi^2/\mathrm{pix}\sim 1$.  

\smallskip
Figure \ref{fig:m2fs_hecto_weird} also reveals that mean $\chi^2$ values rise steadily at S/N ratios $\ga 10$.  One contribution to this behavior comes from the fact that our polynomial model for the continuum spectrum is fixed at order $l=5$ (Section \ref{sec:analysis}), limiting ability to fit details of the continuum structure that become apparent only at high S/N.  In order to flag anomalous spectra despite the steady rise in $\chi^2$ with S/N ratio, we identify outliers above the smooth S/N-dependent curves drawn in both panels of Figure \ref{fig:m2fs_hecto_weird}.  The curves are broken power laws of the form $\chi^2/\mathrm{pix}=a_1(1+(S/N)/a_2)^{3}$, with $(a_1,a_2)=(1.2,25)$ for M2FS and $(4.0,75)$ for Hectochelle.  For all anomalous spectra identified in this way, we set the flag chi2\_flag=True in the data catalogs (Section \ref{sec:dataset}).  We identify \mtwofshireschitwooutliergoodn\ such anomalous M2FS HiRes spectra, \mtwofsmedreschitwooutliergoodn\ anomalous M2FS MedRes spectra and \hectochitwooutliergoodn\ anomalous Hectochelle spectra having median S/N ratio $\geq 1$ per pixel. For sources having at least one observation that passed our quality-control filter, we set the flag `any\_chi2\_flag=True' if the spectrum from any of the individual accepted observations has chi2\_flag=True.  There are \mtwofshiresanychitwon\ such sources in our M2FS HiRes catalog (not necessarily the same as those that have S/N$\geq 1$), \mtwofsmedresanychitwon\ in our M2FS MedRes catalog and \hectoanychitwon\ in our Hectochelle catalog.  

\begin{figure*}[ht] \includegraphics[width=7.5in,trim=0.in 0.in 0in 0.in, clip]{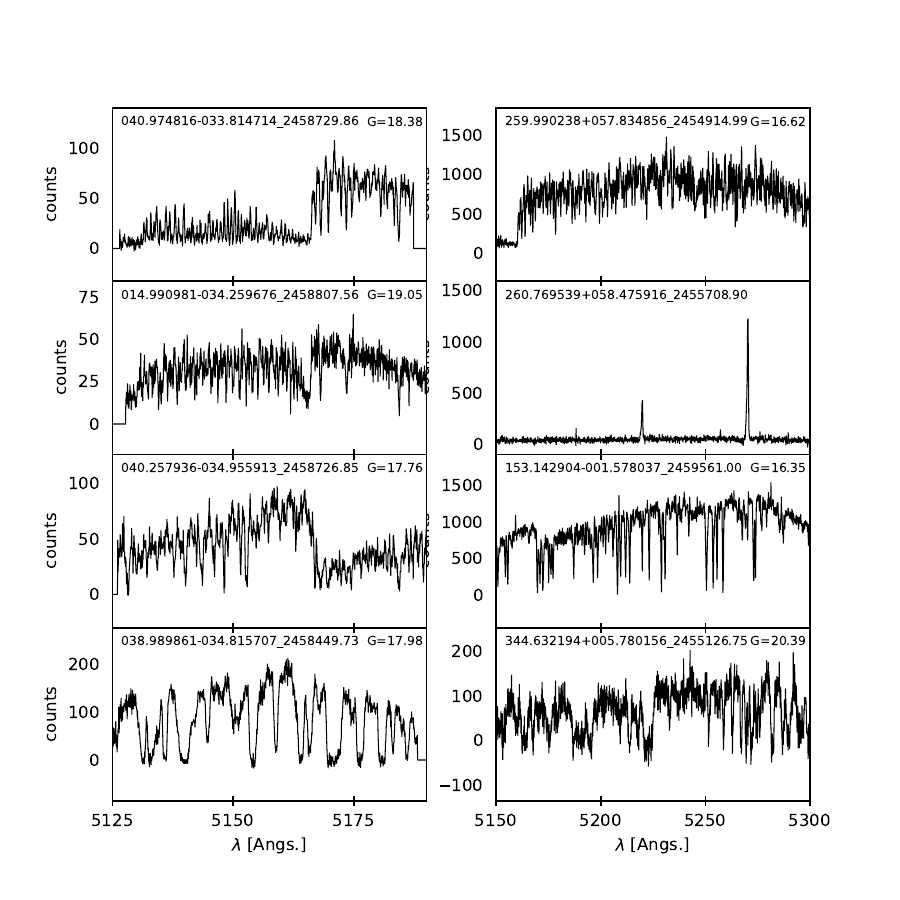}
\caption{Examples of M2FS HiRes (left) and Hectochelle (right) spectra from anomalous sources, with text indicating celestial coordinates, HJD of observation and Gaia G-band magnitude (if available).  The top two M2FS spectra, and the top Hectochelle spectrum, come from stars showing various levels of carbon enhancement, with the prominent \citet{swan_1857} C$_2$ bandhead near 5165~\AA.  The second (from top) Hectochelle spectrum is dominated by emission lines from an extragalactic source.  Spectra in the third row are from cool M giant stars, with the TiO bandhead apparent near 5170~\AA.  Spectra in the bottom row are from known quasars, at redshift $z\sim 3.7$ \citep[][left]{boutsia21} and $z\sim 3.4$ \citep[][right]{paris14}. \label{fig:m2fs_hecto_chi2_outlier_spectra}}
\end{figure*}

\smallskip
Figure \ref{fig:m2fs_hecto_chi2_outlier_spectra} displays representative examples of these anomalous spectra, some types of which have already been identified and discussed in previous M2FS papers by \citet{walker15,song19,song21}.  The top two M2FS spectra (left-hand panels) and the top Hectochelle spectrum (right-hand panels) are from stars showing various levels of carbon enhancement, with the \citet{swan_1857} C$_2$ bandhead clearly visible near 5165 \AA.  The second (from top) Hectochelle spectrum is dominated by emission lines, presumably from a distant star-forming galaxy; a few tens of similar spectra are among the $\chi^2$ outliers in our Hectochelle sample but not, due to our masking of strong emission-like features (Section \ref{sec:cr_rejection}), in our M2FS sample.  The third (from top) row of spectra are from cool M dwarf stars, with the TiO bandhead visible near 5170 \AA.  The bottom row of spectra are from  known quasars, previously measured to have redshifts of $z\sim 3.7$ \citep[][left]{boutsia21} and $z\sim 3.4$ \citep{[][right]paris14}.

\smallskip
Following \citet{song21}, we obtain a cleaner sample of carbon stars by comparing the median flux across the bandpass 5160--5167~\AA, denoted $W_{5163}$ to the median flux across 5176--5183~\AA, denoted $W_{5180}$.  The bottom two panels of Figure \ref{fig:m2fs_hecto_weird} plot the ratio $W_{5163}/W_{5180}$ as a function of median S/N ratio. 
 We identify as candidate carbon stars those sources for which the flux ratio falls below the curves drawn in the bottom two panels of Figure \ref{fig:m2fs_hecto_weird}; spectra that satisfy this criterion have flag carbon\_flag=True in the data catalogs (Section \ref{sec:dataset}).  The M2FS HiRes sample contains \mtwofshirescarbongoodn\ sources that have at least one spectrum that is flagged as carbon enhanced and has S/N$\geq 1$; the M2FS MedRes sampled contains \mtwofsmedrescarbongoodn\ such sources and the Hectochelle sample contains \hectocarbongoodn\ such sources. 
 For sources having at least one observation that passed our quality-control filter, we set the flag `any\_carbon\_flag'=True if the spectrum from any of the individual accepted observations has carbon\_flag=True.  There are \mtwofshiresanycarbonn\ such sources in our M2FS HiRes catalog, \mtwofsmedresanycarbonn\ in our MedRes catalog and \hectoanycarbonn\ in our Hectochelle catalog.    

\smallskip
Our samples also contain sources that the  Gaia (DR3) database flags as photometrically variable (`phot\_variable\_flag=`VARIABLE') in the main source catalog, and/or lists in dedicated variability tables for active galactic nuclei (variability table `vari\_agn') or RR Lyrae (`vari\_rrlyrae').  Our spectroscopic catalogs list for each source the value of Gaia's phot\_variable\_flag, and also sets flags gaia\_agn=True, gaia\_rrl=True if the source appears in the corresponding variability tables.   Considering only those having at least one spectrum with S/N$\geq 1$, our M2FS HiRes, M2FS MedRes and Hectochelle samples contain \mtwofshiresphotvariablegoodn, \mtwofsmedresphotvariablegoodn\ and \hectophotvariablegoodn\ sources, respectively, that Gaia flags as photometric variables in the main source catalog, with \mtwofshiresagngoodn, \mtwofsmedresagngoodn\ and \hectoagngoodn\ sources appearing in Gaia's dedicated AGN table.  For all but \mtwofshiresagnwrong, \mtwofsmedresagnwrong\ and \hectoagnwrong\ of these sources, our M2FS HiRes, M2FS MedRes and Hectochelle observations do not yield measurements that pass our quality-control criteria.  

\begin{figure*}[ht] \includegraphics[width=7.5in,trim=0.5in 3.3in 0in 0.in, clip]{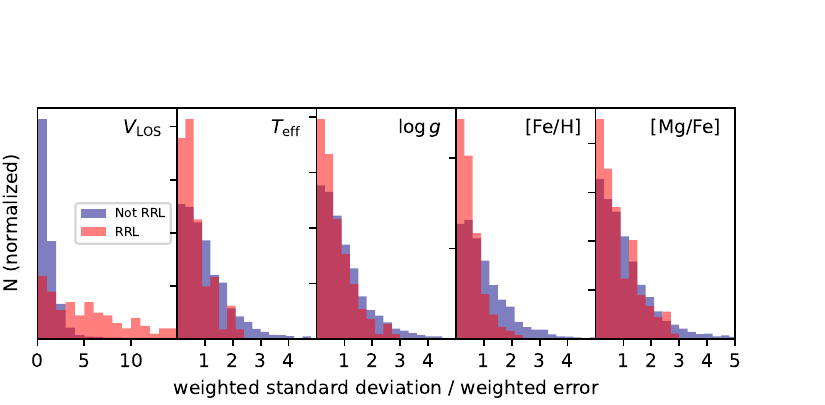}
\caption{Distributions of the ratio of weighted standard deviation to weighted mean uncertainty of spectroscopically-measured parameters, for stars having multiple independent measurements passing our quality-control filter.  Red 
 (resp. blue) histograms correspond to sources that do (do not) appear in Gaia's (DR3) RR Lyrae catalog.  The left-most panel demonstrates the intrinsic variability of \vlos\ for RRL stars.   \label{fig:rrl_scatter}}
\end{figure*}

\smallskip
Finally, considering only those sources having at least one M2FS HiRes, M2FS MedRes or Hectochelle observation that passed our quality control filter, \mtwofshiresgoodrrl, \mtwofsmedresgoodrrl\ and \hectogoodrrl, respectively, are listed in Gaia's dedicated RR Lyrae table.  While we can obtain good fits to the spectra of RR Lyrae, our repeat measurements detect the intrinsic line-of-sight velocity variability of these pulsating stars.  For each of our sources that have multiple spectroscopic measurements that pass our quality-control filter, histograms in Figure \ref{fig:rrl_scatter} show distributions of the ratio of the weighted standard deviation (about the weighted mean) of the measured \vlos, \teff, \logg, \feh and \mgfe\ to the weighted mean error.  This ratio is a measure of intrinsic variability of the source. Red (blue) histograms represent sources that are (are not) listed in Gaia's (DR3) RR Lyrae variability table (vari\_rrlyrae).  The ratios for RRL stars generally track those of the non-RRLs for the atmospheric parameters \teff, \logg, \feh\ and \mgfe.  For \vlos, however (left-most panel of Figure \ref{fig:rrl_scatter}), the RRLs exhibit dramatically larger scatter than the non-RRLs, directly reflecting the rates at which the pulsating stars expand and contract.  Users who are interested in the observed stars as dynamical tracers will need to take into account this source of intrinsic velocity variability.  
\smallskip

\smallskip
Of course, there are sources of intrinsic variability other than pulsation---e.g., binary star systems---for which we do not necessarily have a diagnostic classification \textit{a priori}.  For all stars having multiple independent measurements that pass our quality-control filter, we identify sources exhibiting potentially intrinsic variability as those for which the ratio of weighted standard deviation to weighted mean error exceeds a value of $3$, regardless of whether the source is classified as RRL.  In our data catalogs (Section \ref{sec:dataset}), we set the flag `X\_variable\_flag'=True for such cases, where $X$ can be any of the observables `vlos', `teff', `logg', `feh', `mgfe'.  

\section{M2FS+Hectochelle Dataset}
\label{sec:dataset}

We provide complete data catalogs for our M2FS HiRes, M2FS MedRes and Hectochelle samples.  The catalogs are stored as binary tables in standard `.fits' format, and are available electronically at both the Journal website and the Zenodo database (DOI: 10.5281/zenodo.7837922).  Table \ref{tab:fits_catalog} lists and briefly explains each of the columns listed in these catalogs.  
Most users will need to be mindful of the `obs' and/or `good\_obs' columns, which indicate for a given star the chronologically-ordered observation number.  A star having only one observation will have `obs=1', but for stars observed multiple times, the first observation will have `obs=1', the second will have `obs=2', etc.  The `good\_obs' parameter works the same way, but counts only those observations that pass our crude quality-control filter (velocity error $\sigma_{V_{\rm LOS}}\leq 5$ \kms); all measurements for stars having zero `good' measurements will have good\_obs=0.  This information can be used in tandem with the (inverse variance-weighted) mean parameter estimates that are computed over all `good' observations of a given star, and listed for each individual-epoch measurement (`good' or otherwise) of the star.  So, for example, a user who wants only the mean parameter estimates for each star (as opposed to individual-epoch measurements) can select the mean values (e.g., vlos\_mean, teff\_mean, logg\_mean, feh\_mean, mgfe\_mean, with associated errors vlos\_mean\_error, teff\_mean\_error, logg\_mean\_error, feh\_mean\_error, mgfe\_mean\_error) listed for only observations with good\_obs=1.  

\smallskip
The Zenodo database (DOI: 10.5281/zenodo.7837922) also makes available all of the individual (extracted, 1D, wavelength-calibrated) spectra produced by our processing pipeline.  The spectra are provided in multi-extension .fits files.  A given file contains all (up to 128 for M2FS, up to 240 for Hectochelle) spectra obtained on a given data frame.  In the .fits catalogs discussed above, the `fits\_filename' and `fits\_index' columns specify the filename and array index where the processed spectrum can be found.  Along with the spectra, these multi-extension fits files provide the central wavelength, variance, best-fitting model, mean sky level, and (bad pixel) mask status at each pixel.  %Finally, for each spectrum, we provide the output files from MultiNest that contain random samples from the 16-dimensional posterior PDF.  

%\newpage
\startlongtable
\begin{deluxetable*}{ll}
\tablewidth{0pt}
\tablecaption{Columns in electronic data catalogs
\label{tab:fits_catalog}
}
\tablehead{\colhead{column name}&\colhead{description}}
\startdata
instrument & Instrument used to acquire spectrum (`Hectochelle', `M2FS\_HiRes' or `M2FS\_MedRes')\\
target\_system & Name of target system (name of dwarf galaxy, star cluster, etc.)\\
obs\_id & unique identifier for this observation (R.A.\_Dec\_HJD)\\
exptime & exposure time (s)\\
gaia\_source\_id & source ID in Gaia (DR3) catalog, if available\\
gaia\_gmag & Gaia (DR3) G magnitude, if available\\
gaia\_bpmag & Gaia (DR3) BP magnitude, if available\\
gaia\_rpmag & Gaia (DR3) RP magnitude, if available\\
gaia\_siggmag & Gaia (DR3) error in gaia\_gmag\\
gaia\_sigbpmag & Gaia (DR3) error in gaia\_bpmag\\ 
gaia\_sigrpmag & Gaia (DR3) error in gaia\_rpmag\\
gaia\_gmag\_dered & Gaia (DR3) G magnitude, de-reddened \\
gaia\_bpmag\_dered & Gaia (DR3) BP magnitude, de-reddened \\
gaia\_rpmag\_dered & Gaia (DR3) RP magnitude, de-reddened \\
gaia\_pmra & Gaia (DR3) proper motion, right ascension component, if available (mas yr$^{-1}$)\\
gaia\_pmdec & Gaia (DR3) proper motion, declination component, if available (mas yr$^{-1}$)\\
gaia\_sigpmra & Gaia (DR3) error in gaia\_pmra (mas yr$^{-1}$)\\
gaia\_sigpmdec & Gaia (DR3) error in gaia\_pmdec (mas yr$^{-1}$)\\
gaia\_parallax & Gaia (DR3) parallax, if available  (mas) \\
gaia\_sigparallax & Gaia (DR3) error in gaia\_parallax (mas) \\
ra & Right Ascension (J2000)\\
dec & Declination (J2000)\\
ra\_dec\_source & source catalog from which ra\_deg and dec\_deg are adopted (Gaia DR3 if available)\\
hjd & Heliocentric Julian Date of spectroscopic observation (days)\\
sn\_ratio & median signal-to-noise ratio per pixel\\
vlos\_raw & mean of posterior PDF for \vlos\ (\kms; solar rest frame), without shift to APOGEE zero point\\
vlos\_raw\_error & standard deviation of posterior PDF for \vlos\ (\kms), as sampled by MultiNest\\
vlos\_raw\_skew & skewness of posterior PDF for \vlos, as sampled by MultiNest\\
vlos\_raw\_kurtosis & kurtosis of posterior PDF for \vlos, as sampled by MultiNest\\
vlos & vlos\_raw, shifted to APOGEE zero point\\
vlos\_error & error in vlos\_raw and vlos (\kms), after applying adjustment in Section \ref{sec:internalvalidation}\\
teff\_raw & mean of posterior PDF for \teff\ (K), without shift to APOGEE zero point\\
teff\_raw\_error & standard deviation of posterior PDF for \teff\ (K), as sampled by MultiNest \\
teff\_raw\_skew & skewness of posterior PDF for \teff, as sampled by MultiNest\\
teff\_raw\_kurtosis & kurtosis of posterior PDF for \teff, as sampled by MultiNest\\
teff & teff\_raw, shifted to APOGEE zero point\\
teff\_error & error in teff\_raw and teff (K), after applying adjustment in Section \ref{sec:internalvalidation}\\
logg\_raw & mean of posterior PDF for \logg (cgs units), without shift to APOGEE zero point\\
logg\_raw\_error & standard deviation of posterior PDF for \logg\ (cgs units), as sampled by MultiNest\\
logg\_raw\_skew & skewness of posterior PDF for \logg, as sampled by MultiNest\\
logg\_raw\_kurtosis & kurtosis of posterior PDF for \logg, as sampled by MultiNest\\
logg & logg\_raw, shifted to APOGEE zero point\\
logg\_error & error in logg\_raw and logg, after applying adjustment in Section \ref{sec:internalvalidation} (cgs units)\\
feh\_raw & mean of posterior PDF for \feh\, without shift to APOGEE zero point\\
feh\_raw\_error & standard deviation of posterior PDF for \feh, as sampled by MultiNest\\
feh\_raw\_skew & skewness of posterior PDF for \feh, as sampled by MultiNest\\
feh\_raw\_kurtosis & kurtosis of posterior PDF for \feh, as sampled by MultiNest\\
feh & feh\_raw, shifted to APOGEE zero point\\
feh\_error & error in feh\_raw and feh, after applying adjustment in Section \ref{sec:internalvalidation}\\
mgfe\_raw & mean of posterior PDF for \mgfe\, without shift to APOGEE zero point\\
mgfe\_raw\_error & standard deviation of posterior PDF for \mgfe, as sampled by MultiNest\\
mgfe\_raw\_skew & skewness of posterior PDF for \mgfe, as sampled by MultiNest\\
mgfe\_raw\_kurtosis & kurtosis of posterior PDF for \mgfe, as sampled by MultiNest\\
mgfe & mgfe\_raw, shifted to APOGEE zero point\\
mgfe\_error & error in mgfe\_raw and mgfe, after applying adjustment in Section \ref{sec:internalvalidation}\\
smooth\_raw & bandwidth \lsf\ (Angstroms), of Gaussian smoothing kernel applied to template spectra\\
smooth\_raw\_error & standard deviation of posterior PDF for \lsf\ (Angstroms), as sampled by MultiNest\\
smooth\_raw\_skew & skewness of posterior PDF for \lsf, as sampled by MultiNest\\
smooth\_raw\_kurtosis & kurtosis of posterior PDF for \lsf, as sampled by MultiNest\\
logs1\_raw & base-10 logarithm of error re-scaling parameter $s_1$ (Equation \ref{eq:rescale})\\
logs1\_raw\_error & standard deviation of posterior PDF for $\log_{10}s_1$, as sampled by MultiNest\\
logs1\_raw\_skew & skewness of posterior PDF for $\log_{10}s_1$, as sampled by MultiNest\\
logs1\_raw\_kurtosis & kurtosis of posterior PDF for $\log_{10}s_1$, as sampled by MultiNest\\
logs2\_raw & base-10 logarithm of error floor parameter $s_2$ (Equation \ref{eq:rescale})\\
logs2\_raw\_error & standard deviation of posterior PDF for $\log_{10}s_2$, as sampled by MultiNest\\
logs2\_raw\_skew & skewness of posterior PDF for $\log_{10}s_2$, as sampled by MultiNest\\
logs2\_raw\_kurtosis & kurtosis of posterior PDF for $\log_{10}s_2$, as sampled by MultiNest\\
median\_sky & median count of sky spectrum that was subtracted\\
standard\_deviation\_median\_sky & standard deviation of median\_sky, over spectra acquired in same observation\\
filter\_name & name of filter used for observation\\
chi2 & $\chi^2$ for best-fitting model spectrum, using original variance spectrum\\
chi2\_rescaled & $\chi^2$ for best-fitting model spectrum, using re-scaled variance spectrum from Equation \ref{eq:rescale}\\
npix & number of (unmasked) pixels included in spectrum fit \\
w5163 & median (sky-subtracted) counts over spectral range $5160-5167$ \AA \\
w5180 & median (sky-subtracted) counts over spectral range $5176-5183$ \AA \\
vhelio\_correction & heliocentric correction that was applied (added) to \vlos\ after spectrum model fitting (\kms)\\
fits\_filename & name of multi-extension fits file containing processed spectrum\\
fits\_index & index containing the spectrum of this source (in multi-extension fits frame)\\
obs & (chronological) observation number for this source\\
n\_obs & total number of observations of this source\\
good\_obs & (chronological) observation number for this source, after quality control filter \\
good\_n\_obs & total number of observations of this source, after quality control filter\\
vlos\_raw\_mean & (inverse-variance) weighted mean of vlos\_raw (\kms; solar rest frame) over good\_n\_obs observations\\
vlos\_mean & vlos\_raw\_mean, shifted to APOGEE zero point (\kms)\\
vlos\_mean\_error & error in vlos\_raw\_mean and vlos\_mean (\kms) \\
vlos\_mean\_scatter & (inverse-variance) weighted standard deviation of \vlos\ (\kms) over good\_n\_obs observations\\
teff\_raw\_mean & (inverse-variance) weighted mean of teff\_raw (K) over good\_n\_obs observations\\
teff\_mean & teff\_raw\_mean, shifted to APOGEE zero point (K)\\
teff\_mean\_error & error in teff\_raw\_mean and teff\_mean (K)\\
teff\_mean\_scatter & (inverse-variance) weighted standard deviation of \teff\ (K) over good\_n\_obs observations\\
logg\_raw\_mean & (inverse-variance) weighted mean of logg\_raw over good\_n\_obs observations\\
logg\_mean & logg\_raw\_mean, shifted to APOGEE zero point \\
logg\_mean\_error & error in logg\_raw\_mean and logg\_mean \\
logg\_mean\_scatter & (inverse-variance) weighted standard deviation of \logg\ over good\_n\_obs observations\\
feh\_raw\_mean & (inverse-variance) weighted mean of feh\_raw over good\_n\_obs observations\\
feh\_mean & feh\_raw\_mean, shifted to APOGEE zero point\\
feh\_mean\_error & error in feh\_raw\_mean and feh\_mean\\
feh\_mean\_scatter & (inverse-variance) weighted standard deviation of \feh\ over good\_n\_obs observations\\
mgfe\_raw\_mean & (inverse-variance) weighted mean of mgfe\_raw over good\_n\_obs observations\\
mgfe\_mean & mgfe\_raw\_mean, shifted to APOGEE zero point\\
mgfe\_mean\_error & error in mgfe\_raw\_mean and mgfe\_mean\\
mgfe\_mean\_scatter & (inverse-variance) weighted standard deviation of \mgfe\ over good\_n\_obs observations\\
n\_wav\_cal & (M2FS only) number of ThArNe calibration frames used for wavelength calibration\\
temp\_min & (M2FS only) minimum temperature ($^{\circ}$C) recorded at detector during science sub-exposures\\
temp\_max & (M2FS only) maximum temperature ($^{\circ}$C) recorded at detector during science exposures\\
wav\_cal\_flag& (M2FS only) True if n\_wav\_cal=1 and temp\_max$-$temp\_min $\geq 1$ $^{\circ}$C\\
chi2\_flag & True if chi2 is above curve in top panels of Figure \ref{fig:m2fs_hecto_weird}\\
carbon\_flag & True if flux ratio $W_{5163}/W_{5180}$ is below curve in bottom panels of Figure \ref{fig:m2fs_hecto_weird}\\
any\_chi2\_flag & True if any observations contributing to mean have chi2\_flag=True\\
any\_carbon\_flag & True if any observations contributing to mean have carbon\_flag=True\\
vlos\_variable\_flag & True if vlos\_mean\_scatter $\geq 3$ vlos\_mean\_error \\
teff\_variable\_flag & True if teff\_mean\_scatter $\geq 3$ teff\_mean\_error \\
logg\_variable\_flag & True if logg\_mean\_scatter $\geq 3$ logg\_mean\_error \\
feh\_variable\_flag & True if feh\_mean\_scatter $\geq 3$ feh\_mean\_error \\
mgfe\_variable\_flag & True if mgfe\_mean\_scatter $\geq 3$ mgfe\_mean\_error \\
gaia\_phot\_variable\_flag & Gaia (DR3) phot\_variable\_flag \\
gaia\_rrl & True if source is listed in Gaia DR3 variability RR Lyrae table (vari\_rrlyrae)\\
gaia\_agn & True if source is listed in Gaia DR3 variability AGN catalog (vari\_agn)\\
\enddata
\end{deluxetable*}

\begin{figure*}    \includegraphics[width=7.in,trim=0.25in 3.35in 2.75in 0.5in, clip]{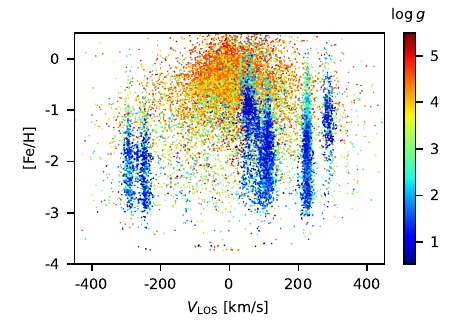}
    \caption{Chemo-dynamic substructure within the Milky Way Halo: Metallicity vs. line-of-sight velocity, from Magellan/M2FS and MMT/Hectochelle spectra acquired for \allgoodstar\ stars observed toward \allsystems\ Galactic Halo objects.  Marker color indicates spectroscopically-estimated surface gravity.  Given our color/magnitude criteria for spectroscopic target selection, redder marker colors tend to identify dwarf stars in the Galactic disk, while bluer marker colors indicate giant stars in the Galactic halo and its substructures.  The halo objects that are most obvious here are the `classical' dwarf spheroidal galaxies Draco (\vlos\ $\sim -290$ \kms), Ursa Minor ($-250$ \kms), Fornax ($+55$ \kms), Leo II ($+80$ \kms), Sculptor ($+110$ \kms), Carina/Sextans (both at $+220$ \kms) and Leo I ($+280$ \kms).  
\label{fig:field_of_halos_v}}
\end{figure*}

\smallskip
We now present some of the macroscopic properties of the M2FS+Hectochelle dataset.  Figure \ref{fig:field_of_halos_v} provides a comprehensive view of chemo-dynamical structure within the Galactic Halo, plotting metallicity against line-of-sight velocity for the entire sample (using inverse-variance-weighted mean values for stars with multiple good measurements), with marker color coded according to surface gravity.  Red giants within dwarf galaxies are conspicuous as bluer (\logg\ $\lesssim 3$) points that tend to have lower mean metallicity (\feh\ $\lesssim -1.5$) and  cluster into narrower velocity distributions (velocity dispersion $\lesssim 10$ \kms) than do foreground stars, which tend to be late-type dwarfs (\logg\ $\gtrsim 4$) contributed by the Galactic disk.  Visually dominating population of substructures traced by red giants are the classical dwarf spheroidals Ursa Minor (\vlos\ $\sim -250$ \kms), Draco (\vlos\ $\sim -290$ \kms), Fornax (\vlos\ $\sim +55$ \kms), Leo II (\vlos\ $\sim +80$ \kms), Sculptor (\vlos\ $\sim +110$ \kms), Carina/Sextans (both at \vlos\ $\sim 220$ \kms) and Leo I (\vlos\ $\sim +280$ \kms).  Many less luminous Halo substructures are present in our sample, but are less obvious against the foreground populations.  Figures \ref{fig:m2fs_cmdmap} and \ref{fig:hecto_cmdmap} display the \feh\ vs \vlos\ scatterplots for individual systems. 

\begin{figure}
    \includegraphics[width=3.5in,trim=0.25in 0in 2.5in 3.5in, clip]{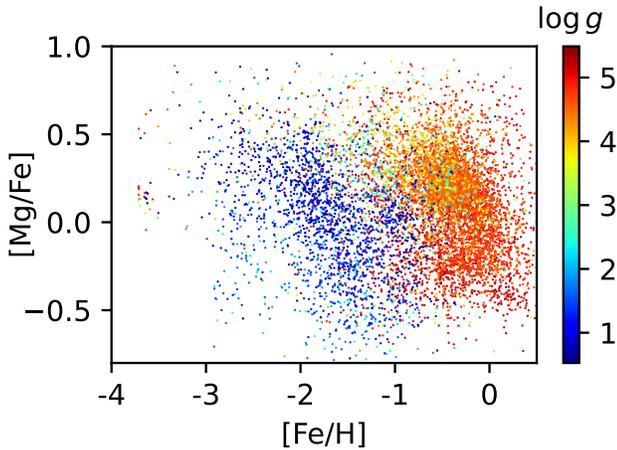}
    \caption{Magnesium abudance vs. metallicity, for the \allgoodzstar\ stars in our M2FS+Hectochelle dataset that have observational errors $\leq 0.5$ in each of \logg, \feh\ and \mgfe.  Marker color indicates spectroscopically-estimated surface gravity.
\label{fig:field_of_halos_alpha}}
\end{figure}

\smallskip
Figure \ref{fig:field_of_halos_alpha} plots \mgfe\ against \feh\ for our M2FS+Hectochelle sample.  For clarity, we display only the \allgoodzstar\ stars for which observational errors in \logg, \feh\ and \mgfe\ are all $\leq 0.5$ dex.  The red giant sample (bluer points), dominated by Halo substructures, is clearly offset toward lower metallicity than the foreground Galactic stellar populations.  Also apparent, although blurred somewhat by the inclusion of all targeted systems simultaneously, is the characteristic `knee' (near \feh\ $\sim -2$), where \mgfe\ declines toward higher metallicities because stars have formed from gas pre-enriched by Type-Ia supernovae.  

\begin{figure}
    \includegraphics[width=3.5in, trim=0.2in 2.8in 2.9in 0.5in, clip]{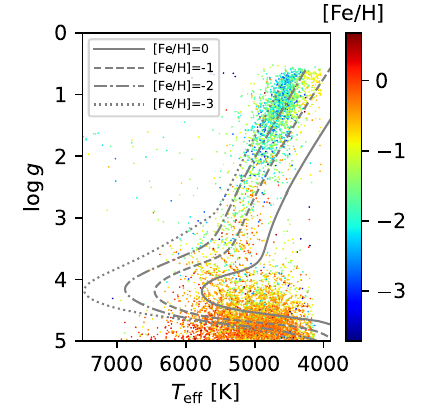}
    \caption{Surface gravity vs. effective temperature (both estimated spectroscopically), for the \allgoodzstar\ stars in our M2FS+Hectochelle dataset that have observational errors $\leq 0.5$ in each of \logg, \feh\ and \mgfe.  Marker color indicates spectroscopically-estimated metallicity.  Overplotted, for comparison, are theoretical isochrones \citep{isochrones,dotter16} calculated for age = 10~Gyr and a range of [Fe/H].
\label{fig:field_of_halos_hr_diagram}}
\end{figure}

\smallskip
Figure \ref{fig:field_of_halos_hr_diagram} plots surface gravity against effective temperature, with marker color indicating \feh.  Again, for clarity, we display only stars for which errors in \logg, \feh\ and \mgfe\ are all $\leq 0.5$ dex.  Overplotted are MESA isochrones \citep{isochrones,dotter16}, calculated for age = 10~Gyr and a range of stellar metallicity. 
 Reassuringly, low-gravity stars within our sample clearly populate the red giant branch expected for low-metallicity stars ($-3\lesssim \mathrm{[Fe/H]}\lesssim -1$).  Higher-gravity stars populate regions near the main sequence expected for the higher-metallicity stars contributed by the Galactic foreground.

 \smallskip
We note the presence in Figures \ref{fig:field_of_halos_v} and \ref{fig:field_of_halos_alpha} of $\sim 10$ sources that are measured to have extremely low metallicity (\feh $\lesssim -3.6$), high surface gravity (\logg\ $\ga 4.5$) and approximately solar \mgfe.  Figure \ref{fig:m2fs_hecto_weird_lowz} of the Appendix displays spectra from each of these sources, with best-fitting models overplotted.  We find that most of these spectra exhibit the broad absorption features characteristic of AGN, suggesting that our measurements for these sources are spurious.  However, none of the sources are listed in Gaia's AGN variability table. 
   
\smallskip
We do not attempt here to evaluate the population (e.g., dwarf galaxy vs Galactic foreground) membership status of individual stars within our sample.  The reason is that a star's probability of membership to a specific population, given the star's observed properties, depends fundamentally on the model invoked to describe the ensemble of populations.  
 We hope and anticipate that our dataset will be used to evaluate a large variety of models.  Therefore we leave to the user any evaluation of membership status for individual stars.

\smallskip
Instead we use our spectroscopic measurements to give rough indications of the mixtures of stellar populations that are present within our samples. 
 As is evident in Figure \ref{fig:field_of_halos_hr_diagram}, our measurements of surface gravity can effectively distinguish red giants from dwarf stars.  While red giant status correlates strongly with membership within most of the dwarf galaxies and Halo substructures targeted by our program, systems at distances $\lesssim 50$~kpc can have observed targets on the sub-giant branch at \logg\ $\ga 3$.  Moreover, we expect red giant samples to include contamination from bona fide red giants within the Galactic halo.  Thus the number of observed red giants is a useful but imperfect proxy for the number of observed member stars within the targeted systems.  
 
 \smallskip
 Therefore, in order to summarize the contents of our spectroscopic samples, we count not just the number of red giant sources, but also the number of sources that have both \vlos\ and proper motion consistent with membership (regardless of \logg).  Text in Figures \ref{fig:m2fs_cmdmap} and \ref{fig:hecto_cmdmap} lists numbers of individual sources observed (denoted $N_{\rm obs}$) with at least one `good' measurement that passes our crude quality-control filter, the number of likely giant stars (denoted $N_{\rm giant}$), identified as sources measured to have \logg\ $\lesssim 3$, and the number of sources that have \vlos\ and Gaia=measured (DR3) proper motion to be within $3\sigma$ of the previously-measured systemic mean values (denoted $N_{\rm mem})$.  For the \vlos\ criterion, we define $\sigma$ to be quadrature sum of formal uncertainties in our measurement of \vlos\ for the source, the measurement of the systemic mean velocity, and the (previously-measured) systemic velocity dispersion.  We take the previously-published mean values from the compilation by \citet{pace22}.  For the proper motion criterion, we take $\sigma$ to be the propagated uncertainty in the separation (in 2D proper motion space, neglecting covariance between the components) between the source and previously-measured  systemic mean  proper motions \citep{pace22}.  
 
 \smallskip
 Our samples contain several hundred members in each of the Milky Way's eight `classical' dSph satellites, ranging from $\sim 200$ in Leo II to $\sim 850$ in Carina and Sculptor.  In the less luminous satellites and star clusters, likely member samples range from zero to a few tens.  These samples extend to larger galactocentric radii than most previously-published counterparts.  We count \nmemberfartotal\ (\nmemberfarthreetotal, \nmemberfarfourtotal, \nmemberfarfivetotal) likely members projected farther than 2 (3, 4, 5) projected halflight radii from the center of their host galaxy, providing information about the stellar populations and dynamical state in the outer parts of these systems.  

\section{Summary}
We have presented new spectroscopic data and catalogs of new measurements of spectroscopic parameters for \allgoodstar\ unique sources toward \allsystems\ 
 target systems.  The sample includes repeat (multi-epoch) measurements for \allrepeated\ sources, with as many as \allmaxgoodobs\ epochs per source.  We have calibrated internal errors and used external data sets to calibrate zero points for each physical parameter.  We have defined criteria for identifying anomalous sources that should be handled carefully in subsequent analysis.  Using simple but crude diagnostic criteria, we estimate that the sample includes $\sim$ \ngianttotal\ red giant stars and $\sim$ \nmembertotal\ members of the target systems, in some many cases pushing the available samples beyond several halflight radii.  Data products include catalogs of measured stellar parameters and all processed and calibrated spectra.

\section*{Acknowledgements}
We thank the anonymous referee, whose feedback and suggestions improved the manuscript.  We thank David Nidever and Paul Cristofari for providing helpful tests and feedback about the M2FS processing pipeline.  We thank Charlie Conroy, Phill Cargile and the H3 team for providing comparison measurements using methodology from the H3 survey, and for providing spectra from H3's observations of globular clusters.  

\smallskip E.O. remembers Jill Bechtold here.

\smallskip 
M.G.W. acknowledges support from National Science Foundation (NSF) grants AST-1813881, AST-1909584 and AST-2206046.  
M.M. acknowledges support from NSF grants AST-0923160, AST-1312997, AST-1815403 and AST-2205847. 
 E.O.  acknowledges support from NSF grants AST-1815767, AST-1313006, AST-0807498.  
Nelson Caldwell acknowledges support from NSF grant AST-1812461.
I.U.R.\ acknowledges support from NSF grants AST-1815403, AST-2205847, and PHYS-1430152 (Physics Frontier Center/JINA-CEE).
A.B.P. acknowledges support from NSF grant AST-1813881.

\smallskip
This work has made use of NASA's Astrophysics
Data System Bibliographic Services.  This paper made
use of the Whole Sky Database (WSDB), created by Sergey Koposov and maintained at the Institute of Astronomy Cambridge by Sergey Koposov, Vasily Belokurov and Wyn Evans with financial support from the Science \& Technology Facilities Council (STFC) and the European Research Council (ERC).  This work made use of Astropy:\footnote{http://www.astropy.org} a community-developed core Python package and an ecosystem of tools and resources for astronomy.

\smallskip
This work has made use of data from the European Space Agency (ESA) mission Gaia (https://www.
cosmos.esa.int/gaia), processed by the Gaia Data Processing and Analysis Consortium (DPAC, https://www.
cosmos.esa.int/web/gaia/dpac/consortium). Funding
for the DPAC has been provided by national institutions, in particular the institutions participating in the Gaia Multilateral Agreement.

\smallskip
Figures \ref{fig:m2fshires_spectra}-\ref{fig:hecto_spectra} use atomic line identifications from the Virtual Atomic and Molecular Data Centre (VAMDC) Consortium \citep{dubernet16}, provided by the BASS2000 website.

\smallskip
This work has made use of data from the Sloan Digital Sky Survey IV.  
Funding for the Sloan Digital Sky 
Survey IV has been provided by the 
Alfred P. Sloan Foundation, the U.S. 
Department of Energy Office of 
Science, and the Participating 
Institutions. 

SDSS-IV acknowledges support and 
resources from the Center for High 
Performance Computing  at the 
University of Utah. The SDSS 
website is www.sdss4.org.

SDSS-IV is managed by the 
Astrophysical Research Consortium 
for the Participating Institutions 
of the SDSS Collaboration including 
the Brazilian Participation Group, 
the Carnegie Institution for Science, 
Carnegie Mellon University, Center for 
Astrophysics | Harvard \& 
Smithsonian, the Chilean Participation 
Group, the French Participation Group, 
Instituto de Astrof\'isica de 
Canarias, The Johns Hopkins 
University, Kavli Institute for the 
Physics and Mathematics of the 
Universe (IPMU) / University of 
Tokyo, the Korean Participation Group, 
Lawrence Berkeley National Laboratory, 
Leibniz Institut f\"ur Astrophysik 
Potsdam (AIP),  Max-Planck-Institut 
f\"ur Astronomie (MPIA Heidelberg), 
Max-Planck-Institut f\"ur 
Astrophysik (MPA Garching), 
Max-Planck-Institut f\"ur 
Extraterrestrische Physik (MPE), 
National Astronomical Observatories of 
China, New Mexico State University, 
New York University, University of 
Notre Dame, Observat\'ario 
Nacional / MCTI, The Ohio State 
University, Pennsylvania State 
University, Shanghai 
Astronomical Observatory, United 
Kingdom Participation Group, 
Universidad Nacional Aut\'onoma 
de M\'exico, University of Arizona, 
University of Colorado Boulder, 
University of Oxford, University of 
Portsmouth, University of Utah, 
University of Virginia, University 
of Washington, University of 
Wisconsin, Vanderbilt University, 
and Yale University.

\smallskip
This project used public archival data from the Dark Energy Survey (DES). Funding for the DES Projects has been provided by the U.S. Department of Energy, the U.S. National Science Foundation, the Ministry of Science and Education of Spain, the Science and Technology FacilitiesCouncil of the United Kingdom, the Higher Education Funding Council for England, the National Center for Supercomputing Applications at the University of Illinois at Urbana-Champaign, the Kavli Institute of Cosmological Physics at the University of Chicago, the Center for Cosmology and Astro-Particle Physics at the Ohio State University, the Mitchell Institute for Fundamental Physics and Astronomy at Texas A\&M University, Financiadora de Estudos e Projetos, Funda{\c c}{\~a}o Carlos Chagas Filho de Amparo {\`a} Pesquisa do Estado do Rio de Janeiro, Conselho Nacional de Desenvolvimento Cient{\'i}fico e Tecnol{\'o}gico and the Minist{\'e}rio da Ci{\^e}ncia, Tecnologia e Inova{\c c}{\~a}o, the Deutsche Forschungsgemeinschaft, and the Collaborating Institutions in the Dark Energy Survey.
The Collaborating Institutions are Argonne National Laboratory, the University of California at Santa Cruz, the University of Cambridge, Centro de Investigaciones Energ{\'e}ticas, Medioambientales y Tecnol{\'o}gicas-Madrid, the University of Chicago, University College London, the DES-Brazil Consortium, the University of Edinburgh, the Eidgen{\"o}ssische Technische Hochschule (ETH) Z{\"u}rich,  Fermi National Accelerator Laboratory, the University of Illinois at Urbana-Champaign, the Institut de Ci{\`e}ncies de l'Espai (IEEC/CSIC), the Institut de F{\'i}sica d'Altes Energies, Lawrence Berkeley National Laboratory, the Ludwig-Maximilians Universit{\"a}t M{\"u}nchen and the associated Excellence Cluster Universe, the University of Michigan, the National Optical Astronomy Observatory, the University of Nottingham, The Ohio State University, the OzDES Membership Consortium, the University of Pennsylvania, the University of Portsmouth, SLAC National Accelerator Laboratory, Stanford University, the University of Sussex, and Texas A\&M University.
Based in part on observations at Cerro Tololo Inter-American Observatory, National Optical Astronomy Observatory, which is operated by the Association of Universities for Research in Astronomy (AURA) under a cooperative agreement with the National Science Foundation.

\smallskip
This project has made use of public data from the Pan-STARRS1 survey.  The Pan-STARRS1 Surveys (PS1) and the PS1 public science archive have been made possible through contributions by the Institute for Astronomy, the University of Hawaii, the Pan-STARRS Project Office, the Max-Planck Society and its participating institutes, the Max Planck Institute for Astronomy, Heidelberg and the Max Planck Institute for Extraterrestrial Physics, Garching, The Johns Hopkins University, Durham University, the University of Edinburgh, the Queen's University Belfast, the Harvard-Smithsonian Center for Astrophysics, the Las Cumbres Observatory Global Telescope Network Incorporated, the National Central University of Taiwan, the Space Telescope Science Institute, the National Aeronautics and Space Administration under Grant No. NNX08AR22G issued through the Planetary Science Division of the NASA Science Mission Directorate, the National Science Foundation Grant No. AST-1238877, the University of Maryland, Eotvos Lorand University (ELTE), the Los Alamos National Laboratory, and the Gordon and Betty Moore Foundation.

\smallskip
For the purpose of open access, the author has applied a Creative Commons
Attribution (CC BY) licence to any Author Accepted Manuscript version arising
from this submission.

\begin{figure*}[ht] \includegraphics[width=7.5in,trim=0.in 0.in 0in 0.in, clip]{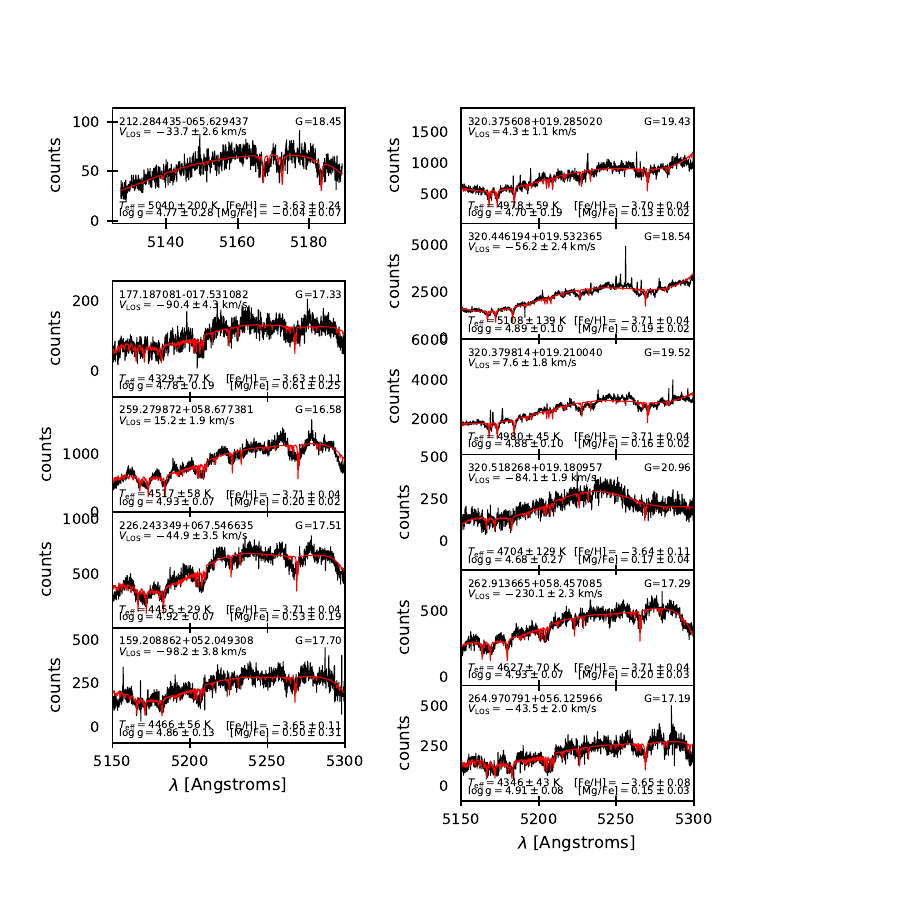}
\caption{Examples of M2FS (top left) and Hectochelle (all other panels) spectra corresponding to spurious measurements of extremely low metallicity (\feh\ $\lesssim -3.6)$, high surface gravity (\logg $\ga 4.5$) and alpha-enhanced \mgfe.  Over-plotted in red are best-fitting models, which tend to find absorption features but fail to reproduce their broadness.  Text indicates target coordinates, Gaia G-band magnitude, and values of spectroscopically-inferred parameters.
 \label{fig:m2fs_hecto_weird_lowz}}
\end{figure*}

\appendix
Figures \ref{fig:m2fs_cmdmap} and \ref{fig:hecto_cmdmap} display sky positions, color-magnitude diagrams (CMDs), proper motions and our measurements of metallicity, \feh,  vs. (heliocentric) line-of-sight velocity, \vlos, for spectroscopic targets toward each observed system.  
Figure \ref{fig:m2fs_hecto_weird_lowz} displays spectra from sources spuriously measured to have extremely low metallicity (\feh $\lesssim -3.6$), high surface gravity (\logg\ $\ga 4.5$) and approximately solar \mgfe, with best-fitting models overplotted.  

\newpage
\begin{figure*}
%\centering
\begin{tabular}{@{}l@{}}
  \includegraphics[width=7.5in,angle=0,trim= 0in 4.1in 0in 0.5in, clip]{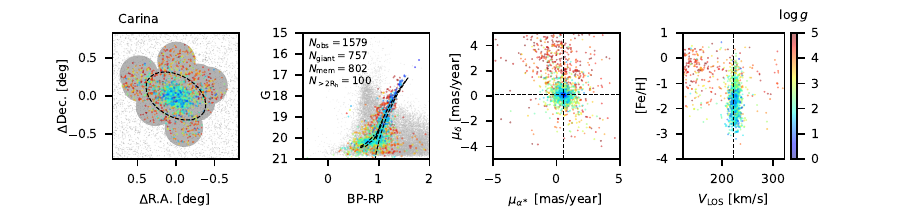}\vspace{-0.0in}\\
\includegraphics[width=7.5in,trim= 0.in 4.1in 0in 0.5in, clip]{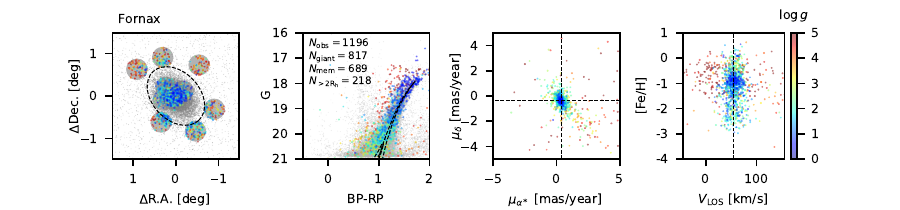}\vspace{-0.0in}\\
\includegraphics[width=7.5in,trim= 0.in 4.1in 0in 0.5in, clip]{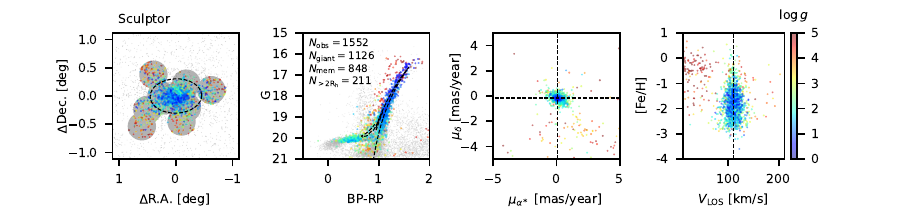}\vspace{-0.0in}\\
\includegraphics[width=7.5in,angle=0,trim= 0in 4.1in 0in 0.5in, clip]{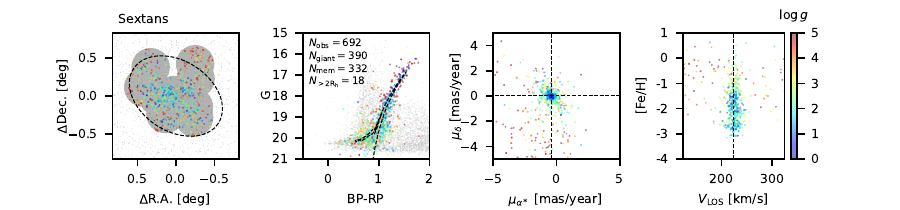}\vspace{-0.0in}\\
\end{tabular}
 \caption{Sky maps, color-magnitude diagrams (CMDs; middle), proper motion coordinates and spectroscopic \feh\ vs \vlos, from \gaia photometry/astrometry and Magellan/M2FS spectroscopy of point sources toward Galactic satellites.  Colored points indicate sources for which we report spectroscopic measurements, with bluer colors identifying likely red giant stars belonging to the satellites.  In sky maps, dashed ellipses have semi-major axis $a=2R_{\rm half}/\sqrt{1-\epsilon}$, where $R_{\rm half}$ is the projected halflight radius  and $\epsilon\equiv 1-b/a$ is the measured ellipticity, both adopted from the compilation by \citet{pace22}.  In CMDs, gray points indicate unobserved point sources within $1^{\circ}$ of the  satellite center; in sky maps, gray points indicate unobserved sources within $\delta=\max(0.15,\sqrt{\sigma_{\rm G}^2+\sigma_{\rm BP}^2+\sigma_{\rm RP}^2})$ magnitudes of the theoretical isochrone \citep{isochrones,dotter16} overplotted in the corresponding CMD (chosen for typical age  =10~Gyr and according to previously published mean metallicity).  Dashed lines indicate previously measured mean systemic proper motions and line-of-sight velocities.
\label{fig:m2fs_cmdmap}}
\end{figure*}

\renewcommand{\thefigure}{\arabic{figure} (Continued)}
%\addtocounter{figure}{-1}
\begin{figure*}
\ContinuedFloat
\captionsetup{list=off,format=cont}
%\centering
\begin{tabular}{@{}l@{}}
\includegraphics[width=7.5in,trim= 0.in 4.1in 0in 0.5in, clip]{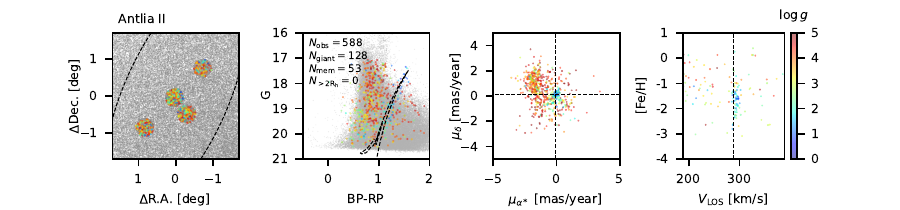}\vspace{-0.0in}\\
\includegraphics[width=7.5in,trim= 0.in 4.1in 0in 0.5in, clip]{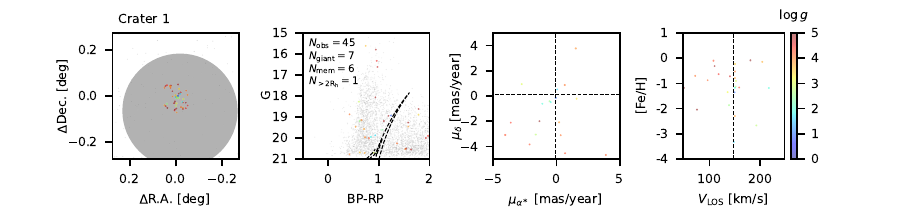}\vspace{-0.0in}\\
\includegraphics[width=7.5in,angle=0,trim= 0in 4.1in 0in 0.5in, clip]{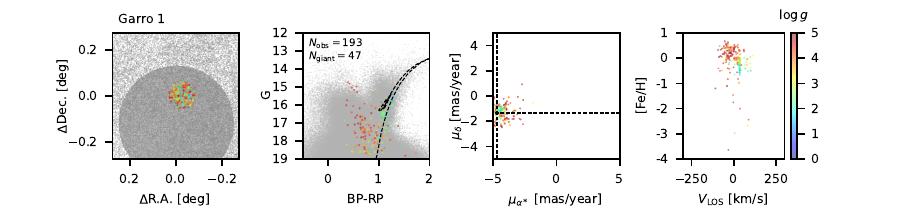}\vspace{-0.0in}\\
\includegraphics[width=7.5in,trim= 0.in 4.1in 0in 0.5in, clip]{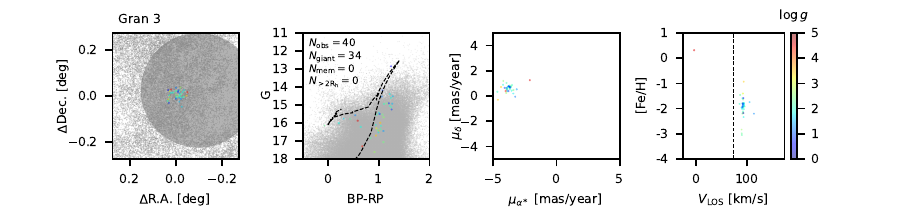}\vspace{-0.0in}\\
\includegraphics[width=7.5in,trim= 0.in 4.1in 0in 0.5in, clip]{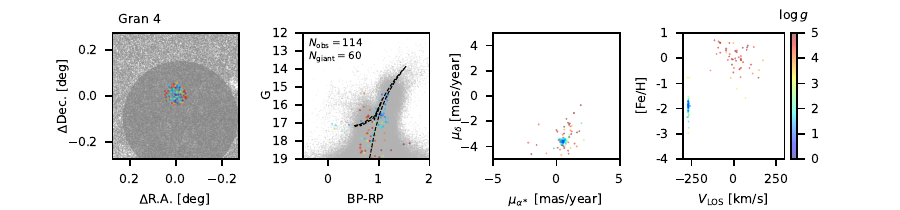}\vspace{-0.0in}\\\end{tabular}
 \caption{ 
%\label{fig:cmd}}
}
\end{figure*}
\renewcommand{\thefigure}{\arabic{figure}}

\renewcommand{\thefigure}{\arabic{figure} (Continued)}
%\addtocounter{figure}{-1}
\begin{figure*}
\ContinuedFloat
\captionsetup{list=off,format=cont}
%\centering
\begin{tabular}{@{}l@{}}
\includegraphics[width=7.5in,angle=0,trim= 0in 4.1in 0in 0.5in, clip]{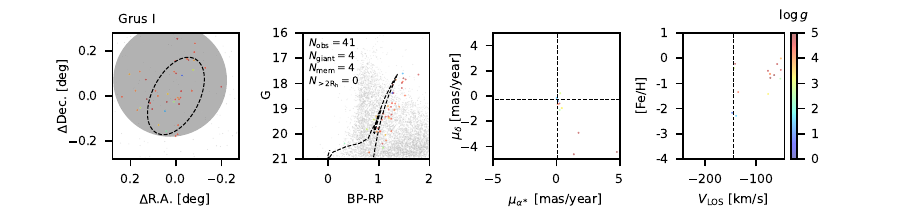}\vspace{-0.0in}\\
\includegraphics[width=7.5in,trim= 0.in 4.1in 0in 0.5in, clip]{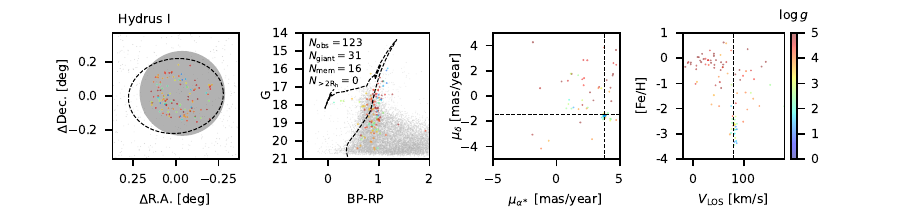}\vspace{-0.0in}\\
\includegraphics[width=7.5in,trim=0.in 4.1in 0in 0.5in, clip]{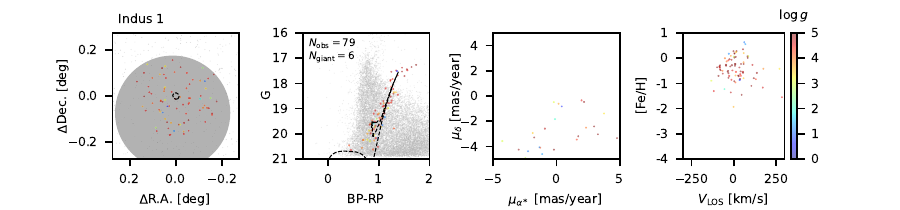}\vspace{-0.0in}\\
\includegraphics[width=7.5in,trim=0.in 4.1in 0in 0.5in, clip]{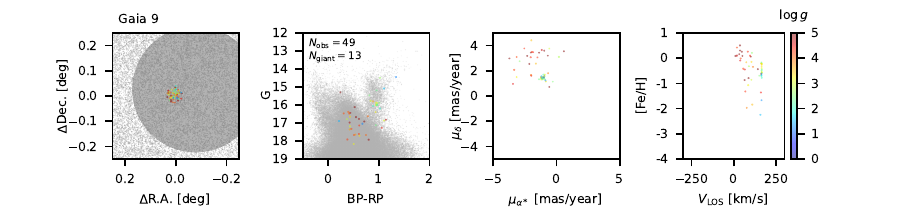}\vspace{-0.0in}\\
\includegraphics[width=7.5in,trim= 0.in 4.1in 0in 0.5in, clip]{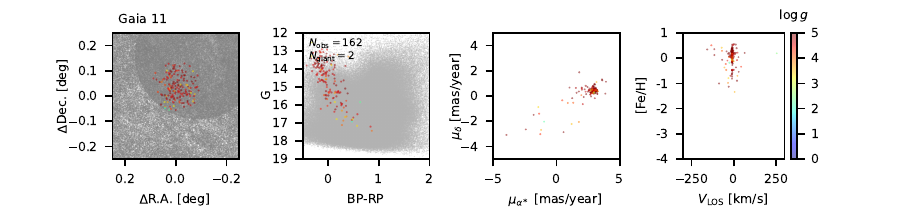}\vspace{-0.0in}\\\end{tabular}
 \caption{ 
%\label{fig:cmd}}
}
\end{figure*}
\renewcommand{\thefigure}{\arabic{figure}}

\renewcommand{\thefigure}{\arabic{figure} (Continued)}
%\addtocounter{figure}{-1}
\begin{figure*}
\ContinuedFloat
\captionsetup{list=off,format=cont}
\begin{tabular}{@{}l@{}}
\includegraphics[width=7.5in,angle=0,trim= 0in 4.1in 0in 0.5in, clip]{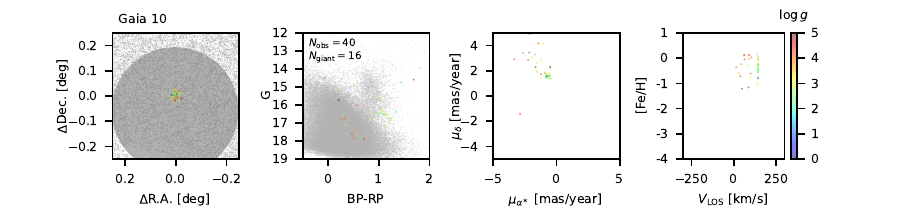}\vspace{-0.0in}\\
\includegraphics[width=7.5in,trim= 0.in 4.1in 0in 0.5in, clip]{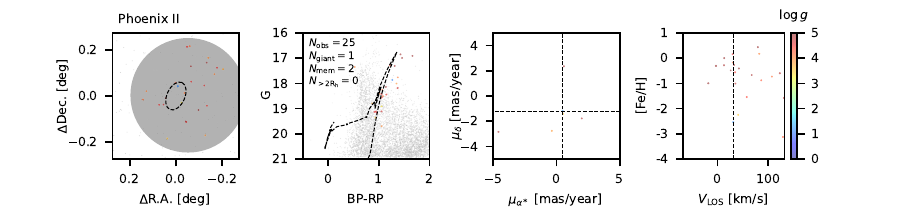}\vspace{-0.0in}\\
\includegraphics[width=7.5in,trim=0.in 4.1in 0in 0.5in, clip]{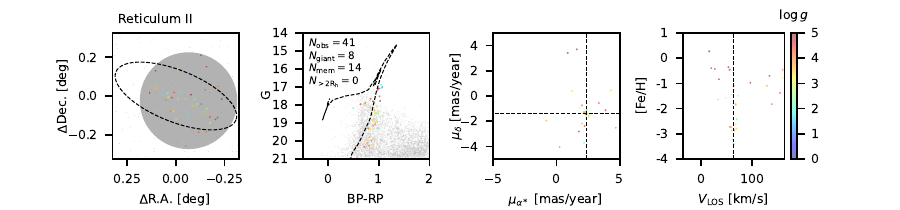}\vspace{-0.0in}\\
\includegraphics[width=7.5in,trim=0.in 4.1in 0in 0.5in, clip]{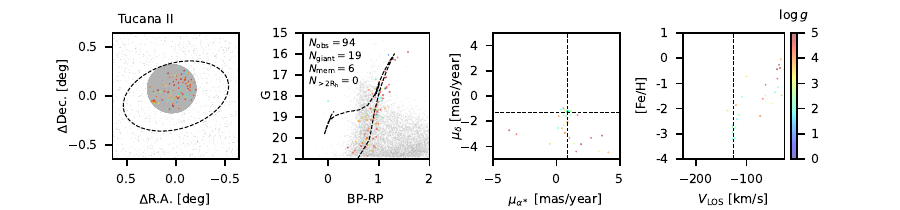}\vspace{-0.0in}\\
\end{tabular}
 \caption{ 
%\label{fig:cmd}}
}
\end{figure*}
\renewcommand{\thefigure}{\arabic{figure}}

\begin{figure*}
%\centering
\begin{tabular}{@{}l@{}}
  \includegraphics[width=7.5in,angle=0,trim= 0in 4.1in 0in 0.5in, clip]{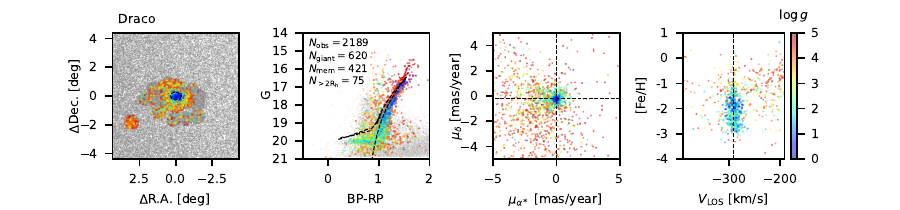}\vspace{-0.0in}\\
  \includegraphics[width=7.5in,trim= 0.in 4.1in 0in 0.5in, clip]{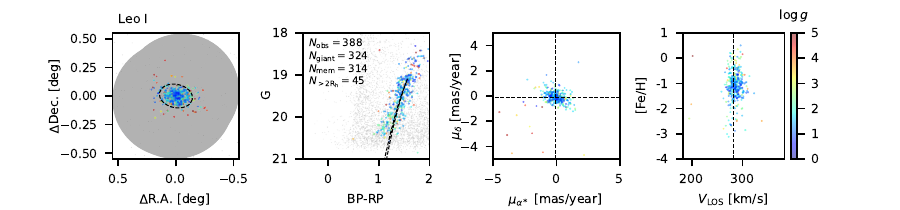}\vspace{-0.0in}\\
  \includegraphics[width=7.5in,trim= 0.in 4.1in 0in 0.5in, clip]{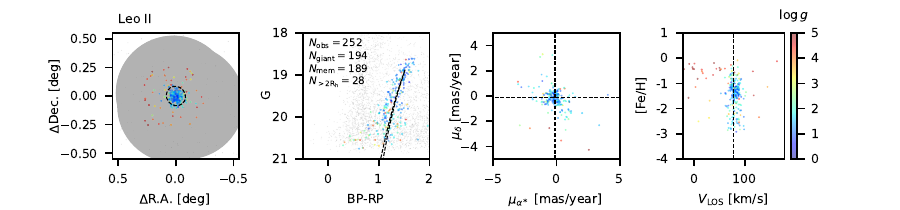}\vspace{-0.0in}\\
\includegraphics[width=7.5in,angle=0,trim= 0in 4.1in 0in 0.5in, clip]{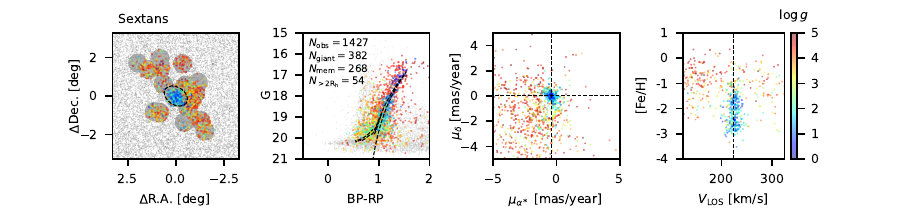}\vspace{-0.0in}\\
\includegraphics[width=7.5in,trim= 0.in 4.1in 0in 0.5in, clip]{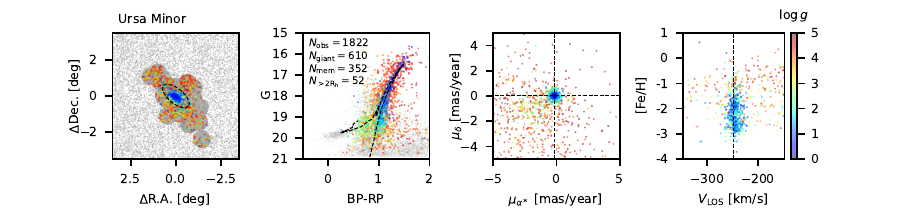}\vspace{-0.0in}\\\end{tabular}
 \caption{Same as Figure \ref{fig:m2fs_cmdmap}, but for Galactic satellites observed with MMT/Hectochelle.  
\label{fig:hecto_cmdmap}}
\end{figure*}

\renewcommand{\thefigure}{\arabic{figure} (Continued)}
%\addtocounter{figure}{-1}
\begin{figure*}
\ContinuedFloat
\captionsetup{list=off,format=cont}
%\centering
\begin{tabular}{@{}l@{}}
\includegraphics[width=7.5in,trim= 0.in 4.1in 0in 0.5in, clip]{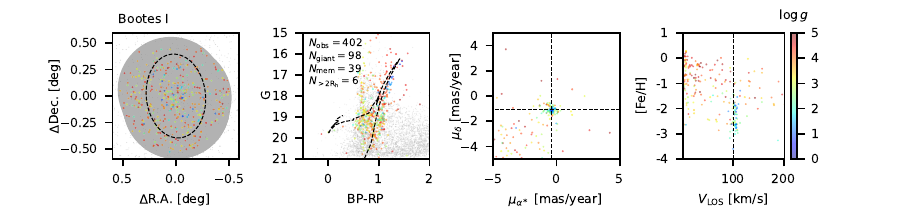}\vspace{-0.0in}\\
\includegraphics[width=7.5in,angle=0,trim= 0in 4.1in 0in 0.5in, clip]{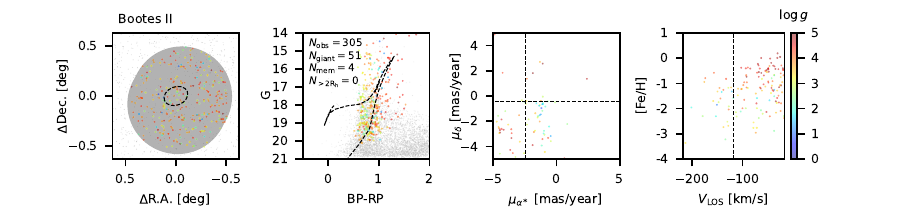}\vspace{-0.0in}\\
\includegraphics[width=7.5in,trim= 0.in 4.1in 0in 0.5in, clip]{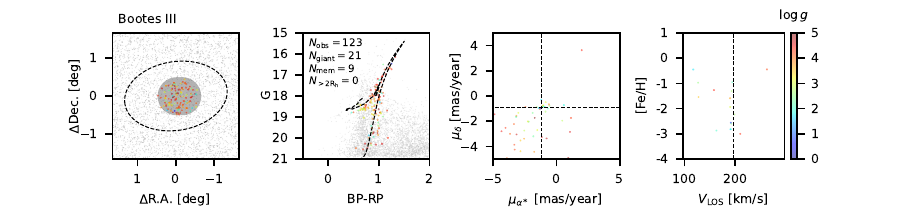}\vspace{-0.0in}\\
\includegraphics[width=7.5in,trim= 0.in 4.1in 0in 0.5in, clip]{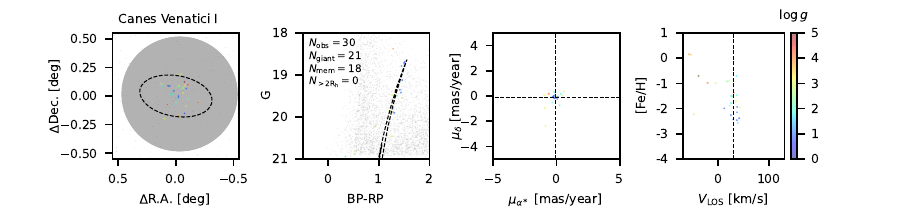}\vspace{-0.0in}\\
\includegraphics[width=7.5in,angle=0,trim= 0in 4.1in 0in 0.5in, clip]{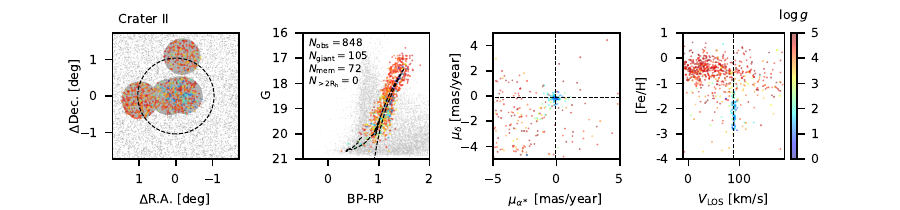}\vspace{-0.0in}\\
\end{tabular}
 \caption{ 
%\label{fig:cmd}}
}
\end{figure*}
\renewcommand{\thefigure}{\arabic{figure}}

\renewcommand{\thefigure}{\arabic{figure} (Continued)}
%\addtocounter{figure}{-1}
\begin{figure*}
\ContinuedFloat
\captionsetup{list=off,format=cont}
%\centering
\begin{tabular}{@{}l@{}}
\includegraphics[width=7.5in,trim= 0.in 4.1in 0in 0.5in, clip]{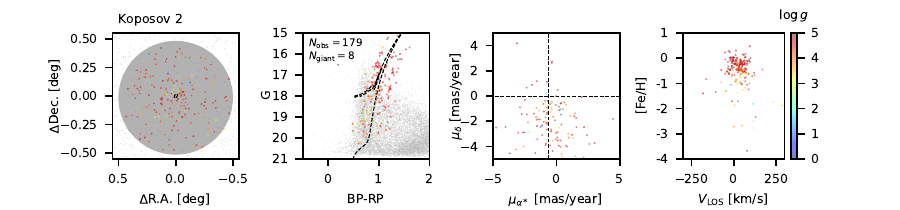}\vspace{-0.0in}\\
\includegraphics[width=7.5in,trim= 0.in 4.1in 0in 0.5in, clip]{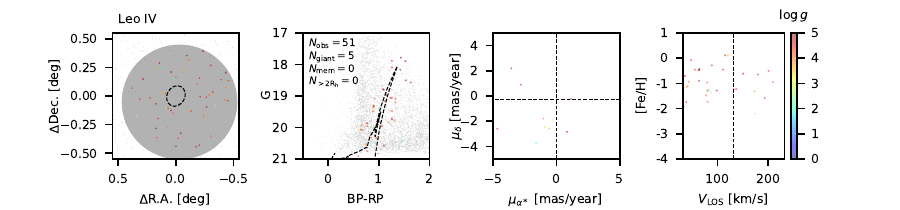}\vspace{-0.0in}\\
\includegraphics[width=7.5in,angle=0,trim= 0in 4.1in 0in 0.5in, clip]{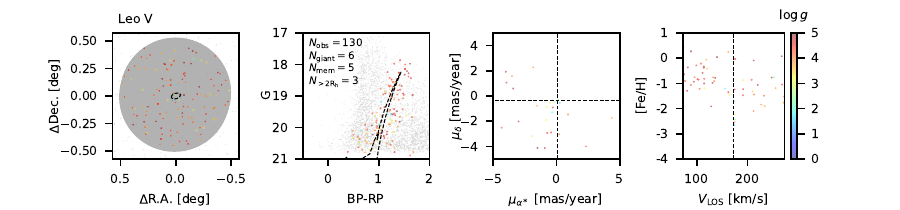}\vspace{-0.0in}\\
\includegraphics[width=7.5in,trim= 0.in 4.1in 0in 0.5in, clip]{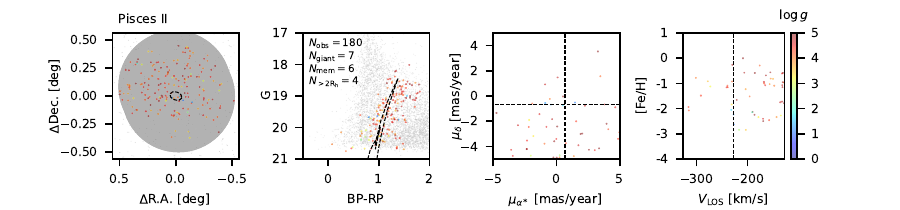}\vspace{-0.0in}\\
\includegraphics[width=7.5in,trim= 0.in 4.1in 0in 0.5in, clip]{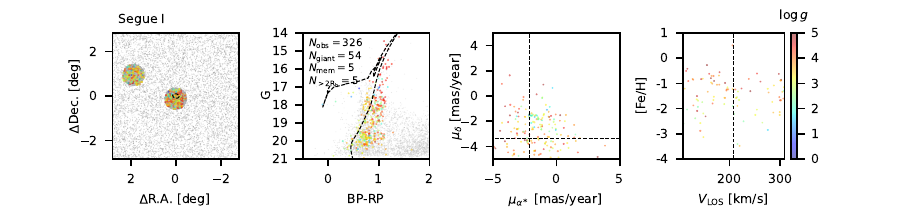}\vspace{-0.0in}\\
\end{tabular}
 \caption{ 
%\label{fig:cmd}}
}
\end{figure*}
\renewcommand{\thefigure}{\arabic{figure}}

\renewcommand{\thefigure}{\arabic{figure} (Continued)}
%\addtocounter{figure}{-1}
\begin{figure*}
\ContinuedFloat
\captionsetup{list=off,format=cont}
%\centering
\begin{tabular}{@{}l@{}}
\includegraphics[width=7.5in,angle=0,trim= 0in 4.1in 0in 0.5in, clip]{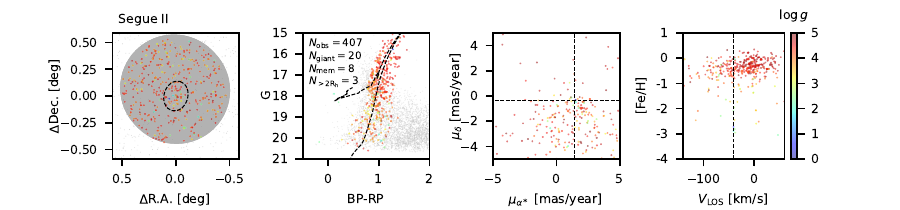}\vspace{-0.0in}\\
\includegraphics[width=7.5in,trim= 0.in 4.1in 0in 0.5in, clip]{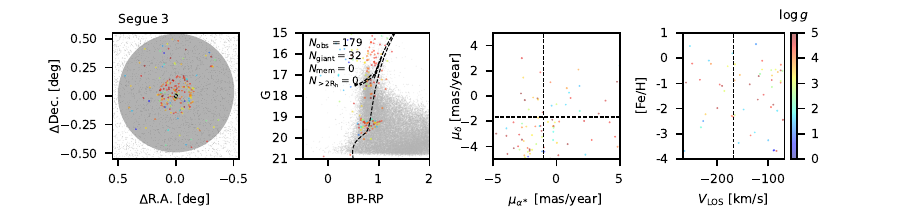}\vspace{-0.0in}\\
\includegraphics[width=7.5in,trim= 0.in 4.1in 0in 0.5in, clip]{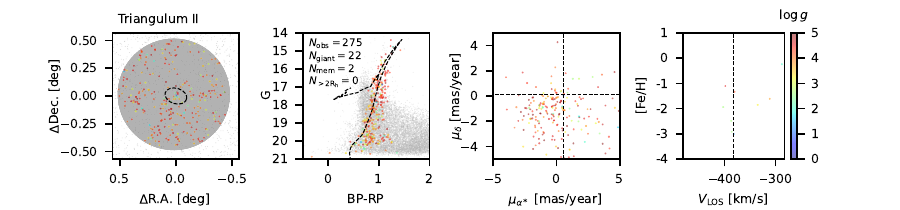}\vspace{-0.0in}\\
\includegraphics[width=7.5in,trim= 0.in 4.1in 0in 0.5in, clip]{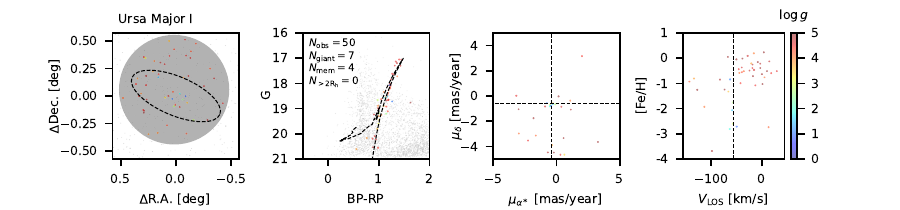}\vspace{-0.0in}\\
\end{tabular}
 \caption{ 
%\label{fig:cmd}}
}
\end{figure*}
\renewcommand{\thefigure}{\arabic{figure}}

\renewcommand{\thefigure}{\arabic{figure} (Continued)}
%\addtocounter{figure}{-1}
\begin{figure*}
\ContinuedFloat
\captionsetup{list=off,format=cont}
%\centering
\begin{tabular}{@{}l@{}}
\includegraphics[width=7.5in,angle=0,trim= 0in 4.1in 0in 0.5in, clip]{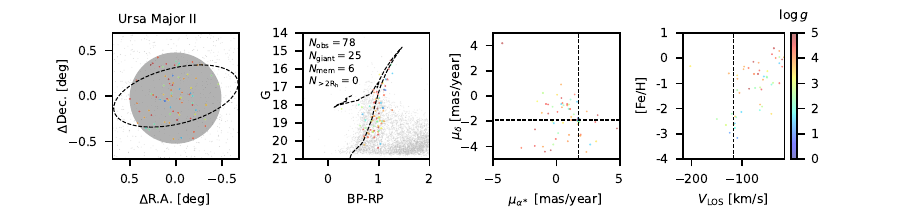}\vspace{-0.0in}\\
\end{tabular}
 \caption{ 
%\label{fig:cmd}}
}
\end{figure*}
\renewcommand{\thefigure}{\arabic{figure}}

\bibliography{ref}
\end{document}

%% file: m2fs_error_adjust.tex
\newcommand{\mtwofsHiReswavrmsn}{$34698$} 
\newcommand{\mtwofsHiReswavrmsmean}{$0.009$} 
\newcommand{\mtwofsHiReswavrmsstd}{$0.001$} 
 
\newcommand{\mtwofssigvfloorHiRes}{$0.57 \pm 0.01$} 
\newcommand{\mtwofssigvscaleHiRes}{$0.86 \pm 0.02$} 
\newcommand{\mtwofssigvmixfracHiRes}{$0.10 \pm 0.00$} 
\newcommand{\mtwofssigvsigoutHiRes}{$24.30 \pm 0.69$} 
\newcommand{\mtwofsvnpairsHiRes}{$6830$}

\newcommand{\mtwofssigtefffloorHiRes}{$58.59 \pm 2.13$} 
\newcommand{\mtwofssigteffscaleHiRes}{$0.91 \pm 0.01$} 
\newcommand{\mtwofssigteffmixfracHiRes}{$0.08 \pm 0.01$} 
\newcommand{\mtwofssigteffsigoutHiRes}{$424.43 \pm 24.79$}

\newcommand{\mtwofssigloggfloorHiRes}{$0.12 \pm 0.02$} 
\newcommand{\mtwofssigloggscaleHiRes}{$0.86 \pm 0.04$} 
\newcommand{\mtwofssigloggmixfracHiRes}{$0.13 \pm 0.06$} 
\newcommand{\mtwofssigloggsigoutHiRes}{$0.61 \pm 0.38$}

\newcommand{\mtwofssigfehfloorHiRes}{$0.06 \pm 0.00$} 
\newcommand{\mtwofssigfehscaleHiRes}{$1.18 \pm 0.02$} 
\newcommand{\mtwofssigfehmixfracHiRes}{$0.17 \pm 0.02$} 
\newcommand{\mtwofssigfehsigoutHiRes}{$0.29 \pm 0.02$}

\newcommand{\mtwofssigmgfefloorHiRes}{$0.04 \pm 0.01$} 
\newcommand{\mtwofssigmgfescaleHiRes}{$0.74 \pm 0.03$} 
\newcommand{\mtwofssigmgfemixfracHiRes}{$0.48 \pm 0.01$} 
\newcommand{\mtwofssigmgfesigoutHiRes}{$0.34 \pm 0.01$}

\newcommand{\mtwofsMedReswavrmsn}{$3248$} 
\newcommand{\mtwofsMedReswavrmsmean}{$0.023$} 
\newcommand{\mtwofsMedReswavrmsstd}{$0.003$} 
 
\newcommand{\mtwofssigvfloorMedRes}{$0.59 \pm 0.78$} 
\newcommand{\mtwofssigvscaleMedRes}{$1.38 \pm 0.14$} 
\newcommand{\mtwofssigvmixfracMedRes}{$0.12 \pm 0.03$} 
\newcommand{\mtwofssigvsigoutMedRes}{$116.61 \pm 170.64$} 
\newcommand{\mtwofsvnpairsMedRes}{$259$}

\newcommand{\mtwofssigtefffloorMedRes}{$10.16 \pm 19.39$} 
\newcommand{\mtwofssigteffscaleMedRes}{$1.00 \pm 0.10$} 
\newcommand{\mtwofssigteffmixfracMedRes}{$0.18 \pm 0.10$} 
\newcommand{\mtwofssigteffsigoutMedRes}{$507.14 \pm 519.59$}

\newcommand{\mtwofssigloggfloorMedRes}{$0.03 \pm 0.02$} 
\newcommand{\mtwofssigloggscaleMedRes}{$0.46 \pm 0.05$} 
\newcommand{\mtwofssigloggmixfracMedRes}{$0.09 \pm 0.05$} 
\newcommand{\mtwofssigloggsigoutMedRes}{$4.24 \pm 2.43$}

\newcommand{\mtwofssigfehfloorMedRes}{$0.12 \pm 0.08$} 
\newcommand{\mtwofssigfehscaleMedRes}{$1.34 \pm 0.15$} 
\newcommand{\mtwofssigfehmixfracMedRes}{$0.19 \pm 0.12$} 
\newcommand{\mtwofssigfehsigoutMedRes}{$0.51 \pm 0.55$}

\newcommand{\mtwofssigmgfefloorMedRes}{$0.03 \pm 0.02$} 
\newcommand{\mtwofssigmgfescaleMedRes}{$0.80 \pm 0.07$} 
\newcommand{\mtwofssigmgfemixfracMedRes}{$0.19 \pm 0.10$} 
\newcommand{\mtwofssigmgfesigoutMedRes}{$0.97 \pm 1.50$}

%% file: hecto_error_adjust.tex
\newcommand{\hectosigvfloor}{$0.39 \pm 0.01$} 
\newcommand{\hectosigvscale}{$0.94 \pm 0.02$} 
\newcommand{\hectosigvmixfrac}{$0.06 \pm 0.00$} 
\newcommand{\hectosigvsigout}{$27.64 \pm 1.14$} 
\newcommand{\hectovnpairs}{$6301$}

\newcommand{\hectosigtefffloor}{$0.61 \pm 1.00$} 
\newcommand{\hectosigteffscale}{$1.17 \pm 0.01$} 
\newcommand{\hectosigteffmixfrac}{$0.06 \pm 0.01$} 
\newcommand{\hectosigteffsigout}{$216.25 \pm 26.55$}

\newcommand{\hectosigloggfloor}{$0.03 \pm 0.01$} 
\newcommand{\hectosigloggscale}{$1.05 \pm 0.02$} 
\newcommand{\hectosigloggmixfrac}{$0.10 \pm 0.02$} 
\newcommand{\hectosigloggsigout}{$0.42 \pm 0.04$}

\newcommand{\hectosigfehfloor}{$0.01 \pm 0.00$} 
\newcommand{\hectosigfehscale}{$1.20 \pm 0.02$} 
\newcommand{\hectosigfehmixfrac}{$0.16 \pm 0.01$} 
\newcommand{\hectosigfehsigout}{$0.37 \pm 0.02$}

\newcommand{\hectosigmgfefloor}{$0.01 \pm 0.00$} 
\newcommand{\hectosigmgfescale}{$1.11 \pm 0.02$} 
\newcommand{\hectosigmgfemixfrac}{$0.23 \pm 0.02$} 
\newcommand{\hectosigmgfesigout}{$0.38 \pm 0.02$}

%% file: offsets_newcommands.tex
\newcommand{\mtwofshireshectovn}{180} 
\newcommand{\mtwofshireshthreevn}{77} 
\newcommand{\mtwofshireswalkervn}{1440} 
\newcommand{\mtwofshiresapogeevn}{117} 
\newcommand{\mtwofshiresmtwofsmedresvn}{26} 
 
\newcommand{\hectohthreevn}{767} 
\newcommand{\hectowalkervn}{194} 
\newcommand{\hectoapogeevn}{94} 
\newcommand{\hectomtwofsmedresvn}{4} 
 
\newcommand{\hthreewalkervn}{91} 
\newcommand{\hthreeapogeevn}{22} 
\newcommand{\hthreemtwofsmedresvn}{1} 
 
\newcommand{\walkerapogeevn}{281} 
\newcommand{\walkermtwofsmedresvn}{10} 
 
\newcommand{\mtwofsmedresapogeevn}{2}

\newcommand{\mtwofshireskirbyteffn}{115}

\newcommand{\hectokirbyteffn}{326}

\newcommand{\kirbyhthreeteffn}{25} 
\newcommand{\kirbyapogeeteffn}{75}

\newcommand{\mtwofshiresvoffset}{$0.47\pm 0.05$} 
\newcommand{\hectovoffset}{$-0.14\pm 0.05$} 
 
\newcommand{\hthreevoffset}{$-0.31\pm 0.05$} 
\newcommand{\walkervoffset}{$-0.19\pm 0.05$} 
\newcommand{\mtwofsmedresvoffset}{$0.07\pm 0.44$} 
\newcommand{\mtwofshiresteffoffset}{$-141\pm 5$} 
\newcommand{\hectoteffoffset}{$-221\pm 4$} 
\newcommand{\kirbyteffoffset}{$-172\pm 3$} 
\newcommand{\hthreeteffoffset}{$-69\pm 4$} 
 
\newcommand{\mtwofsmedresteffoffset}{$-323\pm 16$} 
\newcommand{\mtwofshiresloggoffset}{$-0.53\pm 0.02$} 
\newcommand{\hectologgoffset}{$-0.49\pm 0.01$} 
\newcommand{\kirbyloggoffset}{$-0.26\pm 0.01$} 
\newcommand{\hthreeloggoffset}{$-0.20\pm 0.01$} 
 
\newcommand{\mtwofsmedresloggoffset}{$-0.40\pm 0.05$} 
\newcommand{\mtwofshiresfehoffset}{$-0.26\pm 0.01$} 
\newcommand{\hectofehoffset}{$-0.26\pm 0.01$} 
\newcommand{\kirbyfehoffset}{$-0.18\pm 0.01$} 
\newcommand{\hthreefehoffset}{$-0.01\pm 0.01$} 
 
\newcommand{\mtwofsmedresfehoffset}{$-0.59\pm 0.06$} 
\newcommand{\mtwofshiresalphaoffset}{$0.19\pm 0.01$} 
\newcommand{\hectoalphaoffset}{$-0.01\pm 0.01$} 
\newcommand{\kirbyalphaoffset}{$0.24\pm 0.02$} 
\newcommand{\hthreealphaoffset}{$0.23\pm 0.01$} 
 
\newcommand{\mtwofsmedresalphaoffset}{$0.09\pm 0.06$} 

%% file: summary_stats.tex
\newcommand{\mtwofshiressystems}{$18$} 
 
\newcommand{\mtwofsmedressystems}{$5$}

\newcommand{\mtwofshiresgoodobs}{$8983$} 
 
\newcommand{\mtwofshiresgoodstar}{$6609$} 
\newcommand{\mtwofshiresgoodstark}{$6.6$} 
 
\newcommand{\mtwofshiresrepeated}{$1330$}

\newcommand{\mtwofsmedresgoodobs}{$189$} 
 
\newcommand{\mtwofsmedresgoodstar}{$82$}

\newcommand{\mtwofsmedresrepeated}{$33$}

\newcommand{\mtwofsrepeatedk}{$1.4$} 
\newcommand{\allrepeated}{$3720$} 
 
\newcommand{\mtwofshiresgoodrrl}{$292$} 
 
\newcommand{\mtwofsmedresgoodrrl}{$0$}

\newcommand{\mtwofsmaxgoodobs}{$15$} 
\newcommand{\hectosystems}{$21$}

\newcommand{\hectogoodobs}{$13328$} 
 
\newcommand{\hectogoodstar}{$9678$} 
\newcommand{\hectogoodstark}{$9.7$} 
 
\newcommand{\hectorepeated}{$2357$} 
\newcommand{\hectorepeatedk}{$2.4$} 
\newcommand{\hectogoodrrl}{$40$} 
\newcommand{\hectomaxgoodobs}{$13$} 
\newcommand{\allmaxgoodobs}{$15$} 
\newcommand{\allsystems}{$38$} 
\newcommand{\allgoodstar}{$16369$} 
 
\newcommand{\allgoodzstar}{$8189$}

\newcommand{\mtwofswavflagn}{$221$} 
\newcommand{\mtwofswavflaggoodn}{$42$} 
\newcommand{\vlosmedianerror}{$1.1$} 
\newcommand{\teffmedianerror}{$234$} 
\newcommand{\loggmedianerror}{$0.52$} 
\newcommand{\fehmedianerror}{$0.38$} 
\newcommand{\mgfemedianerror}{$0.24$} 

%% file: sspp_offsets.tex
\newcommand{\mtwofshiresssppvn}{$1265$} 
\newcommand{\mtwofsmedresssppvn}{$33$} 
\newcommand{\hectossppvn}{$3008$} 
\newcommand{\mtwofshiresssppvoffset}{$-0.47\pm 0.03$} 
\newcommand{\mtwofsmedresssppvoffset}{$-2.19\pm 0.63$} 
\newcommand{\hectossppvoffset}{$0.68\pm 0.01$} 
\newcommand{\mtwofshiresssppteffoffset}{$168\pm 3$} 
\newcommand{\mtwofsmedresssppteffoffset}{$123\pm 36$} 
\newcommand{\hectossppteffoffset}{$-180\pm 1$} 
\newcommand{\mtwofshiressspploggoffset}{$0.45\pm 0.01$} 
\newcommand{\mtwofsmedressspploggoffset}{$0.15\pm 0.07$} 
\newcommand{\hectosspploggoffset}{$-0.24\pm 0.00$} 
\newcommand{\mtwofshiresssppfehoffset}{$0.21\pm 0.01$} 
\newcommand{\mtwofsmedresssppfehoffset}{$0.20\pm 0.07$} 
\newcommand{\hectossppfehoffset}{$-0.18\pm 0.00$} 

%% file: m2fs_obs_newcommands.tex
\newcommand{\mtwofsnfieldstot}{92} 
\newcommand{\mtwofsnfieldshires}{74} 
\newcommand{\mtwofsnfieldsmedres}{1} 
\newcommand{\mtwofsnfieldsmix}{17} 
\newcommand{\mtwofsexptimetot}{0.68} 

%% file: hecto_obs_newcommands.tex
\newcommand{\hectoexptimetot}{1.42} 

%% file: flags_newcommands.tex
\newcommand{\mtwofshiresagnwrong}{$3$} 
 
\newcommand{\mtwofshiresphotvariablegoodn}{$551$}

\newcommand{\mtwofshiresagngoodn}{$75$}

\newcommand{\mtwofshireschitwooutliergoodn}{$60$} 
 
\newcommand{\mtwofshirescarbongoodn}{$37$} 
\newcommand{\mtwofshiresanychitwon}{$44$} 
\newcommand{\mtwofshiresanycarbonn}{$37$} 
\newcommand{\mtwofsmedresagnwrong}{$0$} 
 
\newcommand{\mtwofsmedresphotvariablegoodn}{$3$}

\newcommand{\mtwofsmedresagngoodn}{$1$}

\newcommand{\mtwofsmedreschitwooutliergoodn}{$114$} 
 
\newcommand{\mtwofsmedrescarbongoodn}{$1$} 
\newcommand{\mtwofsmedresanychitwon}{$28$} 
\newcommand{\mtwofsmedresanycarbonn}{$0$} 
\newcommand{\hectoagnwrong}{$6$} 
 
\newcommand{\hectophotvariablegoodn}{$764$}

\newcommand{\hectoagngoodn}{$363$}

\newcommand{\hectochitwooutliergoodn}{$131$} 
 
\newcommand{\hectocarbongoodn}{$144$} 
\newcommand{\hectoanychitwon}{$41$} 
\newcommand{\hectoanycarbonn}{$88$} 

%% file: nmember.tex
\newcommand{\ngianttotal}{$6078$} 
\newcommand{\nmembertotal}{$4494$} 
\newcommand{\nmemberfartotal}{$823$} 
\newcommand{\nmemberfarthreetotal}{$124$} 
\newcommand{\nmemberfarfourtotal}{$64$} 
\newcommand{\nmemberfarfivetotal}{$42$}

%% file: m2fs_obs_table_short.tex
\begin{deluxetable*}{lllllcccl} 
\tablewidth{0pt} 
\tablecaption{Log of M2FS Observations of Galactic Halo Objects (abbreviated---see electronic version for full table) \label{tab:m2fs_obs_table}} \tablehead{\colhead{Instrument}&\colhead{Field Center}&&\colhead{UT date\tablenotemark{a}}&\colhead{UT start\tablenotemark{b}}&\colhead{Exp. Time}&\colhead{$N_{\rm exp}$}&\colhead{$N_{\rm target}$}&\colhead{Object}\\ 
&\colhead{$\alpha_{2000}$ [deg.]}&\colhead{$\delta_{2000}$ [deg.]}&&&\colhead{[sec.]}&} 
\startdata 
M2FS HiRes & $153.028333$ & $-001.754667$ & 2014-02-24 & 08:44:35 &$8900$ & $5$ & $218$ & Sextans\\ 
M2FS HiRes & $100.746667$ & $-050.848333$ & 2014-02-25 & 03:01:36 &$5400$ & $3$ & $214$ & Carina\\ 
M2FS HiRes & $100.610000$ & $-051.082361$ & 2014-02-25 & 05:04:05 &$5700$ & $3$ & $214$ & Carina\\ 
M2FS HiRes & $153.684583$ & $-001.500944$ & 2014-02-26 & 08:00:05 &$6600$ & $5$ & $216$ & Sextans\\ 
M2FS HiRes & $153.685000$ & $-001.501000$ & 2014-02-27 & 07:13:52 &$3600$ & $3$ & $216$ & Sextans\\ 
M2FS HiRes & $153.292917$ & $-001.604694$ & 2014-02-28 & 07:46:28 &$3600$ & $3$ & $216$ & Sextans\\ 
M2FS HiRes & $100.399167$ & $-050.947139$ & 2014-12-15 & 06:28:27 &$8100$ & $3$ & $218$ & Carina\\ 
M2FS HiRes & $100.835417$ & $-051.099611$ & 2014-12-19 & 08:05:29 &$4500$ & $3$ & $207$ & Carina\\ 
M2FS HiRes & $099.959583$ & $-050.786417$ & 2014-12-21 & 07:56:08 &$5400$ & $3$ & $187$ & Carina\\ 
M2FS HiRes & $099.961667$ & $-050.784722$ & 2014-12-23 & 08:01:05 &$3600$ & $3$ & $177$ & Carina\\ 
M2FS HiRes & $053.810000$ & $-054.075444$ & 2015-02-19 & 02:24:12 &$7200$ & $3$ & $186$ & Reticulum II\\ 
M2FS HiRes & $153.292500$ & $-001.601639$ & 2015-02-22 & 06:50:05 &$7200$ & $4$ & $214$ & Sextans\\ 
M2FS HiRes/MedRes & $343.062917$ & $-058.493583$ & 2015-07-18 & 09:40:45 &$9000$ & $5$ & $137$ & Tucana II\\ 
\enddata 
\tablenotetext{a}{YYYY-MM-DD format} 
\tablenotetext{b}{Universal time at start of first exposure; HH:MM:SS format} 
\end{deluxetable*}

%% file: hecto_obs_table_short.tex
\begin{deluxetable*}{lllllcccl} 
\tablewidth{0pt} 
\tablecaption{Log of MMT/Hectochelle Observations of Galactic Halo Objects (abbreviated---see electronic version for full table) \label{tab:hecto_obs_table}
} 
\tablehead{\colhead{Instrument}&\colhead{Field Center}&&\colhead{UT date\tablenotemark{a}}&\colhead{UT start\tablenotemark{b}}&\colhead{Exp. Time}&\colhead{$N_{\rm exp}$}&\colhead{$N_{\rm target}$}&\colhead{Object}\\ 
&\colhead{$\alpha_{2000}$ [deg.]}&\colhead{$\delta_{2000}$ [deg.]}&&&\colhead{[sec.]}
} 
\startdata 
Hectochelle & $152.064708$ & $+012.349136$ & 2005-04-01 & 05:06:30 &$9079$ & $3$ & $143$ & Leo I\\ 
Hectochelle & $152.064708$ & $+012.349136$ & 2005-04-02 & 06:13:04 &$14400$ & $4$ & $143$ & Leo I\\ 
Hectochelle & $259.425000$ & $+058.049972$ & 2005-04-02 & 10:52:57 &$8700$ & $3$ & $132$ & Draco\\ 
Hectochelle & $152.166792$ & $+012.274975$ & 2006-04-20 & 05:53:41 &$7500$ & $3$ & $135$ & Leo I\\ 
Hectochelle & $152.107875$ & $+012.309992$ & 2006-04-24 & 05:07:09 &$8100$ & $3$ & $135$ & Leo I\\ 
Hectochelle & $168.355875$ & $+022.149333$ & 2006-04-25 & 05:12:38 &$8100$ & $3$ & $114$ & Leo II\\ 
Hectochelle & $260.958333$ & $+057.870000$ & 2006-04-25 & 08:12:45 &$4846$ & $5$ & $107$ & Draco\\ 
Hectochelle & $210.005667$ & $+014.483664$ & 2006-05-08 & 04:24:44 &$5400$ & $3$ & $191$ & Bootes I\\ 
Hectochelle & $260.102667$ & $+057.885250$ & 2007-02-23 & 12:27:48 &$5400$ & $3$ & $120$ & Draco\\ 
Hectochelle & $257.091792$ & $+057.877306$ & 2007-02-26 & 10:03:25 &$5400$ & $3$ & $139$ & Draco\\ 
Hectochelle & $262.915167$ & $+058.382108$ & 2007-02-26 & 12:22:21 &$7200$ & $4$ & $145$ & Draco\\ 
Hectochelle & $152.765458$ & $-001.052389$ & 2007-02-27 & 09:57:01 &$8400$ & $4$ & $203$ & Sextans\\ 
Hectochelle & $259.407542$ & $+057.775056$ & 2007-02-27 & 12:05:59 &$5400$ & $3$ & $89$ & Draco\\ 
\enddata 
\tablenotetext{a}{YYYY-MM-DD format} 
\tablenotetext{b}{Universal time at start of first exposure; HH:MM:SS format} 
\end{deluxetable*}

%% file: astroph.bbl
\begin{thebibliography}{}
\expandafter\ifx\csname natexlab\endcsname\relax\def\natexlab#1{#1}\fi

\bibitem[{{Aaronson}(1983)}]{aaronson83}
{Aaronson}, M. 1983, \apjl, 266, L11

\bibitem[{{Abbott} {et~al.}(2021){Abbott}, {Adam{\'o}w}, {Aguena}, {Allam},
  {Amon}, {Annis}, {Avila}, {Bacon}, {Banerji}, {Bechtol}, {Becker},
  {Bernstein}, {Bertin}, {Bhargava}, {Bridle}, {Brooks}, {Burke}, {Carnero
  Rosell}, {Carrasco Kind}, {Carretero}, {Castander}, {Cawthon}, {Chang},
  {Choi}, {Conselice}, {Costanzi}, {Crocce}, {da Costa}, {Davis}, {De Vicente},
  {DeRose}, {Desai}, {Diehl}, {Dietrich}, {Drlica-Wagner}, {Eckert},
  {Elvin-Poole}, {Everett}, {Evrard}, {Ferrero}, {Fert{\'e}}, {Flaugher},
  {Fosalba}, {Friedel}, {Frieman}, {Garc{\'\i}a-Bellido}, {Gaztanaga},
  {Gelman}, {Gerdes}, {Giannantonio}, {Gill}, {Gruen}, {Gruendl}, {Gschwend},
  {Gutierrez}, {Hartley}, {Hinton}, {Hollowood}, {Honscheid}, {Huterer},
  {James}, {Jeltema}, {Johnson}, {Kent}, {Kron}, {Kuehn}, {Kuropatkin},
  {Lahav}, {Li}, {Lidman}, {Lin}, {MacCrann}, {Maia}, {Manning}, {Maloney},
  {March}, {Marshall}, {Martini}, {Melchior}, {Menanteau}, {Miquel}, {Morgan},
  {Myles}, {Neilsen}, {Ogando}, {Palmese}, {Paz-Chinch{\'o}n}, {Petravick},
  {Pieres}, {Plazas}, {Pond}, {Rodriguez-Monroy}, {Romer}, {Roodman}, {Rykoff},
  {Sako}, {Sanchez}, {Santiago}, {Scarpine}, {Serrano}, {Sevilla-Noarbe},
  {Smith}, {Smith}, {Soares-Santos}, {Suchyta}, {Swanson}, {Tarle}, {Thomas},
  {To}, {Tremblay}, {Troxel}, {Tucker}, {Turner}, {Varga}, {Walker},
  {Wechsler}, {Weller}, {Wester}, {Wilkinson}, {Yanny}, {Zhang}, {Nikutta},
  {Fitzpatrick}, {Jacques}, {Scott}, {Olsen}, {Huang}, {Herrera}, {Juneau},
  {Nidever}, {Weaver}, {Adean}, {Correia}, {de Freitas}, {Freitas},
  {Singulani}, {Vila-Verde}, \& {Linea Science Server}}]{desdr2}
{Abbott}, T.~M.~C., {Adam{\'o}w}, M., {Aguena}, M., {et~al.} 2021, \apjs, 255,
  20

\bibitem[{{Abdurro'uf} {et~al.}(2022){Abdurro'uf}, {Accetta}, {Aerts}, {Silva
  Aguirre}, {Ahumada}, {Ajgaonkar}, {Filiz Ak}, {Alam}, {Allende Prieto},
  {Almeida}, {Anders}, {Anderson}, {Andrews}, {Anguiano}, {Aquino-Ort{\'\i}z},
  {Arag{\'o}n-Salamanca}, {Argudo-Fern{\'a}ndez}, {Ata}, {Aubert},
  {Avila-Reese}, {Badenes}, {Barb{\'a}}, {Barger}, {Barrera-Ballesteros},
  {Beaton}, {Beers}, {Belfiore}, {Bender}, {Bernardi}, {Bershady}, {Beutler},
  {Bidin}, {Bird}, {Bizyaev}, {Blanc}, {Blanton}, {Boardman}, {Bolton},
  {Boquien}, {Borissova}, {Bovy}, {Brandt}, {Brown}, {Brownstein}, {Brusa},
  {Buchner}, {Bundy}, {Burchett}, {Bureau}, {Burgasser}, {Cabang}, {Campbell},
  {Cappellari}, {Carlberg}, {Wanderley}, {Carrera}, {Cash}, {Chen}, {Chen},
  {Cherinka}, {Chiappini}, {Choi}, {Chojnowski}, {Chung}, {Clerc}, {Cohen},
  {Comerford}, {Comparat}, {da Costa}, {Covey}, {Crane}, {Cruz-Gonzalez},
  {Culhane}, {Cunha}, {Dai}, {Damke}, {Darling}, {Davidson}, {Davies},
  {Dawson}, {De Lee}, {Diamond-Stanic}, {Cano-D{\'\i}az}, {S{\'a}nchez},
  {Donor}, {Duckworth}, {Dwelly}, {Eisenstein}, {Elsworth}, {Emsellem},
  {Eracleous}, {Escoffier}, {Fan}, {Farr}, {Feng}, {Fern{\'a}ndez-Trincado},
  {Feuillet}, {Filipp}, {Fillingham}, {Frinchaboy}, {Fromenteau}, {Galbany},
  {Garc{\'\i}a}, {Garc{\'\i}a-Hern{\'a}ndez}, {Ge}, {Geisler}, {Gelfand},
  {G{\'e}ron}, {Gibson}, {Goddy}, {Godoy-Rivera}, {Grabowski}, {Green},
  {Greener}, {Grier}, {Griffith}, {Guo}, {Guy}, {Hadjara}, {Harding},
  {Hasselquist}, {Hayes}, {Hearty}, {Hern{\'a}ndez}, {Hill}, {Hogg},
  {Holtzman}, {Horta}, {Hsieh}, {Hsu}, {Hsu}, {Huber}, {Huertas-Company},
  {Hutchinson}, {Hwang}, {Ibarra-Medel}, {Chitham}, {Ilha}, {Imig}, {Jaekle},
  {Jayasinghe}, {Ji}, {Johnson}, {Jones}, {J{\"o}nsson}, {Katkov}, {Khalatyan},
  {Kinemuchi}, {Kisku}, {Knapen}, {Kneib}, {Kollmeier}, {Kong}, {Kounkel},
  {Kreckel}, {Krishnarao}, {Lacerna}, {Lane}, {Langgin}, {Lavender}, {Law},
  {Lazarz}, {Leung}, {Leung}, {Lewis}, {Li}, {Li}, {Lian}, {Liang}, {Lin},
  {Lin}, {Lin}, {Lintott}, {Long}, {Longa-Pe{\~n}a}, {L{\'o}pez-Cob{\'a}},
  {Lu}, {Lundgren}, {Luo}, {Mackereth}, {de la Macorra}, {Mahadevan},
  {Majewski}, {Manchado}, {Mandeville}, {Maraston}, {Margalef-Bentabol},
  {Masseron}, {Masters}, {Mathur}, {McDermid}, {Mckay}, {Merloni},
  {Merrifield}, {Meszaros}, {Miglio}, {Di Mille}, {Minniti}, {Minsley},
  {Monachesi}, {Moon}, {Mosser}, {Mulchaey}, {Muna}, {Mu{\~n}oz}, {Myers},
  {Myers}, {Nadathur}, {Nair}, {Nandra}, {Neumann}, {Newman}, {Nidever},
  {Nikakhtar}, {Nitschelm}, {O'Connell}, {Garma-Oehmichen}, {Luan Souza de
  Oliveira}, {Olney}, {Oravetz}, {Ortigoza-Urdaneta}, {Osorio}, {Otter},
  {Pace}, {Padilla}, {Pan}, {Pan}, {Parikh}, {Parker}, {Peirani}, {Pe{\~n}a
  Ram{\'\i}rez}, {Penny}, {Percival}, {Perez-Fournon}, {Pinsonneault},
  {Poidevin}, {Poovelil}, {Price-Whelan}, {B{\'a}rbara de Andrade Queiroz},
  {Raddick}, {Ray}, {Rembold}, {Riddle}, {Riffel}, {Riffel}, {Rix}, {Robin},
  {Rodr{\'\i}guez-Puebla}, {Roman-Lopes}, {Rom{\'a}n-Z{\'u}{\~n}iga}, {Rose},
  {Ross}, {Rossi}, {Rubin}, {Salvato}, {S{\'a}nchez}, {S{\'a}nchez-Gallego},
  {Sanderson}, {Santana Rojas}, {Sarceno}, {Sarmiento}, {Sayres}, {Sazonova},
  {Schaefer}, {Schiavon}, {Schlegel}, {Schneider}, {Schultheis}, {Schwope},
  {Serenelli}, {Serna}, {Shao}, {Shapiro}, {Sharma}, {Shen}, {Shetrone}, {Shu},
  {Simon}, {Skrutskie}, {Smethurst}, {Smith}, {Sobeck}, {Spoo}, {Sprague},
  {Stark}, {Stassun}, {Steinmetz}, {Stello}, {Stone-Martinez},
  {Storchi-Bergmann}, {Stringfellow}, {Stutz}, {Su}, {Taghizadeh-Popp},
  {Talbot}, {Tayar}, {Telles}, {Teske}, {Thakar}, {Theissen}, {Tkachenko},
  {Thomas}, {Tojeiro}, {Hernandez Toledo}, {Troup}, {Trump}, {Trussler},
  {Turner}, {Tuttle}, {Unda-Sanzana}, {V{\'a}zquez-Mata}, {Valentini},
  {Valenzuela}, {Vargas-Gonz{\'a}lez}, {Vargas-Maga{\~n}a}, {Alfaro},
  {Villanova}, {Vincenzo}, {Wake}, {Warfield}, {Washington}, {Weaver},
  {Weijmans}, {Weinberg}, {Weiss}, {Westfall}, {Wild}, {Wilde}, {Wilson},
  {Wilson}, {Wilson}, {Wolf}, {Wood-Vasey}, {Yan}, {Zamora}, {Zasowski},
  {Zhang}, {Zhao}, {Zheng}, {Zheng}, \& {Zhu}}]{apogee_dr17}
{Abdurro'uf}, {Accetta}, K., {Aerts}, C., {et~al.} 2022, \apjs, 259, 35

\bibitem[{{Ahn} {et~al.}(2012){Ahn}, {Alexandroff}, {Allende Prieto},
  {Anderson}, {Anderton}, {Andrews}, {Aubourg}, {Bailey}, {Balbinot}, {Barnes},
  {Bautista}, {Beers}, {Beifiori}, {Berlind}, {Bhardwaj}, {Bizyaev}, {Blake},
  {Blanton}, {Blomqvist}, {Bochanski}, {Bolton}, {Borde}, {Bovy}, {Brandt},
  {Brinkmann}, {Brown}, {Brownstein}, {Bundy}, {Busca}, {Carithers}, {Carnero},
  {Carr}, {Casetti-Dinescu}, {Chen}, {Chiappini}, {Comparat}, {Connolly},
  {Crepp}, {Cristiani}, {Croft}, {Cuesta}, {da Costa}, {Davenport}, {Dawson},
  {de Putter}, {De Lee}, {Delubac}, {Dhital}, {Ealet}, {Ebelke}, {Edmondson},
  {Eisenstein}, {Escoffier}, {Esposito}, {Evans}, {Fan}, {Femen{\'\i}a
  Castell{\'a}}, {Fern{\'a}ndez Alvar}, {Ferreira}, {Filiz Ak}, {Finley},
  {Fleming}, {Font-Ribera}, {Frinchaboy}, {Garc{\'\i}a-Hern{\'a}ndez},
  {Garc{\'\i}a P{\'e}rez}, {Ge}, {G{\'e}nova-Santos}, {Gillespie}, {Girardi},
  {Gonz{\'a}lez Hern{\'a}ndez}, {Grebel}, {Gunn}, {Guo}, {Haggard}, {Hamilton},
  {Harris}, {Hawley}, {Hearty}, {Ho}, {Hogg}, {Holtzman}, {Honscheid},
  {Huehnerhoff}, {Ivans}, {Ivezi{\'c}}, {Jacobson}, {Jiang}, {Johansson},
  {Johnson}, {Kauffmann}, {Kirkby}, {Kirkpatrick}, {Klaene}, {Knapp}, {Kneib},
  {Le Goff}, {Leauthaud}, {Lee}, {Lee}, {Long}, {Loomis}, {Lucatello},
  {Lundgren}, {Lupton}, {Ma}, {Ma}, {MacDonald}, {Mack}, {Mahadevan}, {Maia},
  {Majewski}, {Makler}, {Malanushenko}, {Malanushenko}, {Manchado},
  {Mandelbaum}, {Manera}, {Maraston}, {Margala}, {Martell}, {McBride},
  {McGreer}, {McMahon}, {M{\'e}nard}, {Meszaros}, {Miralda-Escud{\'e}},
  {Montero-Dorta}, {Montesano}, {Morrison}, {Muna}, {Munn}, {Murayama},
  {Myers}, {Neto}, {Nguyen}, {Nichol}, {Nidever}, {Noterdaeme}, {Nuza},
  {Ogando}, {Olmstead}, {Oravetz}, {Owen}, {Padmanabhan},
  {Palanque-Delabrouille}, {Pan}, {Parejko}, {Parihar}, {P{\^a}ris},
  {Pattarakijwanich}, {Pepper}, {Percival}, {P{\'e}rez-Fournon},
  {P{\'e}rez-R{\`a}fols}, {Petitjean}, {Pforr}, {Pieri}, {Pinsonneault}, {Porto
  de Mello}, {Prada}, {Price-Whelan}, {Raddick}, {Rebolo}, {Rich}, {Richards},
  {Robin}, {Rocha-Pinto}, {Rockosi}, {Roe}, {Ross}, {Ross}, {Rossi},
  {Rubi{\~n}o-Martin}, {Samushia}, {Sanchez Almeida}, {S{\'a}nchez},
  {Santiago}, {Sayres}, {Schlegel}, {Schlesinger}, {Schmidt}, {Schneider},
  {Schultheis}, {Schwope}, {Sc{\'o}ccola}, {Seljak}, {Sheldon}, {Shen}, {Shu},
  {Simmerer}, {Simmons}, {Skibba}, {Skrutskie}, {Slosar}, {Sobreira}, {Sobeck},
  {Stassun}, {Steele}, {Steinmetz}, {Strauss}, {Streblyanska}, {Suzuki},
  {Swanson}, {Tal}, {Thakar}, {Thomas}, {Thompson}, {Tinker}, {Tojeiro},
  {Tremonti}, {Vargas Maga{\~n}a}, {Verde}, {Viel}, {Vikas}, {Vogt}, {Wake},
  {Wang}, {Weaver}, {Weinberg}, {Weiner}, {West}, {White}, {Wilson},
  {Wisniewski}, {Wood-Vasey}, {Yanny}, {Y{\`e}che}, {York}, {Zamora},
  {Zasowski}, {Zehavi}, {Zhao}, {Zheng}, {Zhu}, \& {Zinn}}]{sdssdr9}
{Ahn}, C.~P., {Alexandroff}, R., {Allende Prieto}, C., {et~al.} 2012, \apjs,
  203, 21

\bibitem[{{Aoki} {et~al.}(2009){Aoki}, {Arimoto}, {Sadakane}, {Tolstoy},
  {Battaglia}, {Jablonka}, {Shetrone}, {Letarte}, {Irwin}, {Hill}, {Francois},
  {Venn}, {Primas}, {Helmi}, {Kaufer}, {Tafelmeyer}, {Szeifert}, \&
  {Babusiaux}}]{aoki09}
{Aoki}, W., {Arimoto}, N., {Sadakane}, K., {et~al.} 2009, \aap, 502, 569

\bibitem[{{Astropy Collaboration} {et~al.}(2013){Astropy Collaboration},
  {Robitaille}, {Tollerud}, {Greenfield}, {Droettboom}, {Bray}, {Aldcroft},
  {Davis}, {Ginsburg}, {Price-Whelan}, {Kerzendorf}, {Conley}, {Crighton},
  {Barbary}, {Muna}, {Ferguson}, {Grollier}, {Parikh}, {Nair}, {Unther},
  {Deil}, {Woillez}, {Conseil}, {Kramer}, {Turner}, {Singer}, {Fox}, {Weaver},
  {Zabalza}, {Edwards}, {Azalee Bostroem}, {Burke}, {Casey}, {Crawford},
  {Dencheva}, {Ely}, {Jenness}, {Labrie}, {Lim}, {Pierfederici}, {Pontzen},
  {Ptak}, {Refsdal}, {Servillat}, \& {Streicher}}]{astropy1}
{Astropy Collaboration}, {Robitaille}, T.~P., {Tollerud}, E.~J., {et~al.} 2013,
  \aap, 558, A33

\bibitem[{{Astropy Collaboration} {et~al.}(2018){Astropy Collaboration},
  {Price-Whelan}, {Sip{\H{o}}cz}, {G{\"u}nther}, {Lim}, {Crawford}, {Conseil},
  {Shupe}, {Craig}, {Dencheva}, {Ginsburg}, {Vand erPlas}, {Bradley},
  {P{\'e}rez-Su{\'a}rez}, {de Val-Borro}, {Aldcroft}, {Cruz}, {Robitaille},
  {Tollerud}, {Ardelean}, {Babej}, {Bach}, {Bachetti}, {Bakanov}, {Bamford},
  {Barentsen}, {Barmby}, {Baumbach}, {Berry}, {Biscani}, {Boquien}, {Bostroem},
  {Bouma}, {Brammer}, {Bray}, {Breytenbach}, {Buddelmeijer}, {Burke},
  {Calderone}, {Cano Rodr{\'\i}guez}, {Cara}, {Cardoso}, {Cheedella}, {Copin},
  {Corrales}, {Crichton}, {D'Avella}, {Deil}, {Depagne}, {Dietrich}, {Donath},
  {Droettboom}, {Earl}, {Erben}, {Fabbro}, {Ferreira}, {Finethy}, {Fox},
  {Garrison}, {Gibbons}, {Goldstein}, {Gommers}, {Greco}, {Greenfield},
  {Groener}, {Grollier}, {Hagen}, {Hirst}, {Homeier}, {Horton}, {Hosseinzadeh},
  {Hu}, {Hunkeler}, {Ivezi{\'c}}, {Jain}, {Jenness}, {Kanarek}, {Kendrew},
  {Kern}, {Kerzendorf}, {Khvalko}, {King}, {Kirkby}, {Kulkarni}, {Kumar},
  {Lee}, {Lenz}, {Littlefair}, {Ma}, {Macleod}, {Mastropietro}, {McCully},
  {Montagnac}, {Morris}, {Mueller}, {Mumford}, {Muna}, {Murphy}, {Nelson},
  {Nguyen}, {Ninan}, {N{\"o}the}, {Ogaz}, {Oh}, {Parejko}, {Parley}, {Pascual},
  {Patil}, {Patil}, {Plunkett}, {Prochaska}, {Rastogi}, {Reddy Janga},
  {Sabater}, {Sakurikar}, {Seifert}, {Sherbert}, {Sherwood-Taylor}, {Shih},
  {Sick}, {Silbiger}, {Singanamalla}, {Singer}, {Sladen}, {Sooley},
  {Sornarajah}, {Streicher}, {Teuben}, {Thomas}, {Tremblay}, {Turner},
  {Terr{\'o}n}, {van Kerkwijk}, {de la Vega}, {Watkins}, {Weaver}, {Whitmore},
  {Woillez}, {Zabalza}, \& {Astropy Contributors}}]{astropy2}
{Astropy Collaboration}, {Price-Whelan}, A.~M., {Sip{\H{o}}cz}, B.~M., {et~al.}
  2018, \aj, 156, 123

\bibitem[{{Astropy Collaboration} {et~al.}(2022){Astropy Collaboration},
  {Price-Whelan}, {Lim}, {Earl}, {Starkman}, {Bradley}, {Shupe}, {Patil},
  {Corrales}, {Brasseur}, {N{\"o}the}, {Donath}, {Tollerud}, {Morris},
  {Ginsburg}, {Vaher}, {Weaver}, {Tocknell}, {Jamieson}, {van Kerkwijk},
  {Robitaille}, {Merry}, {Bachetti}, {G{\"u}nther}, {Aldcroft},
  {Alvarado-Montes}, {Archibald}, {B{\'o}di}, {Bapat}, {Barentsen},
  {Baz{\'a}n}, {Biswas}, {Boquien}, {Burke}, {Cara}, {Cara}, {Conroy},
  {Conseil}, {Craig}, {Cross}, {Cruz}, {D'Eugenio}, {Dencheva}, {Devillepoix},
  {Dietrich}, {Eigenbrot}, {Erben}, {Ferreira}, {Foreman-Mackey}, {Fox},
  {Freij}, {Garg}, {Geda}, {Glattly}, {Gondhalekar}, {Gordon}, {Grant},
  {Greenfield}, {Groener}, {Guest}, {Gurovich}, {Handberg}, {Hart},
  {Hatfield-Dodds}, {Homeier}, {Hosseinzadeh}, {Jenness}, {Jones}, {Joseph},
  {Kalmbach}, {Karamehmetoglu}, {Ka{\l}uszy{\'n}ski}, {Kelley}, {Kern},
  {Kerzendorf}, {Koch}, {Kulumani}, {Lee}, {Ly}, {Ma}, {MacBride}, {Maljaars},
  {Muna}, {Murphy}, {Norman}, {O'Steen}, {Oman}, {Pacifici}, {Pascual},
  {Pascual-Granado}, {Patil}, {Perren}, {Pickering}, {Rastogi}, {Roulston},
  {Ryan}, {Rykoff}, {Sabater}, {Sakurikar}, {Salgado}, {Sanghi}, {Saunders},
  {Savchenko}, {Schwardt}, {Seifert-Eckert}, {Shih}, {Jain}, {Shukla}, {Sick},
  {Simpson}, {Singanamalla}, {Singer}, {Singhal}, {Sinha}, {Sip{\H{o}}cz},
  {Spitler}, {Stansby}, {Streicher}, {{\v{S}}umak}, {Swinbank}, {Taranu},
  {Tewary}, {Tremblay}, {de Val-Borro}, {Van Kooten}, {Vasovi{\'c}}, {Verma},
  {de Miranda Cardoso}, {Williams}, {Wilson}, {Winkel}, {Wood-Vasey}, {Xue},
  {Yoachim}, {Zhang}, {Zonca}, \& {Astropy Project Contributors}}]{astropy3}
{Astropy Collaboration}, {Price-Whelan}, A.~M., {Lim}, P.~L., {et~al.} 2022,
  \apj, 935, 167

\bibitem[{{Battaglia} \& {Nipoti}(2022)}]{battaglia22}
{Battaglia}, G., \& {Nipoti}, C. 2022, Nature Astronomy, 6, 659

\bibitem[{{Battaglia} {et~al.}(2006){Battaglia}, {Tolstoy}, {Helmi}, {Irwin},
  {Letarte}, {Jablonka}, {Hill}, {Venn}, {Shetrone}, {Arimoto}, {Primas},
  {Kaufer}, {Francois}, {Szeifert}, {Abel}, \& {Sadakane}}]{battaglia06}
{Battaglia}, G., {Tolstoy}, E., {Helmi}, A., {et~al.} 2006, \aap, 459, 423

\bibitem[{{Belokurov} {et~al.}(2018){Belokurov}, {Erkal}, {Evans}, {Koposov},
  \& {Deason}}]{belokurov18}
{Belokurov}, V., {Erkal}, D., {Evans}, N.~W., {Koposov}, S.~E., \& {Deason},
  A.~J. 2018, \mnras, 478, 611

\bibitem[{{Belokurov} \& {Evans}(2022)}]{belokurov22}
{Belokurov}, V., \& {Evans}, N.~W. 2022, Nature Astronomy, 6, 911

\bibitem[{{Boutsia} {et~al.}(2021){Boutsia}, {Grazian}, {Fontanot},
  {Giallongo}, {Menci}, {Calderone}, {Cristiani}, {D'Odorico}, {Cupani},
  {Guarneri}, \& {Omizzolo}}]{boutsia21}
{Boutsia}, K., {Grazian}, A., {Fontanot}, F., {et~al.} 2021, \apj, 912, 111

\bibitem[{{Buttry} {et~al.}(2022){Buttry}, {Pace}, {Koposov}, {Walker},
  {Caldwell}, {Kirby}, {Martin}, {Mateo}, {Olszewski}, {Starkenburg},
  {Badenes}, \& {Daher}}]{buttry22}
{Buttry}, R., {Pace}, A.~B., {Koposov}, S.~E., {et~al.} 2022, \mnras, 514, 1706

\bibitem[{{Caldwell} {et~al.}(2017){Caldwell}, {Walker}, {Mateo}, {Olszewski},
  {Koposov}, {Belokurov}, {Torrealba}, {Geringer-Sameth}, \&
  {Johnson}}]{caldwell17}
{Caldwell}, N., {Walker}, M.~G., {Mateo}, M., {et~al.} 2017, \apj, 839, 20

\bibitem[{{Cargile} {et~al.}(2020){Cargile}, {Conroy}, {Johnson}, {Ting},
  {Bonaca}, {Dotter}, \& {Speagle}}]{cargile20}
{Cargile}, P.~A., {Conroy}, C., {Johnson}, B.~D., {et~al.} 2020, \apj, 900, 28

\bibitem[{{Castelli} \& {Kurucz}(2004)}]{castelli04}
{Castelli}, F., \& {Kurucz}, R.~L. 2004, ArXiv e-prints, arXiv:astro-ph/0405087

\bibitem[{{Cohen} \& {Huang}(2009)}]{cohen09}
{Cohen}, J.~G., \& {Huang}, W. 2009, \apj, 701, 1053

\bibitem[{{Cohen} \& {Huang}(2010)}]{cohen10}
---. 2010, \apj, 719, 931

\bibitem[{{Conroy} {et~al.}(2019){Conroy}, {Bonaca}, {Cargile}, {Johnson},
  {Caldwell}, {Naidu}, {Zaritsky}, {Fabricant}, {Moran}, {Rhee},
  {Szentgyorgyi}, {Berlind}, {Calkins}, {Kattner}, \& {Ly}}]{conroy19}
{Conroy}, C., {Bonaca}, A., {Cargile}, P., {et~al.} 2019, \apj, 883, 107

\bibitem[{Craig {et~al.}(2017)Craig, Crawford, Seifert, Robitaille, Sip{\H
  o}cz, Walawender, Vin{\'{\i}}cius, Ninan, Droettboom, Youn, Tollerud, Bray,
  Walker, Janga, Stotts, G{\"u}nther, Rol, Bach, Bradley, Deil, Price-Whelan,
  Barbary, Horton, Schoenell, Heidt, Gasdia, Nelson, \& Streicher}]{ccdproc}
Craig, M., Crawford, S., Seifert, M., {et~al.} 2017, astropy/ccdproc:
  v1.3.0.post1, doi:10.5281/zenodo.1069648

\bibitem[{{Dotter}(2016)}]{dotter16}
{Dotter}, A. 2016, \apjs, 222, 8

\bibitem[{Dubernet {et~al.}(2016)Dubernet, Antony, Ba, Babikov, Bartschat,
  Boudon, Braams, Chung, Daniel, Delahaye, Zanna, de~Urquijo, Dimitrijević,
  Domaracka, Doronin, Drouin, Endres, Fazliev, Gagarin, Gordon, Gratier,
  Heiter, Hill, Jevremović, Joblin, Kasprzak, Krishnakumar, Leto, Loboda,
  Louge, Maclot, Marinković, Markwick, Marquart, Mason, Mason, Mendoza,
  Mihajlov, Millar, Moreau, Mulas, Pakhomov, Palmeri, Pancheshnyi, Perevalov,
  Piskunov, Postler, Quinet, Quintas-Sánchez, Ralchenko, Rhee, Rixon, Rothman,
  Roueff, Ryabchikova, Sahal-Bréchot, Scheier, Schlemmer, Schmitt, Stempels,
  Tashkun, Tennyson, Tyuterev, Vujčić, Wakelam, Walton, Zatsarinny, Zeippen,
  \& Zwölf}]{dubernet16}
Dubernet, M.~L., Antony, B.~K., Ba, Y.~A., {et~al.} 2016, Journal of Physics B:
  Atomic, Molecular and Optical Physics, 49, 074003

\bibitem[{{Feroz} \& {Hobson}(2008)}]{feroz08}
{Feroz}, F., \& {Hobson}, M.~P. 2008, \mnras, 384, 449

\bibitem[{{Feroz} {et~al.}(2009){Feroz}, {Hobson}, \& {Bridges}}]{feroz09}
{Feroz}, F., {Hobson}, M.~P., \& {Bridges}, M. 2009, \mnras, 398, 1601

\bibitem[{{Flewelling} {et~al.}(2020){Flewelling}, {Magnier}, {Chambers},
  {Heasley}, {Holmberg}, {Huber}, {Sweeney}, {Waters}, {Calamida}, {Casertano},
  {Chen}, {Farrow}, {Hasinger}, {Henderson}, {Long}, {Metcalfe}, {Narayan},
  {Nieto-Santisteban}, {Norberg}, {Rest}, {Saglia}, {Szalay}, {Thakar},
  {Tonry}, {Valenti}, {Werner}, {White}, {Denneau}, {Draper}, {Hodapp},
  {Jedicke}, {Kaiser}, {Kudritzki}, {Price}, {Wainscoat}, {Chastel}, {McLean},
  {Postman}, \& {Shiao}}]{ps1}
{Flewelling}, H.~A., {Magnier}, E.~A., {Chambers}, K.~C., {et~al.} 2020, \apjs,
  251, 7

\bibitem[{{Frebel} {et~al.}(2010){Frebel}, {Simon}, {Geha}, \&
  {Willman}}]{frebel10b}
{Frebel}, A., {Simon}, J.~D., {Geha}, M., \& {Willman}, B. 2010, \apj, 708, 560

\bibitem[{{Fulbright} {et~al.}(2004){Fulbright}, {Rich}, \&
  {Castro}}]{fulbright04}
{Fulbright}, J.~P., {Rich}, R.~M., \& {Castro}, S. 2004, \apj, 612, 447

\bibitem[{{Gaia Collaboration} {et~al.}(2016){Gaia Collaboration}, {Prusti},
  {de Bruijne}, {Brown}, {Vallenari}, {Babusiaux}, {Bailer-Jones}, {Bastian},
  {Biermann}, {Evans}, {Eyer}, {Jansen}, {Jordi}, {Klioner}, {Lammers},
  {Lindegren}, {Luri}, {Mignard}, {Milligan}, {Panem}, {Poinsignon},
  {Pourbaix}, {Randich}, {Sarri}, {Sartoretti}, {Siddiqui}, {Soubiran},
  {Valette}, {van Leeuwen}, {Walton}, {Aerts}, {Arenou}, {Cropper}, {Drimmel},
  {H{\o}g}, {Katz}, {Lattanzi}, {O'Mullane}, {Grebel}, {Holland}, {Huc},
  {Passot}, {Bramante}, {Cacciari}, {Casta{\~n}eda}, {Chaoul}, {Cheek}, {De
  Angeli}, {Fabricius}, {Guerra}, {Hern{\'a}ndez}, {Jean-Antoine-Piccolo},
  {Masana}, {Messineo}, {Mowlavi}, {Nienartowicz}, {Ord{\'o}{\~n}ez-Blanco},
  {Panuzzo}, {Portell}, {Richards}, {Riello}, {Seabroke}, {Tanga},
  {Th{\'e}venin}, {Torra}, {Els}, {Gracia-Abril}, {Comoretto},
  {Garcia-Reinaldos}, {Lock}, {Mercier}, {Altmann}, {Andrae}, {Astraatmadja},
  {Bellas-Velidis}, {Benson}, {Berthier}, {Blomme}, {Busso}, {Carry},
  {Cellino}, {Clementini}, {Cowell}, {Creevey}, {Cuypers}, {Davidson}, {De
  Ridder}, {de Torres}, {Delchambre}, {Dell'Oro}, {Ducourant}, {Fr{\'e}mat},
  {Garc{\'\i}a-Torres}, {Gosset}, {Halbwachs}, {Hambly}, {Harrison}, {Hauser},
  {Hestroffer}, {Hodgkin}, {Huckle}, {Hutton}, {Jasniewicz}, {Jordan},
  {Kontizas}, {Korn}, {Lanzafame}, {Manteiga}, {Moitinho}, {Muinonen},
  {Osinde}, {Pancino}, {Pauwels}, {Petit}, {Recio-Blanco}, {Robin}, {Sarro},
  {Siopis}, {Smith}, {Smith}, {Sozzetti}, {Thuillot}, {van Reeven}, {Viala},
  {Abbas}, {Abreu Aramburu}, {Accart}, {Aguado}, {Allan}, {Allasia},
  {Altavilla}, {{\'A}lvarez}, {Alves}, {Anderson}, {Andrei}, {Anglada Varela},
  {Antiche}, {Antoja}, {Ant{\'o}n}, {Arcay}, {Atzei}, {Ayache}, {Bach},
  {Baker}, {Balaguer-N{\'u}{\~n}ez}, {Barache}, {Barata}, {Barbier}, {Barblan},
  {Baroni}, {Barrado y Navascu{\'e}s}, {Barros}, {Barstow}, {Becciani},
  {Bellazzini}, {Bellei}, {Bello Garc{\'\i}a}, {Belokurov}, {Bendjoya},
  {Berihuete}, {Bianchi}, {Bienaym{\'e}}, {Billebaud}, {Blagorodnova},
  {Blanco-Cuaresma}, {Boch}, {Bombrun}, {Borrachero}, {Bouquillon}, {Bourda},
  {Bouy}, {Bragaglia}, {Breddels}, {Brouillet}, {Br{\"u}semeister},
  {Bucciarelli}, {Budnik}, {Burgess}, {Burgon}, {Burlacu}, {Busonero}, {Buzzi},
  {Caffau}, {Cambras}, {Campbell}, {Cancelliere}, {Cantat-Gaudin}, {Carlucci},
  {Carrasco}, {Castellani}, {Charlot}, {Charnas}, {Charvet}, {Chassat},
  {Chiavassa}, {Clotet}, {Cocozza}, {Collins}, {Collins}, {Costigan}, {Crifo},
  {Cross}, {Crosta}, {Crowley}, {Dafonte}, {Damerdji}, {Dapergolas}, {David},
  {David}, {De Cat}, {de Felice}, {de Laverny}, {De Luise}, {De March}, {de
  Martino}, {de Souza}, {Debosscher}, {del Pozo}, {Delbo}, {Delgado},
  {Delgado}, {di Marco}, {Di Matteo}, {Diakite}, {Distefano}, {Dolding}, {Dos
  Anjos}, {Drazinos}, {Dur{\'a}n}, {Dzigan}, {Ecale}, {Edvardsson}, {Enke},
  {Erdmann}, {Escolar}, {Espina}, {Evans}, {Eynard Bontemps}, {Fabre},
  {Fabrizio}, {Faigler}, {Falc{\~a}o}, {Farr{\`a}s Casas}, {Faye}, {Federici},
  {Fedorets}, {Fern{\'a}ndez-Hern{\'a}ndez}, {Fernique}, {Fienga}, {Figueras},
  {Filippi}, {Findeisen}, {Fonti}, {Fouesneau}, {Fraile}, {Fraser}, {Fuchs},
  {Furnell}, {Gai}, {Galleti}, {Galluccio}, {Garabato}, {Garc{\'\i}a-Sedano},
  {Gar{\'e}}, {Garofalo}, {Garralda}, {Gavras}, {Gerssen}, {Geyer}, {Gilmore},
  {Girona}, {Giuffrida}, {Gomes}, {Gonz{\'a}lez-Marcos},
  {Gonz{\'a}lez-N{\'u}{\~n}ez}, {Gonz{\'a}lez-Vidal}, {Granvik}, {Guerrier},
  {Guillout}, {Guiraud}, {G{\'u}rpide}, {Guti{\'e}rrez-S{\'a}nchez}, {Guy},
  {Haigron}, {Hatzidimitriou}, {Haywood}, {Heiter}, {Helmi}, {Hobbs},
  {Hofmann}, {Holl}, {Holland}, {Hunt}, {Hypki}, {Icardi}, {Irwin}, {Jevardat
  de Fombelle}, {Jofr{\'e}}, {Jonker}, {Jorissen}, {Julbe}, {Karampelas},
  {Kochoska}, {Kohley}, {Kolenberg}, {Kontizas}, {Koposov}, {Kordopatis},
  {Koubsky}, {Kowalczyk}, {Krone-Martins}, {Kudryashova}, {Kull}, {Bachchan},
  {Lacoste-Seris}, {Lanza}, {Lavigne}, {Le Poncin-Lafitte}, {Lebreton},
  {Lebzelter}, {Leccia}, {Leclerc}, {Lecoeur-Taibi}, {Lemaitre}, {Lenhardt},
  {Leroux}, {Liao}, {Licata}, {Lindstr{\o}m}, {Lister}, {Livanou}, {Lobel},
  {L{\"o}ffler}, {L{\'o}pez}, {Lopez-Lozano}, {Lorenz}, {Loureiro},
  {MacDonald}, {Magalh{\~a}es Fernandes}, {Managau}, {Mann}, {Mantelet},
  {Marchal}, {Marchant}, {Marconi}, {Marie}, {Marinoni}, {Marrese},
  {Marschalk{\'o}}, {Marshall}, {Mart{\'\i}n-Fleitas}, {Martino}, {Mary},
  {Matijevi{\v{c}}}, {Mazeh}, {McMillan}, {Messina}, {Mestre}, {Michalik},
  {Millar}, {Miranda}, {Molina}, {Molinaro}, {Molinaro}, {Moln{\'a}r},
  {Moniez}, {Montegriffo}, {Monteiro}, {Mor}, {Mora}, {Morbidelli}, {Morel},
  {Morgenthaler}, {Morley}, {Morris}, {Mulone}, {Muraveva}, {Musella},
  {Narbonne}, {Nelemans}, {Nicastro}, {Noval}, {Ord{\'e}novic},
  {Ordieres-Mer{\'e}}, {Osborne}, {Pagani}, {Pagano}, {Pailler}, {Palacin},
  {Palaversa}, {Parsons}, {Paulsen}, {Pecoraro}, {Pedrosa}, {Pentik{\"a}inen},
  {Pereira}, {Pichon}, {Piersimoni}, {Pineau}, {Plachy}, {Plum}, {Poujoulet},
  {Pr{\v{s}}a}, {Pulone}, {Ragaini}, {Rago}, {Rambaux}, {Ramos-Lerate},
  {Ranalli}, {Rauw}, {Read}, {Regibo}, {Renk}, {Reyl{\'e}}, {Ribeiro},
  {Rimoldini}, {Ripepi}, {Riva}, {Rixon}, {Roelens}, {Romero-G{\'o}mez},
  {Rowell}, {Royer}, {Rudolph}, {Ruiz-Dern}, {Sadowski}, {Sagrist{\`a}
  Sell{\'e}s}, {Sahlmann}, {Salgado}, {Salguero}, {Sarasso}, {Savietto},
  {Schnorhk}, {Schultheis}, {Sciacca}, {Segol}, {Segovia}, {Segransan},
  {Serpell}, {Shih}, {Smareglia}, {Smart}, {Smith}, {Solano}, {Solitro},
  {Sordo}, {Soria Nieto}, {Souchay}, {Spagna}, {Spoto}, {Stampa}, {Steele},
  {Steidelm{\"u}ller}, {Stephenson}, {Stoev}, {Suess}, {S{\"u}veges}, {Surdej},
  {Szabados}, {Szegedi-Elek}, {Tapiador}, {Taris}, {Tauran}, {Taylor},
  {Teixeira}, {Terrett}, {Tingley}, {Trager}, {Turon}, {Ulla}, {Utrilla},
  {Valentini}, {van Elteren}, {Van Hemelryck}, {van Leeuwen}, {Varadi},
  {Vecchiato}, {Veljanoski}, {Via}, {Vicente}, {Vogt}, {Voss}, {Votruba},
  {Voutsinas}, {Walmsley}, {Weiler}, {Weingrill}, {Werner}, {Wevers},
  {Whitehead}, {Wyrzykowski}, {Yoldas}, {{\v{Z}}erjal}, {Zucker}, {Zurbach},
  {Zwitter}, {Alecu}, {Allen}, {Allende Prieto}, {Amorim},
  {Anglada-Escud{\'e}}, {Arsenijevic}, {Azaz}, {Balm}, {Beck}, {Bernstein},
  {Bigot}, {Bijaoui}, {Blasco}, {Bonfigli}, {Bono}, {Boudreault}, {Bressan},
  {Brown}, {Brunet}, {Bunclark}, {Buonanno}, {Butkevich}, {Carret}, {Carrion},
  {Chemin}, {Ch{\'e}reau}, {Corcione}, {Darmigny}, {de Boer}, {de Teodoro}, {de
  Zeeuw}, {Delle Luche}, {Domingues}, {Dubath}, {Fodor}, {Fr{\'e}zouls},
  {Fries}, {Fustes}, {Fyfe}, {Gallardo}, {Gallegos}, {Gardiol}, {Gebran},
  {Gomboc}, {G{\'o}mez}, {Grux}, {Gueguen}, {Heyrovsky}, {Hoar}, {Iannicola},
  {Isasi Parache}, {Janotto}, {Joliet}, {Jonckheere}, {Keil}, {Kim},
  {Klagyivik}, {Klar}, {Knude}, {Kochukhov}, {Kolka}, {Kos}, {Kutka}, {Lainey},
  {LeBouquin}, {Liu}, {Loreggia}, {Makarov}, {Marseille}, {Martayan},
  {Martinez-Rubi}, {Massart}, {Meynadier}, {Mignot}, {Munari}, {Nguyen},
  {Nordlander}, {Ocvirk}, {O'Flaherty}, {Olias Sanz}, {Ortiz}, {Osorio},
  {Oszkiewicz}, {Ouzounis}, {Palmer}, {Park}, {Pasquato}, {Peltzer}, {Peralta},
  {P{\'e}turaud}, {Pieniluoma}, {Pigozzi}, {Poels}, {Prat}, {Prod'homme},
  {Raison}, {Rebordao}, {Risquez}, {Rocca-Volmerange}, {Rosen}, {Ruiz-Fuertes},
  {Russo}, {Sembay}, {Serraller Vizcaino}, {Short}, {Siebert}, {Silva},
  {Sinachopoulos}, {Slezak}, {Soffel}, {Sosnowska}, {Strai{\v{z}}ys}, {ter
  Linden}, {Terrell}, {Theil}, {Tiede}, {Troisi}, {Tsalmantza}, {Tur},
  {Vaccari}, {Vachier}, {Valles}, {Van Hamme}, {Veltz}, {Virtanen}, {Wallut},
  {Wichmann}, {Wilkinson}, {Ziaeepour}, \& {Zschocke}}]{gaia16}
{Gaia Collaboration}, {Prusti}, T., {de Bruijne}, J.~H.~J., {et~al.} 2016,
  \aap, 595, A1

\bibitem[{{Gaia Collaboration} {et~al.}(2018{\natexlab{a}}){Gaia
  Collaboration}, {Helmi}, {van Leeuwen}, {McMillan}, {Massari}, {Antoja},
  {Robin}, {Lindegren}, {Bastian}, {Arenou}, {Babusiaux}, {Biermann},
  {Breddels}, {Hobbs}, {Jordi}, {Pancino}, {Reyl{\'e}}, {Veljanoski}, {Brown},
  {Vallenari}, {Prusti}, {de Bruijne}, {Bailer-Jones}, {Evans}, {Eyer},
  {Jansen}, {Klioner}, {Lammers}, {Luri}, {Mignard}, {Panem}, {Pourbaix},
  {Randich}, {Sartoretti}, {Siddiqui}, {Soubiran}, {Walton}, {Cropper},
  {Drimmel}, {Katz}, {Lattanzi}, {Bakker}, {Cacciari}, {Casta{\~n}eda},
  {Chaoul}, {Cheek}, {De Angeli}, {Fabricius}, {Guerra}, {Holl}, {Masana},
  {Messineo}, {Mowlavi}, {Nienartowicz}, {Panuzzo}, {Portell}, {Riello},
  {Seabroke}, {Tanga}, {Th{\'e}venin}, {Gracia-Abril}, {Comoretto},
  {Garcia-Reinaldos}, {Teyssier}, {Altmann}, {Andrae}, {Audard},
  {Bellas-Velidis}, {Benson}, {Berthier}, {Blomme}, {Burgess}, {Busso},
  {Carry}, {Cellino}, {Clementini}, {Clotet}, {Creevey}, {Davidson}, {De
  Ridder}, {Delchambre}, {Dell'Oro}, {Ducourant},
  {Fern{\'a}ndez-Hern{\'a}ndez}, {Fouesneau}, {Fr{\'e}mat}, {Galluccio},
  {Garc{\'\i}a-Torres}, {Gonz{\'a}lez-N{\'u}{\~n}ez}, {Gonz{\'a}lez-Vidal},
  {Gosset}, {Guy}, {Halbwachs}, {Hambly}, {Harrison}, {Hern{\'a}ndez},
  {Hestroffer}, {Hodgkin}, {Hutton}, {Jasniewicz}, {Jean-Antoine-Piccolo},
  {Jordan}, {Korn}, {Krone-Martins}, {Lanzafame}, {Lebzelter}, {L{\"o}ffler},
  {Manteiga}, {Marrese}, {Mart{\'\i}n-Fleitas}, {Moitinho}, {Mora}, {Muinonen},
  {Osinde}, {Pauwels}, {Petit}, {Recio-Blanco}, {Richards}, {Rimoldini},
  {Sarro}, {Siopis}, {Smith}, {Sozzetti}, {S{\"u}veges}, {Torra}, {van Reeven},
  {Abbas}, {Abreu Aramburu}, {Accart}, {Aerts}, {Altavilla}, {{\'A}lvarez},
  {Alvarez}, {Alves}, {Anderson}, {Andrei}, {Anglada Varela}, {Antiche},
  {Arcay}, {Astraatmadja}, {Bach}, {Baker}, {Balaguer-N{\'u}{\~n}ez}, {Balm},
  {Barache}, {Barata}, {Barbato}, {Barblan}, {Barklem}, {Barrado}, {Barros},
  {Barstow}, {Bartholom{\'e} Mu{\~n}oz}, {Bassilana}, {Becciani}, {Bellazzini},
  {Berihuete}, {Bertone}, {Bianchi}, {Bienaym{\'e}}, {Blanco-Cuaresma}, {Boch},
  {Boeche}, {Bombrun}, {Borrachero}, {Bossini}, {Bouquillon}, {Bourda},
  {Bragaglia}, {Bramante}, {Bressan}, {Brouillet}, {Br{\"u}semeister},
  {Brugaletta}, {Bucciarelli}, {Burlacu}, {Busonero}, {Butkevich}, {Buzzi},
  {Caffau}, {Cancelliere}, {Cannizzaro}, {Cantat-Gaudin}, {Carballo},
  {Carlucci}, {Carrasco}, {Casamiquela}, {Castellani}, {Castro-Ginard},
  {Charlot}, {Chemin}, {Chiavassa}, {Cocozza}, {Costigan}, {Cowell}, {Crifo},
  {Crosta}, {Crowley}, {Cuypers}, {Dafonte}, {Damerdji}, {Dapergolas}, {David},
  {David}, {de Laverny}, {De Luise}, {De March}, {de Martino}, {de Souza}, {de
  Torres}, {Debosscher}, {del Pozo}, {Delbo}, {Delgado}, {Delgado}, {Di
  Matteo}, {Diakite}, {Diener}, {Distefano}, {Dolding}, {Drazinos},
  {Dur{\'a}n}, {Edvardsson}, {Enke}, {Eriksson}, {Esquej}, {Eynard Bontemps},
  {Fabre}, {Fabrizio}, {Faigler}, {Falc{\~a}o}, {Farr{\`a}s Casas}, {Federici},
  {Fedorets}, {Fernique}, {Figueras}, {Filippi}, {Findeisen}, {Fonti},
  {Fraile}, {Fraser}, {Fr{\'e}zouls}, {Gai}, {Galleti}, {Garabato},
  {Garc{\'\i}a-Sedano}, {Garofalo}, {Garralda}, {Gavel}, {Gavras}, {Gerssen},
  {Geyer}, {Giacobbe}, {Gilmore}, {Girona}, {Giuffrida}, {Glass}, {Gomes},
  {Granvik}, {Gueguen}, {Guerrier}, {Guiraud}, {Guti{\'e}rrez-S{\'a}nchez},
  {Hofmann}, {Holland}, {Huckle}, {Hypki}, {Icardi}, {Jan{\ss}en}, {Jevardat de
  Fombelle}, {Jonker}, {Juh{\'a}sz}, {Julbe}, {Karampelas}, {Kewley}, {Klar},
  {Kochoska}, {Kohley}, {Kolenberg}, {Kontizas}, {Kontizas}, {Koposov},
  {Kordopatis}, {Kostrzewa-Rutkowska}, {Koubsky}, {Lambert}, {Lanza}, {Lasne},
  {Lavigne}, {Le Fustec}, {Le Poncin-Lafitte}, {Lebreton}, {Leccia}, {Leclerc},
  {Lecoeur-Taibi}, {Lenhardt}, {Leroux}, {Liao}, {Licata}, {Lindstr{\o}m},
  {Lister}, {Livanou}, {Lobel}, {L{\'o}pez}, {Managau}, {Mann}, {Mantelet},
  {Marchal}, {Marchant}, {Marconi}, {Marinoni}, {Marschalk{\'o}}, {Marshall},
  {Martino}, {Marton}, {Mary}, {Matijevi{\v{c}}}, {Mazeh}, {Messina},
  {Michalik}, {Millar}, {Molina}, {Molinaro}, {Moln{\'a}r}, {Montegriffo},
  {Mor}, {Morbidelli}, {Morel}, {Morris}, {Mulone}, {Muraveva}, {Musella},
  {Nelemans}, {Nicastro}, {Noval}, {O'Mullane}, {Ord{\'e}novic},
  {Ord{\'o}{\~n}ez-Blanco}, {Osborne}, {Pagani}, {Pagano}, {Pailler},
  {Palacin}, {Palaversa}, {Panahi}, {Pawlak}, {Piersimoni}, {Pineau}, {Plachy},
  {Plum}, {Poggio}, {Poujoulet}, {Pr{\v{s}}a}, {Pulone}, {Racero}, {Ragaini},
  {Rambaux}, {Ramos-Lerate}, {Regibo}, {Riclet}, {Ripepi}, {Riva}, {Rivard},
  {Rixon}, {Roegiers}, {Roelens}, {Romero-G{\'o}mez}, {Rowell}, {Royer},
  {Ruiz-Dern}, {Sadowski}, {Sagrist{\`a} Sell{\'e}s}, {Sahlmann}, {Salgado},
  {Salguero}, {Sanna}, {Santana-Ros}, {Sarasso}, {Savietto}, {Schultheis},
  {Sciacca}, {Segol}, {Segovia}, {S{\'e}gransan}, {Shih}, {Siltala}, {Silva},
  {Smart}, {Smith}, {Solano}, {Solitro}, {Sordo}, {Soria Nieto}, {Souchay},
  {Spagna}, {Spoto}, {Stampa}, {Steele}, {Steidelm{\"u}ller}, {Stephenson},
  {Stoev}, {Suess}, {Surdej}, {Szabados}, {Szegedi-Elek}, {Tapiador}, {Taris},
  {Tauran}, {Taylor}, {Teixeira}, {Terrett}, {Teyssandier}, {Thuillot},
  {Titarenko}, {Torra Clotet}, {Turon}, {Ulla}, {Utrilla}, {Uzzi}, {Vaillant},
  {Valentini}, {Valette}, {van Elteren}, {Van Hemelryck}, {van Leeuwen},
  {Vaschetto}, {Vecchiato}, {Viala}, {Vicente}, {Vogt}, {von Essen}, {Voss},
  {Votruba}, {Voutsinas}, {Walmsley}, {Weiler}, {Wertz}, {Wevems},
  {Wyrzykowski}, {Yoldas}, {{\v{Z}}erjal}, {Ziaeepour}, {Zorec}, {Zschocke},
  {Zucker}, {Zurbach}, \& {Zwitter}}]{helmi18}
{Gaia Collaboration}, {Helmi}, A., {van Leeuwen}, F., {et~al.}
  2018{\natexlab{a}}, \aap, 616, A12

\bibitem[{{Gaia Collaboration} {et~al.}(2018{\natexlab{b}}){Gaia
  Collaboration}, {Babusiaux}, {van Leeuwen}, {Barstow}, {Jordi}, {Vallenari},
  {Bossini}, {Bressan}, {Cantat-Gaudin}, {van Leeuwen}, {Brown}, {Prusti}, {de
  Bruijne}, {Bailer-Jones}, {Biermann}, {Evans}, {Eyer}, {Jansen}, {Klioner},
  {Lammers}, {Lindegren}, {Luri}, {Mignard}, {Panem}, {Pourbaix}, {Randich},
  {Sartoretti}, {Siddiqui}, {Soubiran}, {Walton}, {Arenou}, {Bastian},
  {Cropper}, {Drimmel}, {Katz}, {Lattanzi}, {Bakker}, {Cacciari},
  {Casta{\~n}eda}, {Chaoul}, {Cheek}, {De Angeli}, {Fabricius}, {Guerra},
  {Holl}, {Masana}, {Messineo}, {Mowlavi}, {Nienartowicz}, {Panuzzo},
  {Portell}, {Riello}, {Seabroke}, {Tanga}, {Th{\'e}venin}, {Gracia-Abril},
  {Comoretto}, {Garcia-Reinaldos}, {Teyssier}, {Altmann}, {Andrae}, {Audard},
  {Bellas-Velidis}, {Benson}, {Berthier}, {Blomme}, {Burgess}, {Busso},
  {Carry}, {Cellino}, {Clementini}, {Clotet}, {Creevey}, {Davidson}, {De
  Ridder}, {Delchambre}, {Dell'Oro}, {Ducourant},
  {Fern{\'a}ndez-Hern{\'a}ndez}, {Fouesneau}, {Fr{\'e}mat}, {Galluccio},
  {Garc{\'\i}a-Torres}, {Gonz{\'a}lez-N{\'u}{\~n}ez}, {Gonz{\'a}lez-Vidal},
  {Gosset}, {Guy}, {Halbwachs}, {Hambly}, {Harrison}, {Hern{\'a}ndez},
  {Hestroffer}, {Hodgkin}, {Hutton}, {Jasniewicz}, {Jean-Antoine-Piccolo},
  {Jordan}, {Korn}, {Krone-Martins}, {Lanzafame}, {Lebzelter}, {L{\"o}ffler},
  {Manteiga}, {Marrese}, {Mart{\'\i}n-Fleitas}, {Moitinho}, {Mora}, {Muinonen},
  {Osinde}, {Pancino}, {Pauwels}, {Petit}, {Recio-Blanco}, {Richards},
  {Rimoldini}, {Robin}, {Sarro}, {Siopis}, {Smith}, {Sozzetti}, {S{\"u}veges},
  {Torra}, {van Reeven}, {Abbas}, {Abreu Aramburu}, {Accart}, {Aerts},
  {Altavilla}, {{\'A}lvarez}, {Alvarez}, {Alves}, {Anderson}, {Andrei},
  {Anglada Varela}, {Antiche}, {Antoja}, {Arcay}, {Astraatmadja}, {Bach},
  {Baker}, {Balaguer-N{\'u}{\~n}ez}, {Balm}, {Barache}, {Barata}, {Barbato},
  {Barblan}, {Barklem}, {Barrado}, {Barros}, {Bartholom{\'e} Mu{\~n}oz},
  {Bassilana}, {Becciani}, {Bellazzini}, {Berihuete}, {Bertone}, {Bianchi},
  {Bienaym{\'e}}, {Blanco-Cuaresma}, {Boch}, {Boeche}, {Bombrun}, {Borrachero},
  {Bouquillon}, {Bourda}, {Bragaglia}, {Bramante}, {Breddels}, {Brouillet},
  {Br{\"u}semeister}, {Brugaletta}, {Bucciarelli}, {Burlacu}, {Busonero},
  {Butkevich}, {Buzzi}, {Caffau}, {Cancelliere}, {Cannizzaro}, {Carballo},
  {Carlucci}, {Carrasco}, {Casamiquela}, {Castellani}, {Castro-Ginard},
  {Charlot}, {Chemin}, {Chiavassa}, {Cocozza}, {Costigan}, {Cowell}, {Crifo},
  {Crosta}, {Crowley}, {Cuypers}, {Dafonte}, {Damerdji}, {Dapergolas}, {David},
  {David}, {de Laverny}, {De Luise}, {De March}, {de Martino}, {de Souza}, {de
  Torres}, {Debosscher}, {del Pozo}, {Delbo}, {Delgado}, {Delgado}, {Diakite},
  {Diener}, {Distefano}, {Dolding}, {Drazinos}, {Dur{\'a}n}, {Edvardsson},
  {Enke}, {Eriksson}, {Esquej}, {Eynard Bontemps}, {Fabre}, {Fabrizio},
  {Faigler}, {Falc{\~a}o}, {Farr{\`a}s Casas}, {Federici}, {Fedorets},
  {Fernique}, {Figueras}, {Filippi}, {Findeisen}, {Fonti}, {Fraile}, {Fraser},
  {Fr{\'e}zouls}, {Gai}, {Galleti}, {Garabato}, {Garc{\'\i}a-Sedano},
  {Garofalo}, {Garralda}, {Gavel}, {Gavras}, {Gerssen}, {Geyer}, {Giacobbe},
  {Gilmore}, {Girona}, {Giuffrida}, {Glass}, {Gomes}, {Granvik}, {Gueguen},
  {Guerrier}, {Guiraud}, {Guti{\'e}}, {Haigron}, {Hatzidimitriou}, {Hauser},
  {Haywood}, {Heiter}, {Helmi}, {Heu}, {Hilger}, {Hobbs}, {Hofmann}, {Holland},
  {Huckle}, {Hypki}, {Icardi}, {Jan{\ss}en}, {Jevardat de Fombelle}, {Jonker},
  {Juh{\'a}sz}, {Julbe}, {Karampelas}, {Kewley}, {Klar}, {Kochoska}, {Kohley},
  {Kolenberg}, {Kontizas}, {Kontizas}, {Koposov}, {Kordopatis},
  {Kostrzewa-Rutkowska}, {Koubsky}, {Lambert}, {Lanza}, {Lasne}, {Lavigne}, {Le
  Fustec}, {Le Poncin-Lafitte}, {Lebreton}, {Leccia}, {Leclerc},
  {Lecoeur-Taibi}, {Lenhardt}, {Leroux}, {Liao}, {Licata}, {Lindstr{\o}m},
  {Lister}, {Livanou}, {Lobel}, {L{\'o}pez}, {Managau}, {Mann}, {Mantelet},
  {Marchal}, {Marchant}, {Marconi}, {Marinoni}, {Marschalk{\'o}}, {Marshall},
  {Martino}, {Marton}, {Mary}, {Massari}, {Matijevi{\v{c}}}, {Mazeh},
  {McMillan}, {Messina}, {Michalik}, {Millar}, {Molina}, {Molinaro},
  {Moln{\'a}r}, {Montegriffo}, {Mor}, {Morbidelli}, {Morel}, {Morris},
  {Mulone}, {Muraveva}, {Musella}, {Nelemans}, {Nicastro}, {Noval},
  {O'Mullane}, {Ord{\'e}novic}, {Ord{\'o}{\~n}ez-Blanco}, {Osborne}, {Pagani},
  {Pagano}, {Pailler}, {Palacin}, {Palaversa}, {Panahi}, {Pawlak},
  {Piersimoni}, {Pineau}, {Plachy}, {Plum}, {Poggio}, {Poujoulet},
  {Pr{\v{s}}a}, {Pulone}, {Racero}, {Ragaini}, {Rambaux}, {Ramos-Lerate},
  {Regibo}, {Reyl{\'e}}, {Riclet}, {Ripepi}, {Riva}, {Rivard}, {Rixon},
  {Roegiers}, {Roelens}, {Romero-G{\'o}mez}, {Rowell}, {Royer}, {Ruiz-Dern},
  {Sadowski}, {Sagrist{\`a} Sell{\'e}s}, {Sahlmann}, {Salgado}, {Salguero},
  {Sanna}, {Santana-Ros}, {Sarasso}, {Savietto}, {Schultheis}, {Sciacca},
  {Segol}, {Segovia}, {S{\'e}gransan}, {Shih}, {Siltala}, {Silva}, {Smart},
  {Smith}, {Solano}, {Solitro}, {Sordo}, {Soria Nieto}, {Souchay}, {Spagna},
  {Spoto}, {Stampa}, {Steele}, {Steidelm{\"u}ller}, {Stephenson}, {Stoev},
  {Suess}, {Surdej}, {Szabados}, {Szegedi-Elek}, {Tapiador}, {Taris}, {Tauran},
  {Taylor}, {Teixeira}, {Terrett}, {Teyssandier}, {Thuillot}, {Titarenko},
  {Torra Clotet}, {Turon}, {Ulla}, {Utrilla}, {Uzzi}, {Vaillant}, {Valentini},
  {Valette}, {van Elteren}, {Van Hemelryck}, {Vaschetto}, {Vecchiato},
  {Veljanoski}, {Viala}, {Vicente}, {Vogt}, {von Essen}, {Voss}, {Votruba},
  {Voutsinas}, {Walmsley}, {Weiler}, {Wertz}, {Wevers}, {Wyrzykowski},
  {Yoldas}, {{\v{Z}}erjal}, {Ziaeepour}, {Zorec}, {Zschocke}, {Zucker},
  {Zurbach}, \& {Zwitter}}]{babusiaux18}
{Gaia Collaboration}, {Babusiaux}, C., {van Leeuwen}, F., {et~al.}
  2018{\natexlab{b}}, \aap, 616, A10

\bibitem[{{Gaia Collaboration} {et~al.}(2018{\natexlab{c}}){Gaia
  Collaboration}, {Brown}, {Vallenari}, {Prusti}, {de Bruijne}, {Babusiaux},
  {Bailer-Jones}, {Biermann}, {Evans}, {Eyer}, {Jansen}, {Jordi}, {Klioner},
  {Lammers}, {Lindegren}, {Luri}, {Mignard}, {Panem}, {Pourbaix}, {Randich},
  {Sartoretti}, {Siddiqui}, {Soubiran}, {van Leeuwen}, {Walton}, {Arenou},
  {Bastian}, {Cropper}, {Drimmel}, {Katz}, {Lattanzi}, {Bakker}, {Cacciari},
  {Casta{\~n}eda}, {Chaoul}, {Cheek}, {De Angeli}, {Fabricius}, {Guerra},
  {Holl}, {Masana}, {Messineo}, {Mowlavi}, {Nienartowicz}, {Panuzzo},
  {Portell}, {Riello}, {Seabroke}, {Tanga}, {Th{\'e}venin}, {Gracia-Abril},
  {Comoretto}, {Garcia-Reinaldos}, {Teyssier}, {Altmann}, {Andrae}, {Audard},
  {Bellas-Velidis}, {Benson}, {Berthier}, {Blomme}, {Burgess}, {Busso},
  {Carry}, {Cellino}, {Clementini}, {Clotet}, {Creevey}, {Davidson}, {De
  Ridder}, {Delchambre}, {Dell'Oro}, {Ducourant},
  {Fern{\'a}ndez-Hern{\'a}ndez}, {Fouesneau}, {Fr{\'e}mat}, {Galluccio},
  {Garc{\'\i}a-Torres}, {Gonz{\'a}lez-N{\'u}{\~n}ez}, {Gonz{\'a}lez-Vidal},
  {Gosset}, {Guy}, {Halbwachs}, {Hambly}, {Harrison}, {Hern{\'a}ndez},
  {Hestroffer}, {Hodgkin}, {Hutton}, {Jasniewicz}, {Jean-Antoine-Piccolo},
  {Jordan}, {Korn}, {Krone-Martins}, {Lanzafame}, {Lebzelter}, {L{\"o}ffler},
  {Manteiga}, {Marrese}, {Mart{\'\i}n-Fleitas}, {Moitinho}, {Mora}, {Muinonen},
  {Osinde}, {Pancino}, {Pauwels}, {Petit}, {Recio-Blanco}, {Richards},
  {Rimoldini}, {Robin}, {Sarro}, {Siopis}, {Smith}, {Sozzetti}, {S{\"u}veges},
  {Torra}, {van Reeven}, {Abbas}, {Abreu Aramburu}, {Accart}, {Aerts},
  {Altavilla}, {{\'A}lvarez}, {Alvarez}, {Alves}, {Anderson}, {Andrei},
  {Anglada Varela}, {Antiche}, {Antoja}, {Arcay}, {Astraatmadja}, {Bach},
  {Baker}, {Balaguer-N{\'u}{\~n}ez}, {Balm}, {Barache}, {Barata}, {Barbato},
  {Barblan}, {Barklem}, {Barrado}, {Barros}, {Barstow}, {Bartholom{\'e}
  Mu{\~n}oz}, {Bassilana}, {Becciani}, {Bellazzini}, {Berihuete}, {Bertone},
  {Bianchi}, {Bienaym{\'e}}, {Blanco-Cuaresma}, {Boch}, {Boeche}, {Bombrun},
  {Borrachero}, {Bossini}, {Bouquillon}, {Bourda}, {Bragaglia}, {Bramante},
  {Breddels}, {Bressan}, {Brouillet}, {Br{\"u}semeister}, {Brugaletta},
  {Bucciarelli}, {Burlacu}, {Busonero}, {Butkevich}, {Buzzi}, {Caffau},
  {Cancelliere}, {Cannizzaro}, {Cantat-Gaudin}, {Carballo}, {Carlucci},
  {Carrasco}, {Casamiquela}, {Castellani}, {Castro-Ginard}, {Charlot},
  {Chemin}, {Chiavassa}, {Cocozza}, {Costigan}, {Cowell}, {Crifo}, {Crosta},
  {Crowley}, {Cuypers}, {Dafonte}, {Damerdji}, {Dapergolas}, {David}, {David},
  {de Laverny}, {De Luise}, {De March}, {de Martino}, {de Souza}, {de Torres},
  {Debosscher}, {del Pozo}, {Delbo}, {Delgado}, {Delgado}, {Di Matteo},
  {Diakite}, {Diener}, {Distefano}, {Dolding}, {Drazinos}, {Dur{\'a}n},
  {Edvardsson}, {Enke}, {Eriksson}, {Esquej}, {Eynard Bontemps}, {Fabre},
  {Fabrizio}, {Faigler}, {Falc{\~a}o}, {Farr{\`a}s Casas}, {Federici},
  {Fedorets}, {Fernique}, {Figueras}, {Filippi}, {Findeisen}, {Fonti},
  {Fraile}, {Fraser}, {Fr{\'e}zouls}, {Gai}, {Galleti}, {Garabato},
  {Garc{\'\i}a-Sedano}, {Garofalo}, {Garralda}, {Gavel}, {Gavras}, {Gerssen},
  {Geyer}, {Giacobbe}, {Gilmore}, {Girona}, {Giuffrida}, {Glass}, {Gomes},
  {Granvik}, {Gueguen}, {Guerrier}, {Guiraud}, {Guti{\'e}rrez-S{\'a}nchez},
  {Haigron}, {Hatzidimitriou}, {Hauser}, {Haywood}, {Heiter}, {Helmi}, {Heu},
  {Hilger}, {Hobbs}, {Hofmann}, {Holland}, {Huckle}, {Hypki}, {Icardi},
  {Jan{\ss}en}, {Jevardat de Fombelle}, {Jonker}, {Juh{\'a}sz}, {Julbe},
  {Karampelas}, {Kewley}, {Klar}, {Kochoska}, {Kohley}, {Kolenberg},
  {Kontizas}, {Kontizas}, {Koposov}, {Kordopatis}, {Kostrzewa-Rutkowska},
  {Koubsky}, {Lambert}, {Lanza}, {Lasne}, {Lavigne}, {Le Fustec}, {Le
  Poncin-Lafitte}, {Lebreton}, {Leccia}, {Leclerc}, {Lecoeur-Taibi},
  {Lenhardt}, {Leroux}, {Liao}, {Licata}, {Lindstr{\o}m}, {Lister}, {Livanou},
  {Lobel}, {L{\'o}pez}, {Managau}, {Mann}, {Mantelet}, {Marchal}, {Marchant},
  {Marconi}, {Marinoni}, {Marschalk{\'o}}, {Marshall}, {Martino}, {Marton},
  {Mary}, {Massari}, {Matijevi{\v{c}}}, {Mazeh}, {McMillan}, {Messina},
  {Michalik}, {Millar}, {Molina}, {Molinaro}, {Moln{\'a}r}, {Montegriffo},
  {Mor}, {Morbidelli}, {Morel}, {Morris}, {Mulone}, {Muraveva}, {Musella},
  {Nelemans}, {Nicastro}, {Noval}, {O'Mullane}, {Ord{\'e}novic},
  {Ord{\'o}{\~n}ez-Blanco}, {Osborne}, {Pagani}, {Pagano}, {Pailler},
  {Palacin}, {Palaversa}, {Panahi}, {Pawlak}, {Piersimoni}, {Pineau}, {Plachy},
  {Plum}, {Poggio}, {Poujoulet}, {Pr{\v{s}}a}, {Pulone}, {Racero}, {Ragaini},
  {Rambaux}, {Ramos-Lerate}, {Regibo}, {Reyl{\'e}}, {Riclet}, {Ripepi}, {Riva},
  {Rivard}, {Rixon}, {Roegiers}, {Roelens}, {Romero-G{\'o}mez}, {Rowell},
  {Royer}, {Ruiz-Dern}, {Sadowski}, {Sagrist{\`a} Sell{\'e}s}, {Sahlmann},
  {Salgado}, {Salguero}, {Sanna}, {Santana-Ros}, {Sarasso}, {Savietto},
  {Schultheis}, {Sciacca}, {Segol}, {Segovia}, {S{\'e}gransan}, {Shih},
  {Siltala}, {Silva}, {Smart}, {Smith}, {Solano}, {Solitro}, {Sordo}, {Soria
  Nieto}, {Souchay}, {Spagna}, {Spoto}, {Stampa}, {Steele},
  {Steidelm{\"u}ller}, {Stephenson}, {Stoev}, {Suess}, {Surdej}, {Szabados},
  {Szegedi-Elek}, {Tapiador}, {Taris}, {Tauran}, {Taylor}, {Teixeira},
  {Terrett}, {Teyssandier}, {Thuillot}, {Titarenko}, {Torra Clotet}, {Turon},
  {Ulla}, {Utrilla}, {Uzzi}, {Vaillant}, {Valentini}, {Valette}, {van Elteren},
  {Van Hemelryck}, {van Leeuwen}, {Vaschetto}, {Vecchiato}, {Veljanoski},
  {Viala}, {Vicente}, {Vogt}, {von Essen}, {Voss}, {Votruba}, {Voutsinas},
  {Walmsley}, {Weiler}, {Wertz}, {Wevers}, {Wyrzykowski}, {Yoldas},
  {{\v{Z}}erjal}, {Ziaeepour}, {Zorec}, {Zschocke}, {Zucker}, {Zurbach}, \&
  {Zwitter}}]{gaiadr2}
{Gaia Collaboration}, {Brown}, A.~G.~A., {Vallenari}, A., {et~al.}
  2018{\natexlab{c}}, \aap, 616, A1

\bibitem[{{Gaia Collaboration} {et~al.}(2022){Gaia Collaboration}, {Vallenari},
  {Brown}, {Prusti}, {de Bruijne}, {Arenou}, {Babusiaux}, {Biermann},
  {Creevey}, {Ducourant}, {Evans}, {Eyer}, {Guerra}, {Hutton}, {Jordi},
  {Klioner}, {Lammers}, {Lindegren}, {Luri}, {Mignard}, {Panem}, {Pourbaix},
  {Randich}, {Sartoretti}, {Soubiran}, {Tanga}, {Walton}, {Bailer-Jones},
  {Bastian}, {Drimmel}, {Jansen}, {Katz}, {Lattanzi}, {van Leeuwen}, {Bakker},
  {Cacciari}, {Casta{\~n}eda}, {De Angeli}, {Fabricius}, {Fouesneau},
  {Fr{\'e}mat}, {Galluccio}, {Guerrier}, {Heiter}, {Masana}, {Messineo},
  {Mowlavi}, {Nicolas}, {Nienartowicz}, {Pailler}, {Panuzzo}, {Riclet}, {Roux},
  {Seabroke}, {Sordo{\o}rcit}, {Th{\'e}venin}, {Gracia-Abril}, {Portell},
  {Teyssier}, {Altmann}, {Andrae}, {Audard}, {Bellas-Velidis}, {Benson},
  {Berthier}, {Blomme}, {Burgess}, {Busonero}, {Busso}, {C{\'a}novas}, {Carry},
  {Cellino}, {Cheek}, {Clementini}, {Damerdji}, {Davidson}, {de Teodoro},
  {Nu{\~n}ez Campos}, {Delchambre}, {Dell'Oro}, {Esquej},
  {Fern{\'a}ndez-Hern{\'a}ndez}, {Fraile}, {Garabato}, {Garc{\'\i}a-Lario},
  {Gosset}, {Haigron}, {Halbwachs}, {Hambly}, {Harrison}, {Hern{\'a}ndez},
  {Hestroffer}, {Hodgkin}, {Holl}, {Jan{\ss}en}, {Jevardat de Fombelle},
  {Jordan}, {Krone-Martins}, {Lanzafame}, {L{\"o}ffler}, {Marchal}, {Marrese},
  {Moitinho}, {Muinonen}, {Osborne}, {Pancino}, {Pauwels}, {Recio-Blanco},
  {Reyl{\'e}}, {Riello}, {Rimoldini}, {Roegiers}, {Rybizki}, {Sarro}, {Siopis},
  {Smith}, {Sozzetti}, {Utrilla}, {van Leeuwen}, {Abbas}, {{\'A}brah{\'a}m},
  {Abreu Aramburu}, {Aerts}, {Aguado}, {Ajaj}, {Aldea-Montero}, {Altavilla},
  {{\'A}lvarez}, {Alves}, {Anders}, {Anderson}, {Anglada Varela}, {Antoja},
  {Baines}, {Baker}, {Balaguer-N{\'u}{\~n}ez}, {Balbinot}, {Balog}, {Barache},
  {Barbato}, {Barros}, {Barstow}, {Bartolom{\'e}}, {Bassilana}, {Bauchet},
  {Becciani}, {Bellazzini}, {Berihuete}, {Bernet}, {Bertone}, {Bianchi},
  {Binnenfeld}, {Blanco-Cuaresma}, {Blazere}, {Boch}, {Bombrun}, {Bossini},
  {Bouquillon}, {Bragaglia}, {Bramante}, {Breedt}, {Bressan}, {Brouillet},
  {Brugaletta}, {Bucciarelli}, {Burlacu}, {Butkevich}, {Buzzi}, {Caffau},
  {Cancelliere}, {Cantat-Gaudin}, {Carballo}, {Carlucci}, {Carnerero},
  {Carrasco}, {Casamiquela}, {Castellani}, {Castro-Ginard}, {Chaoul},
  {Charlot}, {Chemin}, {Chiaramida}, {Chiavassa}, {Chornay}, {Comoretto},
  {Contursi}, {Cooper}, {Cornez}, {Cowell}, {Crifo}, {Cropper}, {Crosta},
  {Crowley}, {Dafonte}, {Dapergolas}, {David}, {David}, {de Laverny}, {De
  Luise}, {De March}, {De Ridder}, {de Souza}, {de Torres}, {del Peloso}, {del
  Pozo}, {Delbo}, {Delgado}, {Delisle}, {Demouchy}, {Dharmawardena}, {Di
  Matteo}, {Diakite}, {Diener}, {Distefano}, {Dolding}, {Edvardsson}, {Enke},
  {Fabre}, {Fabrizio}, {Faigler}, {Fedorets}, {Fernique}, {Fienga}, {Figueras},
  {Fournier}, {Fouron}, {Fragkoudi}, {Gai}, {Garcia-Gutierrez},
  {Garcia-Reinaldos}, {Garc{\'\i}a-Torres}, {Garofalo}, {Gavel}, {Gavras},
  {Gerlach}, {Geyer}, {Giacobbe}, {Gilmore}, {Girona}, {Giuffrida}, {Gomel},
  {Gomez}, {Gonz{\'a}lez-N{\'u}{\~n}ez}, {Gonz{\'a}lez-Santamar{\'\i}a},
  {Gonz{\'a}lez-Vidal}, {Granvik}, {Guillout}, {Guiraud},
  {Guti{\'e}rrez-S{\'a}nchez}, {Guy}, {Hatzidimitriou}, {Hauser}, {Haywood},
  {Helmer}, {Helmi}, {Sarmiento}, {Hidalgo}, {Hilger}, {H{\l}adczuk}, {Hobbs},
  {Holland}, {Huckle}, {Jardine}, {Jasniewicz}, {Jean-Antoine Piccolo},
  {Jim{\'e}nez-Arranz}, {Jorissen}, {Juaristi Campillo}, {Julbe}, {Karbevska},
  {Kervella}, {Khanna}, {Kontizas}, {Kordopatis}, {Korn}, {K{\'o}sp{\'a}l},
  {Kostrzewa-Rutkowska}, {Kruszy{\'n}ska}, {Kun}, {Laizeau}, {Lambert},
  {Lanza}, {Lasne}, {Le Campion}, {Lebreton}, {Lebzelter}, {Leccia}, {Leclerc},
  {Lecoeur-Taibi}, {Liao}, {Licata}, {Lindstr{\o}m}, {Lister}, {Livanou},
  {Lobel}, {Lorca}, {Loup}, {Madrero Pardo}, {Magdaleno Romeo}, {Managau},
  {Mann}, {Manteiga}, {Marchant}, {Marconi}, {Marcos}, {Marcos Santos},
  {Mar{\'\i}n Pina}, {Marinoni}, {Marocco}, {Marshall}, {Polo},
  {Mart{\'\i}n-Fleitas}, {Marton}, {Mary}, {Masip}, {Massari},
  {Mastrobuono-Battisti}, {Mazeh}, {McMillan}, {Messina}, {Michalik}, {Millar},
  {Mints}, {Molina}, {Molinaro}, {Moln{\'a}r}, {Monari}, {Mongui{\'o}},
  {Montegriffo}, {Montero}, {Mor}, {Mora}, {Morbidelli}, {Morel}, {Morris},
  {Muraveva}, {Murphy}, {Musella}, {Nagy}, {Noval}, {Oca{\~n}a}, {Ogden},
  {Ordenovic}, {Osinde}, {Pagani}, {Pagano}, {Palaversa}, {Palicio},
  {Pallas-Quintela}, {Panahi}, {Payne-Wardenaar}, {Pe{\~n}alosa Esteller},
  {Penttil{\"a}}, {Pichon}, {Piersimoni}, {Pineau}, {Plachy}, {Plum}, {Poggio},
  {Pr{\v{s}}a}, {Pulone}, {Racero}, {Ragaini}, {Rainer}, {Raiteri}, {Rambaux},
  {Ramos}, {Ramos-Lerate}, {Re Fiorentin}, {Regibo}, {Richards}, {Rios Diaz},
  {Ripepi}, {Riva}, {Rix}, {Rixon}, {Robichon}, {Robin}, {Robin}, {Roelens},
  {Rogues}, {Rohrbasser}, {Romero-G{\'o}mez}, {Rowell}, {Royer}, {Ruz Mieres},
  {Rybicki}, {Sadowski}, {S{\'a}ez N{\'u}{\~n}ez}, {Sagrist{\`a} Sell{\'e}s},
  {Sahlmann}, {Salguero}, {Samaras}, {Sanchez Gimenez}, {Sanna},
  {Santove{\~n}a}, {Sarasso}, {Schultheis}, {Sciacca}, {Segol}, {Segovia},
  {S{\'e}gransan}, {Semeux}, {Shahaf}, {Siddiqui}, {Siebert}, {Siltala},
  {Silvelo}, {Slezak}, {Slezak}, {Smart}, {Snaith}, {Solano}, {Solitro},
  {Souami}, {Souchay}, {Spagna}, {Spina}, {Spoto}, {Steele},
  {Steidelm{\"u}ller}, {Stephenson}, {S{\"u}veges}, {Surdej}, {Szabados},
  {Szegedi-Elek}, {Taris}, {Taylo}, {Teixeira}, {Tolomei}, {Tonello}, {Torra},
  {Torra}, {Torralba Elipe}, {Trabucchi}, {Tsounis}, {Turon}, {Ulla}, {Unger},
  {Vaillant}, {van Dillen}, {van Reeven}, {Vanel}, {Vecchiato}, {Viala},
  {Vicente}, {Voutsinas}, {Weiler}, {Wevers}, {Wyrzykowski}, {Yoldas}, {Yvard},
  {Zhao}, {Zorec}, {Zucker}, \& {Zwitter}}]{gaiadr3}
{Gaia Collaboration}, {Vallenari}, A., {Brown}, A.~G.~A., {et~al.} 2022, arXiv
  e-prints, arXiv:2208.00211

\bibitem[{{Gilmore} {et~al.}(2007){Gilmore}, {Wilkinson}, {Wyse}, {Kleyna},
  {Koch}, {Evans}, \& {Grebel}}]{gilmore07}
{Gilmore}, G., {Wilkinson}, M.~I., {Wyse}, R.~F.~G., {et~al.} 2007, \apj, 663,
  948

\bibitem[{{Guy} {et~al.}(2022){Guy}, {Bailey}, {Kremin}, {Alam}, {Alexander},
  {Allende Prieto}, {BenZvi}, {Bolton}, {Brooks}, {Chaussidon}, {Cooper},
  {Dawson}, {de la Macorra}, {Dey}, {Dey}, {Dhungana}, {Eisenstein},
  {Font-Ribera}, {Forero-Romero}, {Gazta{\~n}aga}, {Gontcho}, {Green},
  {Honscheid}, {Ishak}, {Kehoe}, {Kirkby}, {Kisner}, {Koposov}, {Lan},
  {Landriau}, {Le Guillou}, {Levi}, {Magneville}, {Manser}, {Martini},
  {Meisner}, {Miquel}, {Moustakas}, {Myers}, {Newman}, {Nie},
  {Palanque-Delabrouille}, {Percival}, {Poppett}, {Prada}, {Raichoor},
  {Ravoux}, {Ross}, {Schlafly}, {Schlegel}, {Schubnell}, {Sharples},
  {Tarl{\'e}}, {Weaver}, {Y{\`e}che}, {Zhou}, {Zhou}, \& {Zou}}]{guy22}
{Guy}, J., {Bailey}, S., {Kremin}, A., {et~al.} 2022, arXiv e-prints,
  arXiv:2209.14482

\bibitem[{{Hargreaves} {et~al.}(1994{\natexlab{a}}){Hargreaves}, {Gilmore},
  {Irwin}, \& {Carter}}]{hargreaves94b}
{Hargreaves}, J.~C., {Gilmore}, G., {Irwin}, M.~J., \& {Carter}, D.
  1994{\natexlab{a}}, \mnras, 269, 957

\bibitem[{{Hargreaves} {et~al.}(1994{\natexlab{b}}){Hargreaves}, {Gilmore},
  {Irwin}, \& {Carter}}]{hargreaves94a}
---. 1994{\natexlab{b}}, \mnras, 271, 693

\bibitem[{{Hargreaves} {et~al.}(1996){Hargreaves}, {Gilmore}, {Irwin}, \&
  {Carter}}]{hargreaves96a}
---. 1996, \mnras, 282, 305

\bibitem[{{Harris}(1996)}]{harris96}
{Harris}, W.~E. 1996, \aj, 112, 1487

\bibitem[{{Helmi}(2020)}]{helmi20}
{Helmi}, A. 2020, \araa, 58, 205

\bibitem[{{Helmi} {et~al.}(2018){Helmi}, {Babusiaux}, {Koppelman}, {Massari},
  {Veljanoski}, \& {Brown}}]{helmi18b}
{Helmi}, A., {Babusiaux}, C., {Koppelman}, H.~H., {et~al.} 2018, \nat, 563, 85

\bibitem[{{Hinkle} {et~al.}(2000){Hinkle}, {Wallace}, {Valenti}, \&
  {Harmer}}]{hinkle00}
{Hinkle}, K., {Wallace}, L., {Valenti}, J., \& {Harmer}, D. 2000, {Visible and
  Near Infrared Atlas of the Arcturus Spectrum 3727-9300 A}

\bibitem[{{Hinkle} {et~al.}(2013){Hinkle}, {Wallace}, {Ram}, {Bernath},
  {Sneden}, \& {Lucatello}}]{hinkle13}
{Hinkle}, K.~H., {Wallace}, L., {Ram}, R.~S., {et~al.} 2013, \apjs, 207, 26

\bibitem[{{Holweger} \& {M\"{u}ller}(1974)}]{holweger74}
{Holweger}, H., \& {M\"{u}ller}, E.~A. 1974, \solphys, 39, 19

\bibitem[{{Horne}(1986)}]{horne86}
{Horne}, K. 1986, \pasp, 98, 609

\bibitem[{{Kirby} {et~al.}(2010){Kirby}, {Guhathakurta}, {Simon}, {Geha},
  {Rockosi}, {Sneden}, {Cohen}, {Sohn}, {Majewski}, \& {Siegel}}]{kirby10}
{Kirby}, E.~N., {Guhathakurta}, P., {Simon}, J.~D., {et~al.} 2010, \apjs, 191,
  352

\bibitem[{{Kleyna} {et~al.}(2002){Kleyna}, {Wilkinson}, {Evans}, {Gilmore}, \&
  {Frayn}}]{kleyna02}
{Kleyna}, J., {Wilkinson}, M.~I., {Evans}, N.~W., {Gilmore}, G., \& {Frayn}, C.
  2002, \mnras, 330, 792

\bibitem[{{Kleyna} {et~al.}(2005){Kleyna}, {Wilkinson}, {Evans}, \&
  {Gilmore}}]{kleyna05}
{Kleyna}, J.~T., {Wilkinson}, M.~I., {Evans}, N.~W., \& {Gilmore}, G. 2005,
  \apjl, 630, L141

\bibitem[{{Koch} {et~al.}(2006){Koch}, {Grebel}, {Wyse}, {Kleyna}, {Wilkinson},
  {Harbeck}, {Gilmore}, \& {Evans}}]{koch06}
{Koch}, A., {Grebel}, E.~K., {Wyse}, R.~F.~G., {et~al.} 2006, \aj, 131, 895

\bibitem[{{Koch} {et~al.}(2007{\natexlab{a}}){Koch}, {Kleyna}, {Wilkinson},
  {Grebel}, {Gilmore}, {Evans}, {Wyse}, \& {Harbeck}}]{koch07b}
{Koch}, A., {Kleyna}, J.~T., {Wilkinson}, M.~I., {et~al.} 2007{\natexlab{a}},
  \aj, 134, 566

\bibitem[{{Koch} {et~al.}(2007{\natexlab{b}}){Koch}, {Wilkinson}, {Kleyna},
  {Gilmore}, {Grebel}, {Mackey}, {Evans}, \& {Wyse}}]{koch07}
{Koch}, A., {Wilkinson}, M.~I., {Kleyna}, J.~T., {et~al.} 2007{\natexlab{b}},
  \apj, 657, 241

\bibitem[{{Koleva} {et~al.}(2009){Koleva}, {Prugniel}, {Bouchard}, \&
  {Wu}}]{koleva09}
{Koleva}, M., {Prugniel}, P., {Bouchard}, A., \& {Wu}, Y. 2009, \aap, 501, 1269

\bibitem[{{Koposov} {et~al.}(2011){Koposov}, {Gilmore}, {Walker}, {Belokurov},
  {Wyn Evans}, {Fellhauer}, {Gieren}, {Geisler}, {Monaco}, {Norris}, {Okamoto},
  {Pe{\~n}arrubia}, {Wilkinson}, {Wyse}, \& {Zucker}}]{koposov11}
{Koposov}, S.~E., {Gilmore}, G., {Walker}, M.~G., {et~al.} 2011, \apj, 736, 146

\bibitem[{{Koposov} {et~al.}(2018){Koposov}, {Walker}, {Belokurov}, {Casey},
  {Geringer-Sameth}, {Mackey}, {Da Costa}, {Erkal}, {Jethwa}, {Mateo},
  {Olszewski}, \& {Bailey}}]{koposov18}
{Koposov}, S.~E., {Walker}, M.~G., {Belokurov}, V., {et~al.} 2018, \mnras, 479,
  5343

\bibitem[{{Kurucz}(2011)}]{kurucz11}
{Kurucz}, R.~L. 2011, Canadian Journal of Physics, 89, 417

\bibitem[{{Kurucz} {et~al.}(1984){Kurucz}, {Furenlid}, {Brault}, \&
  {Testerman}}]{kurucz84}
{Kurucz}, R.~L., {Furenlid}, I., {Brault}, J., \& {Testerman}, L. 1984, {Solar
  flux atlas from 296 to 1300 nm}

\bibitem[{{Lawler} {et~al.}(2017){Lawler}, {Sneden}, {Cowan}, {Den Hartog}, \&
  {Wood}}]{lawler17review}
{Lawler}, J.~E., {Sneden}, C., {Cowan}, J.~J., {Den Hartog}, E.~A., \& {Wood},
  M.~P. 2017, Canadian Journal of Physics, 95, 783

\bibitem[{{Lawler} {et~al.}(2009){Lawler}, {Sneden}, {Cowan}, {Ivans}, \& {Den
  Hartog}}]{lawler09}
{Lawler}, J.~E., {Sneden}, C., {Cowan}, J.~J., {Ivans}, I.~I., \& {Den Hartog},
  E.~A. 2009, \apjs, 182, 51

\bibitem[{{Lee et al.}(2008)}]{lee08}
{Lee et al.} 2008, \aj, 136, 2050

\bibitem[{{Letarte} {et~al.}(2009){Letarte}, {Chapman}, {Collins}, {Ibata},
  {Irwin}, {Ferguson}, {Lewis}, {Martin}, {McConnachie}, \&
  {Tanvir}}]{letarte07}
{Letarte}, B., {Chapman}, S.~C., {Collins}, M., {et~al.} 2009, ArXiv e-prints,
  arXiv:0901.0820

\bibitem[{{Li} {et~al.}(2019){Li}, {Koposov}, {Zucker}, {Lewis}, {Kuehn},
  {Simpson}, {Ji}, {Shipp}, {Mao}, {Geha}, {Pace}, {Mackey}, {Allam}, {Tucker},
  {Da Costa}, {Erkal}, {Simon}, {Mould}, {Martell}, {Wan}, {De Silva},
  {Bechtol}, {Balbinot}, {Belokurov}, {Bland-Hawthorn}, {Casey}, {Cullinane},
  {Drlica-Wagner}, {Sharma}, {Vivas}, {Wechsler}, {Yanny}, \& {S5
  Collaboration}}]{li19}
{Li}, T.~S., {Koposov}, S.~E., {Zucker}, D.~B., {et~al.} 2019, \mnras, 490,
  3508

\bibitem[{{Lucchesi} {et~al.}(2020){Lucchesi}, {Lardo}, {Primas}, {Jablonka},
  {North}, {Battaglia}, {Starkenburg}, {Hill}, {Irwin}, {Francois}, {Shetrone},
  {Tolstoy}, \& {Venn}}]{lucchesi20}
{Lucchesi}, R., {Lardo}, C., {Primas}, F., {et~al.} 2020, \aap, 644, A75

\bibitem[{{Martin} {et~al.}(2008){Martin}, {de Jong}, \& {Rix}}]{martin08}
{Martin}, N.~F., {de Jong}, J.~T.~A., \& {Rix}, H.-W. 2008, \apj, 684, 1075

\bibitem[{{Martin} {et~al.}(2007){Martin}, {Ibata}, {Chapman}, {Irwin}, \&
  {Lewis}}]{martin07}
{Martin}, N.~F., {Ibata}, R.~A., {Chapman}, S.~C., {Irwin}, M., \& {Lewis},
  G.~F. 2007, \mnras, 380, 281

\bibitem[{{Masseron} {et~al.}(2014){Masseron}, {Plez}, {Van Eck}, {Colin},
  {Daoutidis}, {Godefroid}, {Coheur}, {Bernath}, {Jorissen}, \&
  {Christlieb}}]{masseron14}
{Masseron}, T., {Plez}, B., {Van Eck}, S., {et~al.} 2014, \aap, 571, A47

\bibitem[{{Mateo} {et~al.}(2012){Mateo}, {Bailey}, {Crane}, {Shectman},
  {Thompson}, {Roederer}, {Bigelow}, \& {Gunnels}}]{mateo12}
{Mateo}, M., {Bailey}, J.~I., {Crane}, J., {et~al.} 2012, in Society of
  Photo-Optical Instrumentation Engineers (SPIE) Conference Series, Vol. 8446,
  Ground-based and Airborne Instrumentation for Astronomy IV, ed. I.~S.
  {McLean}, S.~K. {Ramsay}, \& H.~{Takami}, 84464Y

\bibitem[{{Mateo} {et~al.}(1991){Mateo}, {Olszewski}, {Welch}, {Fischer}, \&
  {Kunkel}}]{mateo91}
{Mateo}, M., {Olszewski}, E., {Welch}, D.~L., {Fischer}, P., \& {Kunkel}, W.
  1991, \aj, 102, 914

\bibitem[{{Mateo} {et~al.}(1993){Mateo}, {Olszewski}, {Pryor}, {Welch}, \&
  {Fischer}}]{mateo93}
{Mateo}, M., {Olszewski}, E.~W., {Pryor}, C., {Welch}, D.~L., \& {Fischer}, P.
  1993, \aj, 105, 510

\bibitem[{{Mateo} {et~al.}(2008){Mateo}, {Olszewski}, \& {Walker}}]{mateo08}
{Mateo}, M., {Olszewski}, E.~W., \& {Walker}, M.~G. 2008, \apj, 675, 201

\bibitem[{{Mateo}(1998)}]{mateo98}
{Mateo}, M.~L. 1998, \araa, 36, 435

\bibitem[{{McConnachie}(2012)}]{mcconnachie12}
{McConnachie}, A.~W. 2012, \aj, 144, 4

\bibitem[{{McConnachie} {et~al.}(2009){McConnachie}, {Irwin}, {Ibata},
  {Dubinski}, {Widrow}, {Martin}, {C{\^o}t{\'e}}, {Dotter}, {Navarro},
  {Ferguson}, {Puzia}, {Lewis}, {Babul}, {Barmby}, {Bienaym{\'e}}, {Chapman},
  {Cockcroft}, {Collins}, {Fardal}, {Harris}, {Huxor}, {Mackey},
  {Pe{\~n}arrubia}, {Rich}, {Richer}, {Siebert}, {Tanvir}, {Valls-Gabaud}, \&
  {Venn}}]{mcconnachie09}
{McConnachie}, A.~W., {Irwin}, M.~J., {Ibata}, R.~A., {et~al.} 2009, \nat, 461,
  66

\bibitem[{{Morton}(2015)}]{isochrones}
{Morton}, T.~D. 2015, {isochrones: Stellar model grid package}, Astrophysics
  Source Code Library, record ascl:1503.010, ascl:1503.010

\bibitem[{{Mu{\~n}oz} {et~al.}(2006){Mu{\~n}oz}, {Carlin}, {Frinchaboy},
  {Nidever}, {Majewski}, \& {Patterson}}]{munoz06b}
{Mu{\~n}oz}, R.~R., {Carlin}, J.~L., {Frinchaboy}, P.~M., {et~al.} 2006, \apjl,
  650, L51

\bibitem[{{Naidu} {et~al.}(2020){Naidu}, {Conroy}, {Bonaca}, {Johnson}, {Ting},
  {Caldwell}, {Zaritsky}, \& {Cargile}}]{naidu20}
{Naidu}, R.~P., {Conroy}, C., {Bonaca}, A., {et~al.} 2020, \apj, 901, 48

\bibitem[{{Norris} {et~al.}(2010){Norris}, {Yong}, {Gilmore}, \&
  {Wyse}}]{norris10}
{Norris}, J.~E., {Yong}, D., {Gilmore}, G., \& {Wyse}, R. F.~G. 2010, \apj,
  711, 350

\bibitem[{{Olszewski} \& {Aaronson}(1985)}]{olszewski85}
{Olszewski}, E.~W., \& {Aaronson}, M. 1985, \aj, 90, 2221

\bibitem[{{Olszewski} {et~al.}(1995){Olszewski}, {Aaronson}, \&
  {Hill}}]{olszewski95}
{Olszewski}, E.~W., {Aaronson}, M., \& {Hill}, J.~M. 1995, \aj, 110, 2120

\bibitem[{{Pace} {et~al.}(2022){Pace}, {Erkal}, \& {Li}}]{pace22}
{Pace}, A.~B., {Erkal}, D., \& {Li}, T.~S. 2022, \apj, 940, 136

\bibitem[{{Pace} \& {Li}(2019)}]{pace19}
{Pace}, A.~B., \& {Li}, T.~S. 2019, \apj, 875, 77

\bibitem[{{Pace} {et~al.}(2021){Pace}, {Walker}, {Koposov}, {Caldwell},
  {Mateo}, {Olszewski}, {Bailey}, \& {Wang}}]{pace21}
{Pace}, A.~B., {Walker}, M.~G., {Koposov}, S.~E., {et~al.} 2021, \apj, 923, 77

\bibitem[{{Pace} {et~al.}(2023){Pace}, {Koposov}, {Walker}, {Caldwell},
  {Mateo}, {Olszewski}, {Roederer}, {Bailey}, {Belokurov}, {Kuehn}, {Li}, \&
  {Zucker}}]{pace23}
{Pace}, A.~B., {Koposov}, S.~E., {Walker}, M.~G., {et~al.} 2023, arXiv
  e-prints, arXiv:2304.06904

\bibitem[{{Pace} {et~al.}(2008){Pace}, {Pasquini}, \&
  {Fran{\c{c}}ois}}]{pace08}
{Pace}, G., {Pasquini}, L., \& {Fran{\c{c}}ois}, P. 2008, \aap, 489, 403

\bibitem[{{Palmer} \& {Engleman}(1983)}]{palmer83}
{Palmer}, B.~A., \& {Engleman}, R. 1983, {Atlas of the Thorium spectrum}

\bibitem[{{P{\^a}ris} {et~al.}(2014){P{\^a}ris}, {Petitjean}, {Aubourg},
  {Ross}, {Myers}, {Streblyanska}, {Bailey}, {Hall}, {Strauss}, {Anderson},
  {Bizyaev}, {Borde}, {Brinkmann}, {Bovy}, {Brandt}, {Brewington},
  {Brownstein}, {Cook}, {Ebelke}, {Fan}, {Filiz Ak}, {Finley}, {Font-Ribera},
  {Ge}, {Hamann}, {Ho}, {Jiang}, {Kinemuchi}, {Malanushenko}, {Malanushenko},
  {Marchante}, {McGreer}, {McMahon}, {Miralda-Escud{\'e}}, {Muna},
  {Noterdaeme}, {Oravetz}, {Palanque-Delabrouille}, {Pan}, {Perez-Fournon},
  {Pieri}, {Riffel}, {Schlegel}, {Schneider}, {Simmons}, {Viel}, {Weaver},
  {Wood-Vasey}, {Y{\`e}che}, \& {York}}]{paris14}
{P{\^a}ris}, I., {Petitjean}, P., {Aubourg}, {\'E}., {et~al.} 2014, \aap, 563,
  A54

\bibitem[{{Placco} {et~al.}(2021){Placco}, {Sneden}, {Roederer}, {Lawler}, {Den
  Hartog}, {Hejazi}, {Maas}, \& {Bernath}}]{placco21linemake}
{Placco}, V.~M., {Sneden}, C., {Roederer}, I.~U., {et~al.} 2021, Research Notes
  of the American Astronomical Society, 5, 92

\bibitem[{{Ram} {et~al.}(2014){Ram}, {Brooke}, {Bernath}, {Sneden}, \&
  {Lucatello}}]{ram14}
{Ram}, R.~S., {Brooke}, J.~S.~A., {Bernath}, P.~F., {Sneden}, C., \&
  {Lucatello}, S. 2014, \apjs, 211, 5

\bibitem[{{Ram{\'\i}rez} \& {Allende Prieto}(2011)}]{ramirez11}
{Ram{\'\i}rez}, I., \& {Allende Prieto}, C. 2011, \apj, 743, 135

\bibitem[{{Sadakane} {et~al.}(2004){Sadakane}, {Arimoto}, {Ikuta}, {Aoki},
  {Jablonka}, \& {Tajitsu}}]{sadakane04}
{Sadakane}, K., {Arimoto}, N., {Ikuta}, C., {et~al.} 2004, \pasj, 56, 1041

\bibitem[{{Sestito} {et~al.}(2023){Sestito}, {Roediger}, {Navarro}, {Jensen},
  {Venn}, {Smith}, {Hayes}, \& {McConnachie}}]{sestito23}
{Sestito}, F., {Roediger}, J., {Navarro}, J.~F., {et~al.} 2023, \mnras, 523,
  123

\bibitem[{{Shetrone} {et~al.}(2003){Shetrone}, {Venn}, {Tolstoy}, {Primas},
  {Hill}, \& {Kaufer}}]{shetrone03}
{Shetrone}, M., {Venn}, K.~A., {Tolstoy}, E., {et~al.} 2003, \aj, 125, 684

\bibitem[{{Shetrone} {et~al.}(2001){Shetrone}, {C{\^o}t{\'e}}, \&
  {Sargent}}]{shetrone01}
{Shetrone}, M.~D., {C{\^o}t{\'e}}, P., \& {Sargent}, W.~L.~W. 2001, \apj, 548,
  592

\bibitem[{{Simon}(2019)}]{simon19}
{Simon}, J.~D. 2019, \araa, 57, 375

\bibitem[{{Simon} \& {Geha}(2007)}]{simon07}
{Simon}, J.~D., \& {Geha}, M. 2007, \apj, 670, 313

\bibitem[{{Simon} {et~al.}(2015){Simon}, {Jacobson}, {Frebel}, {Thompson},
  {Adams}, \& {Shectman}}]{simon15}
{Simon}, J.~D., {Jacobson}, H.~R., {Frebel}, A., {et~al.} 2015, \apj, 802, 93

\bibitem[{{Skilling}(2004)}]{skilling04}
{Skilling}, J. 2004, in American Institute of Physics Conference Series, Vol.
  735, Bayesian Inference and Maximum Entropy Methods in Science and
  Engineering: 24th International Workshop on Bayesian Inference and Maximum
  Entropy Methods in Science and Engineering, ed. R.~{Fischer}, R.~{Preuss}, \&
  U.~V. {Toussaint}, 395--405

\bibitem[{{Sneden} {et~al.}(2014){Sneden}, {Lucatello}, {Ram}, {Brooke}, \&
  {Bernath}}]{sneden14}
{Sneden}, C., {Lucatello}, S., {Ram}, R.~S., {Brooke}, J.~S.~A., \& {Bernath},
  P. 2014, \apjs, 214, 26

\bibitem[{{Sneden}(1973)}]{sneden73}
{Sneden}, C.~A. 1973, PhD thesis, The University of Texas at Austin.

\bibitem[{{Sobeck} {et~al.}(2011){Sobeck}, {Kraft}, {Sneden}, {Preston},
  {Cowan}, {Smith}, {Thompson}, {Shectman}, \& {Burley}}]{sobeck11}
{Sobeck}, J.~S., {Kraft}, R.~P., {Sneden}, C., {et~al.} 2011, \aj, 141, 175

\bibitem[{{Song} {et~al.}(2021){Song}, {Mateo}, {Bailey}, {Walker}, {Roederer},
  {Olszewski}, {Reiter}, \& {Kremin}}]{song21}
{Song}, Y.-Y., {Mateo}, M., {Bailey}, J.~I., {et~al.} 2021, \mnras, 504, 4160

\bibitem[{{Song} {et~al.}(2019){Song}, {Mateo}, {Mackey}, {Olszewski},
  {Roederer}, {Walker}, \& {Bailey}}]{song19}
{Song}, Y.-Y., {Mateo}, M., {Mackey}, A.~D., {et~al.} 2019, \mnras, 490, 385

\bibitem[{{Spencer} {et~al.}(2018){Spencer}, {Mateo}, {Olszewski}, {Walker},
  {McConnachie}, \& {Kirby}}]{spencer18}
{Spencer}, M.~E., {Mateo}, M., {Olszewski}, E.~W., {et~al.} 2018, \aj, 156, 257

\bibitem[{{Spencer} {et~al.}(2017){Spencer}, {Mateo}, {Walker}, {Olszewski},
  {McConnachie}, {Kirby}, \& {Koch}}]{spencer17}
{Spencer}, M.~E., {Mateo}, M., {Walker}, M.~G., {et~al.} 2017, \aj, 153, 254

\bibitem[{{Starkenburg} {et~al.}(2013){Starkenburg}, {Hill}, {Tolstoy},
  {Fran{\c{c}}ois}, {Irwin}, {Boschman}, {Venn}, {de Boer}, {Lemasle},
  {Jablonka}, {Battaglia}, {Groot}, \& {Kaper}}]{starkenburg13}
{Starkenburg}, E., {Hill}, V., {Tolstoy}, E., {et~al.} 2013, \aap, 549, A88

\bibitem[{{Starkenburg} {et~al.}(2017){Starkenburg}, {Martin}, {Youakim},
  {Aguado}, {Allende Prieto}, {Arentsen}, {Bernard}, {Bonifacio}, {Caffau},
  {Carlberg}, {C{\^o}t{\'e}}, {Fouesneau}, {Fran{\c{c}}ois}, {Franke},
  {Gonz{\'a}lez Hern{\'a}ndez}, {Gwyn}, {Hill}, {Ibata}, {Jablonka},
  {Longeard}, {McConnachie}, {Navarro}, {S{\'a}nchez-Janssen}, {Tolstoy}, \&
  {Venn}}]{starkenburg17}
{Starkenburg}, E., {Martin}, N., {Youakim}, K., {et~al.} 2017, \mnras, 471,
  2587

\bibitem[{Swan(1857)}]{swan_1857}
Swan, W. 1857, Transactions of the Royal Society of Edinburgh, 21, 411–429

\bibitem[{{Szentgyorgyi}(2006)}]{szentgyorgyi06}
{Szentgyorgyi}, A. 2006, New Astronomy Review, 50, 326

\bibitem[{{Tafelmeyer} {et~al.}(2010){Tafelmeyer}, {Jablonka}, {Hill},
  {Shetrone}, {Tolstoy}, {Irwin}, {Battaglia}, {Helmi}, {Starkenburg}, {Venn},
  {Abel}, {Francois}, {Kaufer}, {North}, {Primas}, \&
  {Szeifert}}]{tafelmeyer10}
{Tafelmeyer}, M., {Jablonka}, P., {Hill}, V., {et~al.} 2010, ArXiv:1008.3721,
  arXiv:1008.3721

\bibitem[{{The DES Collaboration} {et~al.}(2015){The DES Collaboration},
  {Bechtol}, {Drlica-Wagner}, {Balbinot}, {Pieres}, {Simon}, {Yanny},
  {Santiago}, {Wechsler}, {Frieman}, {Walker}, {Williams}, {Rozo}, {Rykoff},
  {Queiroz}, {Luque}, {Benoit-Levy}, {Bernstein}, {Tucker}, {Sevilla},
  {Gruendl}, {da Costa}, {Fausti Neto}, {Maia}, {Abbott}, {Allam}, {Armstrong},
  {Bauer}, {Bernstein}, {Bertin}, {Brooks}, {Buckley-Geer}, {Burke}, {Carnero
  Rosell}, {Castander}, {D'Andrea}, {DePoy}, {Desai}, {Diehl}, {Eifler},
  {Estrada}, {Evrard}, {Fernandez}, {Finley}, {Flaugher}, {Gaztanaga},
  {Gerdes}, {Girardi}, {Gladders}, {Gruen}, {Gutierrez}, {Hao}, {Honscheid},
  {Jain}, {James}, {Kent}, {Kron}, {Kuehn}, {Kuropatkin}, {Lahav}, {Li}, {Lin},
  {Makler}, {March}, {Marshall}, {Martini}, {Merritt}, {Miller}, {Miquel},
  {Mohr}, {Neilsen}, {Nichol}, {Nord}, {Ogando}, {Peoples}, {Petravick},
  {Plazas}, {Romer}, {Roodman}, {Sako}, {Sanchez}, {Scarpine}, {Schubnell},
  {Smith}, {Soares-Santos}, {Sobreira}, {Suchyta}, {Swanson}, {Tarle},
  {Thaler}, {Thomas}, {Wester}, \& {Zuntz}}]{des15}
{The DES Collaboration}, {Bechtol}, K., {Drlica-Wagner}, A., {et~al.} 2015,
  ArXiv:1503.02584, arXiv:1503.02584

\bibitem[{{Tody}(1986)}]{iraf}
{Tody}, D. 1986, in Society of Photo-Optical Instrumentation Engineers (SPIE)
  Conference Series, Vol. 627, \procspie, ed. D.~L. {Crawford}, 733

\bibitem[{{Tolstoy} {et~al.}(2009){Tolstoy}, {Hill}, \& {Tosi}}]{tolstoy09}
{Tolstoy}, E., {Hill}, V., \& {Tosi}, M. 2009, \araa, 47, 371

\bibitem[{{Tolstoy} {et~al.}(2023){Tolstoy}, {Sk{\'u}lad{\'o}ttir},
  {Battaglia}, {Brown}, {Massari}, {Irwin}, {Starkenburg}, {Salvadori}, {Hill},
  {Jablonka}, {Salaris}, {van Essen}, {Olsthoorn}, {Helmi}, \&
  {Pritchard}}]{tolstoy23}
{Tolstoy}, E., {Sk{\'u}lad{\'o}ttir}, {\'A}., {Battaglia}, G., {et~al.} 2023,
  arXiv e-prints, arXiv:2304.11980

\bibitem[{{Tolstoy et al.}(2004)}]{tolstoy04}
{Tolstoy et al.} 2004, \apjl, 617, L119

\bibitem[{{Walker} {et~al.}(2015){Walker}, {Mateo}, {Olszewski}, {Bailey},
  {Koposov}, {Belokurov}, \& {Evans}}]{walker15b}
{Walker}, M.~G., {Mateo}, M., {Olszewski}, E.~W., {et~al.} 2015, \apj, 808, 108

\bibitem[{{Walker} {et~al.}(2007){Walker}, {Mateo}, {Olszewski}, {Bernstein},
  {Sen}, \& {Woodroofe}}]{walker07c}
---. 2007, \apjs, 171, 389

\bibitem[{{Walker} {et~al.}(2016){Walker}, {Mateo}, {Olszewski}, {Koposov},
  {Belokurov}, {Jethwa}, {Nidever}, {Bonnivard}, {Bailey}, {Bell}, \&
  {Loebman}}]{walker16}
---. 2016, \apj, 819, 53

\bibitem[{{Walker, Mateo \& Olszewski}(2009)}]{walker09a}
{Walker, Mateo \& Olszewski}. 2009, \aj, 137, 3100

\bibitem[{{Walker, Olszewski \& Mateo}(2015)}]{walker15}
{Walker, Olszewski \& Mateo}. 2015, (accepted for publication in MNRAS), 0, 0

\bibitem[{{Waller} {et~al.}(2023){Waller}, {Venn}, {Sestito}, {Jensen},
  {Kielty}, {Borukhovetskaya}, {Hayes}, {McConnachie}, \& {Navarro}}]{waller23}
{Waller}, F., {Venn}, K.~A., {Sestito}, F., {et~al.} 2023, \mnras, 519, 1349

\bibitem[{{Weisz} \& {Boylan-Kolchin}(2017)}]{weisz17}
{Weisz}, D.~R., \& {Boylan-Kolchin}, M. 2017, \mnras, 469, L83

\bibitem[{{Wilkinson} {et~al.}(2004){Wilkinson}, {Kleyna}, {Evans}, {Gilmore},
  {Irwin}, \& {Grebel}}]{wilkinson04}
{Wilkinson}, M.~I., {Kleyna}, J.~T., {Evans}, N.~W., {et~al.} 2004, \apjl, 611,
  L21

\bibitem[{{Willman} {et~al.}(2011){Willman}, {Geha}, {Strader}, {Strigari},
  {Simon}, {Kirby}, {Ho}, \& {Warres}}]{willman11}
{Willman}, B., {Geha}, M., {Strader}, J., {et~al.} 2011, \aj, 142, 128

\end{thebibliography}
